\documentclass[a4paper,11pt]{article}
\pdfoutput=1 

\usepackage{jheppub} 

\usepackage[T1]{fontenc} 
\usepackage[utf8]{inputenc}
\usepackage{amsmath,amssymb,amstext,amsfonts} 

\usepackage[dvipsnames]{xcolor}
\usepackage{graphicx}
\usepackage{subfig}
\usepackage{hyperref}
\usepackage{cancel} 
\usepackage{float}
\usepackage{bbm}
\usepackage{mathtools}
\usepackage{cleveref}
\usepackage{lscape}
\usepackage{siunitx}
\usepackage{slashed}
\usepackage{physics}

\makeatletter
\renewcommand*\env@matrix[1][\arraystretch]{%
  \edef\arraystretch{#1}%
  \hskip -\arraycolsep
  \let\@ifnextchar\new@ifnextchar
  \array{*\c@MaxMatrixCols c}}
\makeatother

\usepackage{tikz}
    \usetikzlibrary{positioning}
    \usetikzlibrary{arrows}
    \usetikzlibrary{shapes}
    \usepackage{pgfplots}
    \usetikzlibrary{calc}
    \usetikzlibrary{decorations.markings}
     \usetikzlibrary{decorations.pathmorphing}
    \tikzset{snake it/.style={decorate, decoration=snake}}
\def\centerarc[#1](#2)(#3:#4:#5) 
    { \draw[#1] ($(#2)+({#5*cos(#3)},{#5*sin(#3)})$) arc (#3:#4:#5); }
    \usepackage{booktabs}

\makeatletter
\g@addto@macro\bfseries{\boldmath}
\makeatother

\usepackage{caption}
\usepackage{tabularx}
\usepackage{array}
\usepackage{hhline}

\definecolor{cnblue}{RGB}{7,82,154}

\newcommand{\oF}{\mathcal{F}}
\newcommand{\mt}{m}

\allowdisplaybreaks[4]

\title{Analytic two-loop amplitudes for $q\bar{q}\to \gamma \gamma$ and 
$gg \to \gamma \gamma$ mediated by a heavy-quark loop}

\author[a]{Matteo Becchetti,}
\author[b,c]{Federico Coro,}
\author[d]{Christoph Nega,}
\author[d]{Lorenzo Tancredi,}
\author[d]{Fabian J. Wagner}
\affiliation[a]{Dipartimento di Fisica e Astronomia, Università di Bologna e INFN, Sezione di Bologna, via Irnerio 46, I-40126 Bologna, Italy}
\affiliation[b]{Department of Physics and Astronomy, Ghent University, 9000 Ghent, Belgium}
\affiliation[c]{Instituto de Fisica Corpuscular, Universitat de Valencia – Consejo Superior de Investigaciones Cientificas, Parc Cientific, E-46980 Paterna, Valencia, Spain}
\affiliation[d]{Technical University of Munich, TUM School of Natural Sciences, Physics Department, James-Franck-Straße 1, 85748 Garching, Germany}
\emailAdd{matteo.becchetti@unibo.it}
\emailAdd{federico.coro@ugent.be}
\emailAdd{c.nega@tum.de}
\emailAdd{lorenzo.tancredi@tum.de}
\emailAdd{fabianjohannes.wagner@tum.de}

\preprint{{\raggedleft
            TUM-HEP 1554/25  
}}

\abstract{We address the analytic computation of  
the two-loop scattering amplitudes for the production
of two photons in parton-parton scattering, mediated by loops of heavy quarks.
Due to the presence of integrals of elliptic type, both partonic channels have been previously computed using semi-numerical methods. In this paper, leveraging new advances in the theory of differential equations for elliptic Feynman integrals, we derive a canonical basis for all integrals involved and compute them in terms of independent iterated integrals over elliptic and polylogarithmic differential forms. We use this representation to showcase interesting cancellations in the physical expressions for the scattering amplitudes.
Furthermore, we address their numerical evaluation by producing series expansion representations for the whole amplitudes, which we demonstrate to be fast and numerically reliable across a large region of the phase space.
}

\begin{document}

\maketitle

\section{Introduction}
\label{sec:intro}

Scattering amplitudes for the production of two final state particles in partonic collisions play a crucial role in the study of particle interactions.
This is true both phenomenologically, where such amplitudes provide essential building blocks to model events observed at particle colliders, and formally, since starting at four points, their non-trivial dependence on the external kinematics provides important theoretical data for the study of general properties of perturbative Quantum Field Theory.

The complexity of scattering amplitudes strongly scales with the increase in the perturbative order (i.e., the number of virtual loops) and the number of different mass scales involved. More mass scales are generated when considering the production of more particles (as in $2 \to n$ scattering with $n \geq 2$), but also when one allows more massive virtual states to circulate in the loops (top quarks, electroweak vector bosons, etc.).
From this perspective, amplitudes for the production of two photons or two jets in massless QCD are among the simplest. 
In fact, due to momentum conservation, they depend only on one non-trivial dimensionless ratio. 
This class of amplitudes was computed up to second order in perturbation theory long ago~\cite{Anastasiou:2002zn, Bern:2001df}, recently reaching the three-loop order both for the production of two photons~\cite{Caola:2020dfu, Bargiela:2021wuy}, two jets~\cite{Caola:2021izf, Caola:2021rqz, Caola:2022dfa} as well as a photon and a jet~\cite{Bargiela:2022lxz}.
Thanks to the simplicity of the underlying kinematics and the absence of any massive particles in the loops, they can be expressed in terms of a class of very well-understood special functions, namely multiple polylogarithms~\cite{Kummer, Goncharov:1995, Goncharov:1998kja, Vollinga:2004sn} (MPLs). Even more impressively, the subclass of so-called Harmonic Polylogarithms~\cite{Remiddi:1999ew} (HPLs) only turns out to be sufficient for their analytical representation up to the three-loop order.

Modeling the production of two photons and two jets in hadron collisions to a precision consistent with the three-loop order requires many more ingredients in addition to the virtual amplitudes. 
First of all, defining infrared finite observables for any QCD process to N$^3$LO requires the inclusion of real and real-virtual amplitudes, i.e., amplitudes for the production of additional final state partons at a lower loop order. 
Recently, all these quantities have become available, including the last outstanding building blocks constituted by the two-loop amplitudes for the production of two photons and 
a jet~\cite{Agarwal:2021grm, Agarwal:2021vdh, Badger:2021imn} and three jets~\cite{DeLaurentis:2023izi, DeLaurentis:2023nss, Agarwal:2023suw}. While all amplitudes are there, devising a scheme to properly organize and cancel the infrared divergences between real and virtual contributions to N$^3$LO remains an outstanding task, particularly when the production of colored partons is considered.

In addition to the ingredients necessary for a complete N$^3$LO description of diphoton and dijet production, when working at this precision, more effects can potentially have a sizable impact on these observables. 
One such contribution is generated by heavy quarks (especially the top quarks) circulating in the loops. 
These mass effects have the potential to become sizable when one considers the production of photons and jets at high energies and transverse momenta, which are of the same order as the top quark mass. For diphoton production, in particular, these effects have been included up to the two-loop order in a series of recent calculations~\cite{Maltoni:2018zvp, Becchetti:2023wev, Becchetti:2023yat}. 
Interestingly, from a more formal point of view, allowing for the propagation of massive virtual states deeply 
changes the analytic properties of the underlying two-loop amplitudes. 
Not only do they become functions of two independent dimensionless ratios, but also a new type of special functions becomes relevant for their analytic calculation: iterated integrals over kernels related to an elliptic curve.
Integrals of this type have been known to appear in perturbative quantum field theory (QFT) 
for a long time, and the first place where they become relevant is arguably the calculation of the 
electron self-energy to two loops~\cite{Sabry}. The last two decades have led to the discovery of a large number 
of physically relevant scattering amplitudes involving not only these special functions, but also further 
higher-dimensional and higher-genus generalization thereof. This, in turn, has motivated intense work 
to understand their properties and devise methodologies for their calculations. 
Similar functions describe scattering amplitudes in string theory, and methods exported from this field 
and even from pure mathematics have provided the basis to generalize our knowledge of multiple polylogarithms 
to iterated integrals defined on elliptic geometries~\cite{Brown:2011wfj, Broedel:2014vla, Broedel:2017kkb, Broedel:2018qkq} and beyond. 

Despite the progress in understanding these functions analytically, the complexity of non-polylogarithmic amplitudes 
has also stimulated the development of powerful numerical and semi-numerical methods that aim to bypass the problem of having to deal with the specific properties of the geometries involved~\cite{Liu:2022chg, Moriello:2019yhu, Hidding:2020ytt, Prisco:2025wqs}. In fact, most of the results in the literature involving amplitudes of elliptic type have been obtained leveraging these numerical methods, including the recent studies for diphoton production quoted above~\cite{Maltoni:2018zvp,Becchetti:2023yat} and similar ones for the production of a Higgs boson and a jet~\cite{Bonciani:2022jmb}. While numerical methods are powerful, there remain strong reasons to study 
the structure of these amplitudes analytically, especially in view of the calculation of more complex 
$2 \to 3$ massive amplitudes whose large kinematical phase-space can constitute a roadblock for 
automated numerical methods. Moreover, understanding the analytic structure 
of scattering amplitudes can lead to significant simplifications at the amplitude level, which might be 
crucial for phenomenology applications. 
 
In recent years, growing evidence has been found of
large cancellations among complex combinations of iterated integrals of elliptic or even Calabi-Yau type, 
when physical quantities are considered.
These have been demonstrated explicitly for two-point correlators~\cite{Duhr:2024bzt, Forner:2024ojj, Marzucca:2025eak}, 
while hints of similar simplifications can be observed also in elliptic higher-point amplitudes~\cite{Abreu:2022cco,Delto:2023kqv}, suggesting that getting analytic control over 
physical quantities might be simpler than naively expected. 
A crucial tool in these investigations has been the differential equations method~\cite{Kotikov:1990kg, Remiddi:1997ny, Gehrmann:1999as}, which is based on the existence of integration-by-parts identities~\cite{Tkachov:1981wb, Chetyrkin:1981qh, Laporta:2000dsw} among Feynman integrals. Once augmented by the concept of a canonical basis of Feynman integrals~\cite{Arkani-Hamed:2010pyv, Henn:2013pwa}, differential equations allow, in fact, to expose the full analytic structure of the corresponding Feynman integrals and naturally express
them in terms of Chen iterated integrals~\cite{ChenSymbol} over integration kernels related to the relevant geometries.
While canonical bases are well understood only for the case of iterated integrals defined over dlog differential forms~\cite{Lee:2014ioa, Gehrmann:2014bfa, Meyer:2017joq, Dlapa:2020cwj, Lee:2020zfb, Henn:2020lye, Chen:2020uyk, Chen:2022lzr}, in the past two years a new point of view has been developed to generalize their construction, at least in principle, to
differential forms defined on elliptic curves and their higher-dimensional Calabi-Yau~\cite{Gorges:2023zgv} and higher-genus generalizations~\cite{Duhr:2024uid}. 
The existence of these bases is essential to demonstrate the simplifications hinted at above. Moreover, starting from differential equations in this form, one can easily produce series expansion results for the corresponding integrals, which in turn can be used 
for their precise and fast numerical evaluation across the whole kinematical space. 
Having control over the functional relations among these iterated integrals and being able to expand them in series to arbitrarily high orders in all relevant kinematical regions constitutes a potentially very powerful starting point for their fast and precise numerical evaluation.

As a matter of fact, very few examples have been worked out of non-polylogarithmic amplitudes with non-trivial dependence on more than one scale, with the recent calculation of Bhabha scattering~\cite{Delto:2023kqv} and the amplitudes for pseudo-scalar quarkonium production~\cite{Abreu:2022cco} among the few exceptions. It is, therefore, crucial to apply these new methodologies to more multi-scale problems, with the goal of both investigating the universality of the simplifications observed for two-point correlators and also to stress-test the limits of these methods in realistic examples and improving their practical implementation.
Diphoton and dijet amplitudes in QCD with a heavy top quark provide the perfect playground. They provide realistic scattering amplitudes, whose analytic structure is non-trivial due to the appearance of iterated integrals over mixtures of kernels of elliptic type and complicated dlog-forms with algebraic argument. At the same time, they remain simple enough to allow us to disentangle the problem of the analytic evaluation of these functions from 
the swell in algebraic complexity, which is typical of non-trivial amplitudes.

In this paper, we focus on all partonic channels for the production of a pair of photons through a loop of top quarks, which include $q\bar{q}\to \gamma \gamma$ and $gg \to \gamma \gamma$. The amplitudes for producing different combinations of massless partons and phtons will be considered elsewhere. While both channels have been computed with semi-numerical methods and used to do phenomenology, only the former are publicly available. The rest of this paper is organized as follows. In~\cref{sec:setup}, we establish the notation and provide our calculational set-up, 
including the tensor decomposition for the two partonic channels. 
In~\cref{sec:red}, we define the two-loop integral families used for the calculation and discuss the reduction to master integrals. 
The differential equations satisfied by these integrals are discussed in~\cref{sec:diffeq}, where we also elaborate on the elliptic curve relevant for some of these masters and on our construction of a canonical basis. 
In~\cref{sec:solution}, we discuss the solution for the integrals and for the amplitude in terms of iterated integrals, showcasing important cancellations of large groups of elliptic differential forms. 
We continue then in~\cref{sec:ren}, where we discuss UV renormalization and the subtraction of IR poles. 
Our main numerical result is then provided in~\cref{sec:exp}, where we explain how series expansion solutions for the amplitudes can be obtained from our differential equations. We focus in particular 
to the two cases of large-mass expansion, $\mt \to \infty$, and threshold expansion, i.e., $s \to 4 \mt^2$. 
This second point is interesting because the elliptic curve does not degenerate. As we will demonstrate, 
the convergence of these two series exceeds what one might naively expect based on the structure of the singularities of the differential equations. Finally, we conclude in~\cref{sec:conc}. Additional useful formulas are provided in the appendices.

\section{Notation and computational setup}
\label{sec:setup}
We consider the two-loop amplitudes for diphoton production through a heavy-quark loop
\begin{align} 
    q(p_1) + \bar{q}(p_2) \ &\longrightarrow \ \gamma(-p_3) + \gamma(-p_4)  \, , \nonumber \\
    g(p_1) + g(p_2) \ &\longrightarrow \ \gamma(-p_3) + \gamma(-p_4) \, . \label{eq:channels}
\end{align}
The usual Mandelstam variables describe the kinematics of the process
\begin{equation}
    s = (p_1+p_2)^2\,, \hspace{0.3cm} t = (p_1+p_3)^2\,, \hspace{0.3cm} u = (p_2+p_3)^2\,, \hspace{0.3cm} \text{with} \hspace{0.3cm} s+t+u = 0\,,
\end{equation}
where the external particles are on-shell, i.e. $p_i^2 = 0$.
In the physical scattering region, one has $s>0$ and $t<0,\; u<0$. The kinematical
constraints above imply that in the physical scattering region
\begin{equation}
0 \leq -t \leq s \, .
\end{equation}
We also use the symbol $\mt$ for the mass of the heavy quark. 
As it is well known, any amplitude with this kinematics will depend 
in a  non-trivial way only on two dimensionless ratios, which for definiteness, we choose here to be 
\begin{equation}
\label{eq:variabledefs}
    x = \frac{-t}{s}\,, \qquad y = \frac{m^2}{s} \,, \qquad \mbox{with} \quad 0 \leq x \leq 1\,.
\end{equation}
We will see later on that, depending on the region in the two-parameter phase space, it might be useful to parametrize the kinematics of the problem using different dimensionless ratios in order to disentangle the structure of the branch cuts.

To represent the scattering amplitudes, we work in 't Hooft-Veltman dimensional regularization scheme~\cite{tHooft:1972tcz}, where all internal momenta are continued to $d=4 - 2\epsilon$ space-time dimensions, while external states remain exactly four-dimensional. In this scheme, it was shown~\cite{Peraro:2019cjj, Peraro:2020sfm} that scattering amplitudes can be consistently decomposed using a basis of four-dimensional tensors. For the two channels, we write in particular\footnote{For simplicity, we omit color indices on the left side of~\cref{eq:ampqq,eq:ampgg}.}
 \begin{align}
 \mathcal{A}_{qq}(s,t) &= \delta^{kl} (4 \pi \alpha) \,\left[ \sum_{i=1}^4 \mathcal{F}_i(s,t)\, \bar{u}(p_2) \, \Gamma_i^{\mu \nu} \,u(p_1) \right] \epsilon_{3,\mu}(p_3) \epsilon_{4,\nu}(p_4) \,,
\label{eq:ampqq} \\
     \mathcal{A}_{gg}(s,t) &= 
 \delta^{a_1 a_2} (4 \pi \alpha) 
  \,\left[ \sum_{i=1}^8 \mathcal{G}_i(s,t)
 T^{\mu \nu \rho \sigma} \right] \epsilon_{1,\mu}(p_1) \epsilon_{2,\nu}(p_2)
\epsilon_{3,\rho}(p_3) \epsilon_{4,\sigma}(p_4) \,,
\label{eq:ampgg}
 \end{align}
 where $(k,l)$ are the color indices of the two initial state quarks represented by the spinors $u(p_1)$ and $\bar{u}(p_2)$, while $a_i$ is the color index of the gluon of momentum $p_i$ and $\epsilon_{i,\mu}(p_i)$ is its polarization vector. In addition,
 $\alpha$ is the fine structure constant, and we will indicate with $e_q$ the charge of the quark $q$ 
 in units of the electron charge $e = \sqrt{4 \pi \alpha}$.
 Following~\cite{Caola:2020dfu, Bargiela:2021wuy}, we  make a choice for the polarization vectors such that
 \begin{align}
     \epsilon_i \cdot p_i = 0 \,, \quad i=1,...,4\,,
 \end{align}
 and
 \begin{align}
   \left\{   \begin{array}{llc}
         \epsilon_3 \cdot p_2 &= \epsilon_4 \cdot p_1 = 0\,,  & \quad  \mbox{for} \ q\bar{q} \to \gamma \gamma  \\
        \epsilon_i \cdot p_{i+1} &= 0\,,  & \quad \mbox{for}  \ gg \to \gamma \gamma 
     \end{array} \right. \label{eq:gaugeeps}
 \end{align}
 with $p_5 = p_1$\,. 
With this choice, a convenient four-dimensional basis for the tensor structures relevant for \cref{eq:ampqq,eq:ampgg} is
\begin{align}
    \Gamma_1^{\mu \nu} &= \gamma^\mu p_2^{\nu}\,,  &&\Gamma_2^{\mu \nu} = \gamma^\nu p_1^{\mu}\,, \nonumber \\
    \Gamma_3^{\mu \nu} &=  \slashed{p}_3\,p_1^\mu p_2^\nu \,, 
    &&\Gamma_4^{\mu \nu} =  \slashed{p}_3\,g^{\mu \nu}\,, \label{eq:tensqq}\\[2ex]
    T_1^{\mu \nu \rho \sigma} &= p_3^{\mu}p_1^{\nu}p_1^{\rho}p_2^{\sigma}\,, 
    && T_2^{\mu \nu \rho \sigma} = p_3^{\mu}p_1^{\nu}g^{\rho\sigma}\,, \nonumber \\
    T_3^{\mu \nu \rho \sigma} &= p_3^{\mu}p_1^{\rho}g^{\nu\sigma} \,, 
    && T_4^{\mu \nu \rho \sigma} = p_3^{\mu}p_2^{\sigma}g^{\nu\rho}\,, \nonumber \\
    T_5^{\mu \nu \rho \sigma} &= p_1^{\nu}p_1^{\rho}g^{\mu\sigma} \,, 
    && T_6^{\mu \nu \rho \sigma} = p_1^{\nu}p_2^{\sigma}g^{\mu\rho}\,, \nonumber \\
    T_7^{\mu \nu \rho \sigma} &= p_1^{\rho}p_2^{\sigma}g^{\mu\nu} \,, 
    && T_8^{\mu \nu \rho \sigma} = g^{\mu\nu}g^{\rho\sigma}+g^{\mu\sigma}g^{\nu\rho}+g^{\mu\rho}g^{\nu\sigma}\,.
\label{eq:tensgg}
\end{align}

We further fix the helicities of the external states and define
a set of independent helicity amplitudes for the two processes as follows
\begin{align}
    A_{qq}^{\lambda_q \lambda_3 \lambda_4} &=  \delta^{kl} (4 \pi \alpha) \,\left[ \sum_{i=1}^4 \mathcal{F}_i(s,t)\, \bar{u}^{\lambda_q}(p_2) \, \Gamma_i^{\mu \nu} \,u^{\lambda_q}(p_1) \right] \epsilon^{\lambda_3}_{3,\mu}(p_3) \epsilon^{\lambda_4}_{4,\nu}(p_4) \, , \\
    A_{gg}^{\lambda_1 \lambda_2 \lambda_3 \lambda_4} &= \delta^{a_1 a_2} (4 \pi \alpha) 
  \,\left[ \sum_{i=1}^8 \mathcal{G}_i(s,t)
 T^{\mu \nu \rho \sigma} \right] \epsilon^{\lambda_1}_{1,\mu}(p_1) \epsilon^{\lambda_2}_{2,\nu}(p_2)
\epsilon^{\lambda_3}_{3,\rho}(p_3) \epsilon^{\lambda_4}_{4,\sigma}(p_4) \, ,
\end{align}
where $\lambda_q = \{L,R\}$ is the helicity along the massless fermion line while $\lambda_i = \pm$ is the helicity of the vector boson of momentum $p_i$\,.
We fix our conventions for the helicities by picking 
left-handed spinors as
\begin{equation}
    \bar{u}_L(p_2) = \langle 2 | \qquad \text{and} \qquad u_L(p_1) = | 1 ]\,,
\end{equation}
and the vector boson $j$ of momentum $p_j$ and gauge vector $q_j$ as
\begin{equation}
    \epsilon^\mu_{j,+}(q_j) = \frac{\langle q_j | \gamma^\mu | j ] }{ \sqrt{2} \langle q_j j \rangle} \qquad \text{and} \qquad \epsilon^\mu_{j,-}(q_j) = \frac{\langle j | \gamma^\mu | q_j ] }{ \sqrt{2} [ j q_j  ]}\,.
\end{equation}

With these, we find that the helicity amplitudes for the two processes can be written in a very compact form. Specifically, for the quark-induced channel, we write
\begin{equation}
\begin{aligned}
     A_{qq}^{L++} &= \frac{2[34]^2}{\langle 1 3 \rangle [23]} 
    \alpha(x,y) \,, \quad 
     &A_{qq}^{L+-} &= \frac{2\langle 24 \rangle [13]}{\langle 2 3 \rangle [24]} 
     \beta(x,y)\,, \\
     A_{qq}^{L-+} &= \frac{2\langle 23 \rangle [41]}{\langle 2 4 \rangle [32]} 
    \gamma(x,y) \,, \quad
     &A_{qq}^{L--} &= \frac{2\langle 34\rangle^2}{\langle 3 1 \rangle [23]} 
    \delta(x,y) \,.
    \label{eq:helampqq}
\end{aligned}
\end{equation}
with 
\begin{equation}
\begin{aligned}
\alpha(x,y) &=  \frac{t}{2}\,\left( \oF_2 - \frac{t}{2} \oF_3 +  \oF_4 \right) , \qquad
&\beta(x,y) &= \frac{t}{2}\, \left( \frac{s}{2} \oF_3  +  \oF_4 \right) ,
\\
\gamma(x,y) &=  \frac{s\,t}{2u}\left( \oF_2 - \oF_1  - \frac{t}{2} \oF_3 - \frac{t}{s}\oF_4 \right)\,, \qquad
&\delta(x,y) &=\frac{t}{2}\left( \oF_1 + \frac{t}{2} \oF_3 - \oF_4 \right)\,.
\label{eq:alphabetaqq}
\end{aligned}
\end{equation}
Instead, for the gluon fusion channel, we find
\begin{align}
 A_{gg}^{++++} &= \frac{[1 2][3 4]}{\langle1 2\rangle\langle3
    4\rangle} f_{++++}(x,y) \,, & %
    A_{gg}^{-+++} &=
  \frac{\langle1 2\rangle\langle1 4\rangle[2 4]}{\langle3
    4\rangle\langle2 3\rangle\langle2 4\rangle} f_{-+++}(x,y)\,,  \notag\\ 
  A_{gg}^{+-++} &= \frac{\langle2 1\rangle\langle2 4\rangle[1
      4]}{\langle3 4\rangle\langle1 3\rangle\langle1 4\rangle} f_{+-++}(x,y) \,, &
  A_{gg}^{++-+} &= \frac{\langle3 2\rangle\langle3 4\rangle[2
      4]}{\langle1 4\rangle\langle2 1\rangle\langle2 4\rangle} f_{++-+}(x,y) \,,  \notag\\
  A_{gg}^{+++-} &= \frac{\langle4 2\rangle\langle4 3\rangle[2
      3]}{\langle1 3\rangle\langle2 1\rangle\langle2 3\rangle} f_{+++-}(x,y) \,,&
  A_{gg}^{--++} &= \frac{\langle1 2\rangle[3 4]}{[1
      2]\langle3 4\rangle}\, f_{--++}(x,y) \,, \notag\\
  A_{gg}^{-+-+} &= \frac{\langle1 3\rangle[2 4]}{[1 3]\langle2
    4\rangle} f_{-+-+}(x,y) \,,&
    A_{gg}^{+--+} &= \frac{\langle2 3\rangle[1 4]}{[2 3]\langle1 4\rangle}\, f_{+--+}(x,y) \,,
\label{eq:helampgg}
\end{align}
with the helicity coefficients given by the following combinations of form factors
\begin{align}
 f_{++++}(x,y) &=  \frac{t^2}{4}\left(\frac{2\mathcal{G}_{6}}{u}-\frac{2\mathcal{G}_{3}}{s}-\mathcal{G}_{1}\right)+\mathcal{G}_{8}\left(\frac{s}{u}+\frac{u}{s}+4\right)+\frac{t}{2}(\mathcal{G}_{2}-\mathcal{G}_{4}+\mathcal{G}_{5}-\mathcal{G}_{7})\,, \notag\\ 
 f_{-+++}(x,y) &=  \,\,\,\, \frac{t^2}{4}\left(\frac{2\mathcal{G}_{3}}{s}+\mathcal{G}_{1}\right)+t\left(\frac{\mathcal{G}_{8}}{s}+\frac{1}{2}(\mathcal{G}_{4}+\mathcal{G}_{6}-\mathcal{G}_{2})\right)\,, \notag\\ 
 f_{+-++}(x,y) &=  -\frac{t^2}{4}\left(\frac{2\mathcal{G}_{6}}{u}-\mathcal{G}_{1}\right)+t\left(\frac{\mathcal{G}_{8}}{u}-\frac{1}{2}(\mathcal{G}_{2}+\mathcal{G}_{3}+\mathcal{G}_{5})\right)\,, \notag\\ 
 f_{++-+}(x,y) &= \,\,\,\, \frac{t^2}{4}\left(\frac{2\mathcal{G}_{3}}{s}+\mathcal{G}_{1}\right)+t\left(\frac{\mathcal{G}_{8}}{s}+\frac{1}{2}(\mathcal{G}_{6}+\mathcal{G}_{7}-\mathcal{G}_{5})\right)\,, \notag\\ 
 f_{+++-}(x,y) &=  -\frac{t^2}{4}\left(\frac{2\mathcal{G}_{6}}{u}-\mathcal{G}_{1}\right)+t\left(\frac{\mathcal{G}_{8}}{u}+\frac{1}{2}(\mathcal{G}_{4}+\mathcal{G}_{7}-\mathcal{G}_{3})\right)\,, \notag\\ 
 f_{--++}(x,y) &=  -\frac{t^2}{4}\mathcal{G}_{1}+\frac{1}{2}t(\mathcal{G}_{2}+\mathcal{G}_{3}-\mathcal{G}_{6}-\mathcal{G}_{7})+2\mathcal{G}_{8}\,, \notag\\ 
 f_{-+-+}(x,y) &=  t^2\left(\frac{\mathcal{G}_{8}}{su}-\frac{\mathcal{G}_{3}}{2s}+\frac{\mathcal{G}_{6}}{2u}-\frac{\mathcal{G}_{1}}{4}\right)\,, \notag\\ 
 f_{+--+}(x,y) &=  -\frac{t^2}{4}\mathcal{G}_{1}+\frac{1}{2}t(\mathcal{G}_{3}-\mathcal{G}_{4}+\mathcal{G}_{5}-\mathcal{G}_{6})+2\mathcal{G}_{8}
\,.
\label{eq:alphabetagg}
\end{align}

Clearly, not all these helicity amplitudes are independent.
First of all, the remaining four amplitudes for right-handed quarks and the eight missing ones for $gg \to \gamma \gamma$
can  be obtained from the  amplitudes provided in \cref{eq:helampqq} by a charge-conjugation and parity-invariance transformation, respectively
\begin{equation}
    \mathcal{A}_{qq}^{R \lambda_3 \lambda_4} = \mathcal{A}_{qq}^{L \lambda^*_3 \lambda^*_4}\left( \langle ij \rangle \leftrightarrow [ji] \right)\,,
\end{equation}
\begin{equation}
A_{gg}^{\lambda_1 \lambda_2 \lambda_3 \lambda_4} =
A_{gg}^{\lambda_1^*\lambda_2^*\lambda_3^*\lambda_4^*}\left(
\langle ij \rangle \leftrightarrow [ji] \right)\,,
\end{equation}
where $\lambda_i^*$ indicates the opposite helicity of $\lambda_i$, i.e., $\pm^* = \mp$.
Since we can swap the two photons and the amplitudes must be invariant, the helicity coefficients cannot be independent 
under the transformation $p_3 \leftrightarrow p_4$, which
is equivalent to $t \leftrightarrow u$ or $x \leftrightarrow 1-x$, where $x=-t/s$. 
By working out the transformation of the helicity amplitudes explicitly, one can easily verify that it implies 
\begin{equation}
\label{eq:bosesymmqq}
\gamma(x,y)=\beta(1-x,y)\,, \qquad \delta(x,y)=-\alpha(1-x,y)\,, \qquad \alpha(x,y) = -\alpha(1-x,y)\,,
\end{equation}
and
\begin{equation}
\begin{split}
f_{\lambda_2 \lambda_1 \lambda_3 \lambda_4}(x,y) = 
f_{\lambda_1\lambda_2 \lambda_3 \lambda_4}(1-x,y)\,, \qquad 
f_{\lambda_1 \lambda_2 \lambda_4 \lambda_3}(x,y) = 
f_{\lambda_1 \lambda_2 \lambda_3 \lambda_4}(1-x,y)\,.
\end{split}
\label{eq:bosesymmgg}
\end{equation}

The bare helicity coefficients so introduced can be expanded in the bare strong coupling constant $\alpha_{s,b}$. Up to two loops, we write
\begin{align}
    \Omega_{qq} &= \delta_{kl} (4 \pi \alpha)\,\sum_{\ell=0}^{2} \left(\frac{\alpha_{s,b}}{2\pi}\right)^\ell \Omega_{qq}^{(\ell,b)}\,, \\
    \Omega_{gg} &= \delta_{a_1 a_2} (4 \pi \alpha)\, \sum_{\ell=1}^{2} \left(\frac{\alpha_{s,b}}{2\pi}\right)^\ell\Omega_{gg}^{(\ell,b)}\,,
\end{align}
with $\Omega_{qq} = \{ \alpha, \beta, \gamma, \delta\}$ and 
 $\Omega_{gg} = \{ f_{++++},\cdots ,f_{+--+} \}$\, the helicity coefficients defined in \cref{eq:alphabetaqq,eq:alphabetagg}.
In the formulas above, $\ell$ represents the number of loops, and we used the fact that the gluon-fusion amplitudes are one-loop induced, and so for this channel, the sum starts with $\ell = 1$.
To explicitly compute the helicity coefficients $\Omega_{qq}^{(\ell,b)}$, $\Omega_{gg}^{(\ell,b)}$ we follow~\cite{Peraro:2020sfm} and define a suitable set of projector operators which act directly on the Feynman diagrams
representing the amplitudes to the corresponding loop order.
Explicitly, we define projector operators for the quark and gluon channels as
\begin{align}
    &\mathcal{P}_{qq}^{(i)} = \sum_{j=1}^4 c_{ij}\, \bar{u}(p_1) \Gamma_j^{\mu \nu} u(p_2) \epsilon^*_{3, \mu}(p_3)\epsilon^*_{4, \nu}(p_4)\,,\\
    &\mathcal{P}_{gg}^{(i)} = \sum_{j=1}^8 d_{ij}\, T_j^{\mu \nu \rho \sigma} \epsilon^*_{1, \mu}(p_3)\epsilon^*_{2, \nu}(p_4)\epsilon^*_{3, \rho}(p_3)\epsilon^*_{4, \sigma}(p_4)
\end{align}
with the $\Gamma_j^{\mu \nu}$ and $T_j^{\mu \nu \rho \sigma}$ defined in \cref{eq:tensqq,eq:tensgg}.
The projectors are constructed such that
\begin{align}
    &P_{qq}^{(i)} \cdot \mathcal{A}_{qq} \coloneqq \sum_{pol} P_{qq}^{(i)} \,\mathcal{A}_{qq} = \delta^{kl} (4 \pi \alpha) e_q^2\, \mathcal{F}_i \,,\\
    &P_{gg}^{(i)} \cdot \mathcal{A}_{gg} \coloneqq \sum_{pol} P_{qq}^{(i)} \,\mathcal{A}_{gg} = \delta^{a_1 a_2} (4 \pi \alpha) \mathcal{G}_i\,,
\end{align}
where their action, represented by the dot operator ``$\cdot$'', is realized explicitly by summing over the polarization of all external particles (i.e., quarks, gluons, and photons). The polarization sums have to be performed consistently with the choice made in \cref{eq:gaugeeps} for the polarization vectors.
The explicit form of the coefficients $c_{ij}$ and $d_{ij}$ to define the projectors can be found in~\cite{Peraro:2020sfm}
and we do not report them here for simplicity. By suitably combining these projectors, we can obtain helicity projectors that single out directly the helicity coefficients $\Omega_{qq}$ and $\Omega_{gg}$. We provide explicit expressions for our helicity projectors in~\cref{app:helproj}.

\section{Integral families and reduction to master integrals}
\label{sec:red}
\begin{figure}[h!]
    \centering
    \subfloat[PLA]{\includegraphics[width=0.3\textwidth]{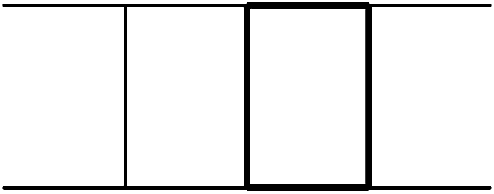}\label{gr:PLA}}
    \hspace{0.03\textwidth}
    \subfloat[PLB]{\includegraphics[width=0.3\textwidth]{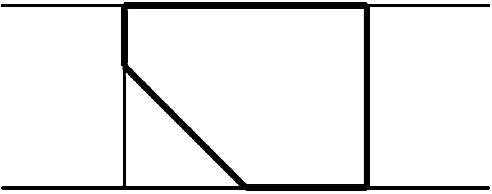}\label{gr:PLB}} 
    \hspace{0.03\textwidth}
    \subfloat[PLC]{\includegraphics[width=0.3\textwidth]{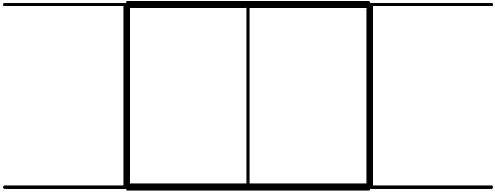}\label{gr:PLC}}
    \\
    
    \subfloat[NPA]{\includegraphics[width=0.3\textwidth]{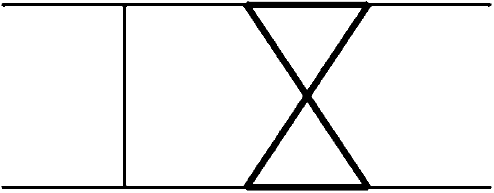}\label{gr:NPA}}
    \hspace{0.03\textwidth}
    \subfloat[NPB]{\includegraphics[width=0.3\textwidth]{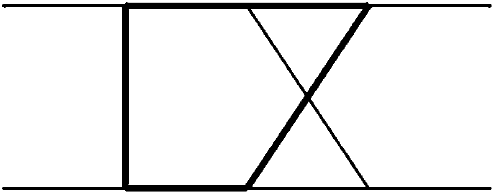}\label{gr:NPB}}
    
    \caption{Representative set of two-loop graphs with internal heavy-quark loops. Thick and thin lines represent massive and massless propagators, respectively.}
    \label{fig:gg_aa_diagrams}
\end{figure}

To compute the scattering amplitudes, we generate all Feynman diagrams that contribute to the two partonic channels in \cref{eq:channels} with \texttt{QGRAF}~\cite{Nogueira:1991ex}.
Here, we focus on the calculation of those diagrams that contain (at least) 
one heavy fermion loop. At two loops, there are 14 such diagrams for $q\bar{q} \to \gamma \gamma$ and 166 for $gg \to \gamma \gamma$.
We use \texttt{FORM}~\cite{Vermaseren:2000nd, Kuipers:2012rf, Ruijl:2017dtg} to apply the projector operators on each diagram and perform color and Dirac algebra. After these manipulations, all two-loop helicity coefficients can be expressed as linear combinations of scalar Feynman integrals. These integrals can be categorized into five integral topologies, which we define as follows
\begin{equation}
\mathcal{I}_{\texttt{topo}}(n_1,...,n_9) = 
\int \prod_{\ell=1}^2 \left[\frac{\mu_0^{2\epsilon}}{ C_\epsilon}\frac{\mathrm d^d k_\ell}{(2 \pi)^d}\right]
\frac{1}{D_{1}^{n_1} \cdots D_{9}^{n_9}}\,,  \label{eq:measure}
\end{equation}
where $\mu_0$ is the dimensional regularization scale,
\begin{equation}
C_\epsilon = i\,(4 \pi)^\epsilon \Gamma(1+\epsilon)\,,\label{eq:norm}
\end{equation}
and
\texttt{topo} labels the different integral families. The corresponding propagators $D_1,\hdots,D_9$ are given in~\cref{fig:toposPL} for the planar families and in~\cref{fig:toposNP} for the non-planar ones.
\begin{table}[H]
\begin{center}
\begin{tabular}{| m{2.1cm} || m{3.7cm} | m{3.7cm} | m{3.7cm} |} 
 \hline
 Denominator & Integral family PLA & Integral family PLB & Integral family PLC \\[5pt]
 \hline
$D_1$ & $k_1^2$ & $k_1^2$ & $k_1^2 - \mt^2$ \\
$D_2$ & $k_2^2 - \mt^2$ & $k_2^2 - \mt^2$ & $k_2^2 - \mt^2$ \\
$D_3$ & $(k_1-k_2)^2 - \mt^2$ & $(k_1-k_2)^2 - \mt^2$ & $(k_1-k_2)^2$ \\
$D_4$ & $(k_1-p_1)^2$ & $(k_1-p_1)^2$ & $(k_1-p_1)^2 - \mt^2$ \\
$D_5$ & $(k_2-p_1)^2$ & $(k_2-p_1)^2 - \mt^2$ & $(k_2-p_1)^2 - \mt^2$ \\
$D_6$ & $(k_1-p_1-p_2)^2$ & $(k_1-p_1-p_2)^2$ & $(k_1-p_1-p_2)^2 - \mt^2$ \\
$D_7$ & $(k_2-p_1-p_2)^2 - \mt^2$ & $(k_2-p_1-p_2)^2 - \mt^2$ & $(k_2-p_1-p_2)^2 - \mt^2$ \\
$D_8$ & $(k_1-p_1-p_2-p_3)^2$ & $(k_1-p_1-p_2-p_3)^2$ & $(k_1-p_1-p_2-p_3)^2 - \mt^2$ \\
$D_9$ & $(k_2-p_1-p_2-p_3)^2 - \mt^2$ & $(k_2-p_1-p_2-p_3)^2 - \mt^2$ & $(k_2-p_1-p_2-p_3)^2 - \mt^2$ \\
 \hline
\end{tabular}
\caption{\label{fig:toposPL} Routing definition for the three planar scalar integrals families PLA, PLB and PLC.}
\end{center}
\end{table}
\begin{table}[H]
\begin{center}
\begin{tabular}{| m{2.1cm} || m{4.5cm} | m{4.5cm} |} 
 \hline
 Denominator & Integral family NPA & Integral family NPB \\[5pt]
 \hline
$D_1$ & $k_1^2$ & $k_1^2 - \mt^2$  \\
$D_2$ & $k_2^2 - \mt^2$ & $k_2^2$  \\
$D_3$ & $(k_1-k_2)^2 - \mt^2$ & $(k_1-k_2)^2 - \mt^2$  \\
$D_4$ & $(k_1-p_1)^2$ & $(k_1-p_1)^2 - \mt^2$  \\
$D_5$ & $(k_2-p_1)^2$ & $(k_2-p_1)^2$  \\
$D_6$ & $(k_1-p_1-p_2)^2$ & $(k_1-p_1-p_2)^2 - \mt^2$  \\
$D_7$ & $(k_1-k_2+p_3)^2 - \mt^2$ & $(k_1-k_2+p_3)^2 - \mt^2$ \\
$D_8$ & $(k_2-p_1-p_2-p_3)^2 - \mt^2$ & $(k_2-p_1-p_2-p_3)^2$ \\
$D_9$ & $(k_1-k_2-p_1-p_2)^2$ & $(k_1-k_2-p_1-p_2)^2$ \\
 \hline
\end{tabular}
\caption{\label{fig:toposNP} Routing definition for the two non-planar scalar integrals families NPA and NPB.}
\end{center}
\end{table}
In \cref{fig:gg_aa_diagrams}, we provide representative graphs for the relevant top sectors. In addition to the five previous integral families, we also have to consider their permutation with respect to the exchange of the external photons ($p_3\leftrightarrow p_4$) to map all Feynman integrals that contribute to the helicity coefficients. If we allow for crossings under any of the external momenta, these five families are even sufficient to map all Feynman diagrams that contribute to dijet production. 

All integrals in the families above can be reduced to a minimal subset of master integrals using integration-by-parts identities~\cite{Tkachov:1981wb, Chetyrkin:1981qh}. We use, in particular, the publicly available implementation of the Laporta algorithm~\cite{Laporta:2000dsw} provided in \texttt{Reduze2}~\cite{Studerus:2009ye,vonManteuffel:2012np} and \texttt{KIRA2}~\cite{Maierhofer:2017gsa, Klappert:2019emp, Klappert:2020nbg, Klappert:2020aqs}. 
For diphoton production, we find a total of $165$ master integrals, which we choose to arrange in the five families above and all their crossings\footnote{Note that this counting includes all relevant crossings of the master integrals.}. Interestingly, while family PLB is useful for matching all Feynman diagrams, after reduction, it contains no extra master integrals, so we ignore it from here on.
As we will see below, there are actually 11 extra relations among these integrals, which we can prove to be valid to all orders in $\epsilon$ from the differential equations they satisfy, reducing the number of independent master integrals to 154. We will describe these relations in the next section. 
The complete list of master integrals in a convenient basis can be found in~\cref{app:masters}.

To compute these master integrals, we resort to the differential equations method.
To avoid having to discuss how to cross the various integrals, we prefer to consider all master integrals, including their relevant crossings, at once. In the next section, we describe how to derive a set of differential equations for these integrals and
how to put the system in an $\epsilon$-factorized form, which will make the analytic properties of the scattering amplitudes manifest.
\section{Differential equations and $\epsilon$-factorized basis}
\label{sec:diffeq}

It is convenient to introduce the vector notation for the master integrals 
$\vec{\mathcal{J}} = \{\mathcal{J}_1,...,\mathcal{J}_{165} \}$, where for simplicity we left the dependence on the kinematics and on the dimensional regulator implicit, i.e., $\mathcal{J}_i = \mathcal{J}_i(x,y;\epsilon)$ for $i=1,...,165$. For now, we consider the full set of $165$ integrals,
even if, as hinted to above, $11$ extra relations can be discovered a posteriori by studying their system of differential equations.

More explicitly, by leveraging integration-by-parts identities, we can easily derive two systems of differential equations for the master integrals in the two dimensionless ratios $x,y$, which take the general form 
\begin{align}
    \frac{\partial}{\partial x} \vec{\mathcal{J}} = A_x(x,y; \epsilon) \vec{\mathcal{J}}
    \,, \qquad 
    \frac{\partial}{\partial y} \vec{\mathcal{J}} = A_y(x,y; \epsilon) \vec{\mathcal{J}}\,,
\end{align}
where the entries of the two matrices $A_x(x,y; \epsilon)$ and $A_y(x,y; \epsilon)$ are rational functions in $x,y$ and the dimensional regulator $\epsilon$.

We are interested in solving these equations as a Laurent series in $\epsilon$.
In principle, this requires first solving all coupled homogeneous blocks in the limit $\epsilon \to 0$ and then obtaining an inhomogeneous solution using Euler's variation of constants. This should then be iterated order by order in $\epsilon$.
While this procedure is, in principle, always possible, 
for a generic choice of master integrals the dependence on $\epsilon$ of the matrices $A_x(x,y;\epsilon)$ and $A_y(x,y;\epsilon)$ can be arbitrarily complicated, which
makes it very difficult to read off the analytic properties of the result  directly from 
the differential equations. For this reason, it is often convenient to derive a so-called $\epsilon$-factorized basis, i.e., one attempts to find a rotation of the basis of master integrals such that all dependence on the dimensional regulator $\epsilon$ factorizes, i.e.,
\begin{align}
    \mathcal{I} = T(x,y; \epsilon) \mathcal{J} \qquad \to \qquad 
      \frac{\partial}{\partial x} \vec{\mathcal{I}} = \epsilon \, B_x(x,y) \vec{\mathcal{I}}
    \,, \qquad 
    \frac{\partial}{\partial y} \vec{\mathcal{I}} = \epsilon\, B_y(x,y) \vec{\mathcal{I}}\,,
\end{align}
with
\begin{equation}
    \epsilon\, B_z(x,y) = T A T^{-1} + \frac{\partial T}{\partial z} T^{-1} \qquad \mbox{and}
\qquad z \in\{x,y\}\,.
\end{equation}
Using the formalism of differential forms, the equations can then be written  as
\begin{align} \label{eq:deqs}
    \mathrm d \mathcal{I} = \epsilon \, \mathrm dB(x,y) \, \mathcal{I}\,, \qquad \mbox{with} \qquad
    \mathrm dB = B_x \mathrm dx + B_y \mathrm dy = \frac{\partial B}{\partial x} \mathrm dx + \frac{\partial B}{\partial y} \mathrm dy\,,
\end{align}
such that their solutions can be written formally as a path-ordered exponential
\begin{align}
    \mathcal{I}(x,y) = \mathbb{P}  \exp\left[ \epsilon \int_\gamma \mathrm dB \right]   \mathcal{I}_0\,.
    \label{eq:cansol}
\end{align}
In \cref{eq:cansol}, $\mathcal{I}_0$ is a suitably chosen boundary condition and $\gamma$ is a one-dimensional path connecting the boundary point to the generic point $(x,y)$.
Writing 
\begin{align}
    \mathrm dB = \sum_{i=1}^N G_i\, \omega_i
\end{align}
with $G_i$ numerical matrices and $\omega_i$  differential forms, it is then obvious that upon expanding in $\epsilon$, the solution of \cref{eq:deqs} can be written in terms of
Chen iterated integrals defined over the differential forms $\omega_i$
\begin{align}
    I(\omega_n,...,\omega_1; \gamma) &\coloneqq 
    \int_\gamma \omega_1 \cdots \omega_n\,\nonumber \\
    &= \int_{t_0}^t f_n(t_n) \mathrm dt_n \int_{t_0}^{t_n} f_{n-1}(t_{n-1}) \mathrm dt_{n-1}\cdots  \int_{t_0}^{t_2} f_{1}(t_{1}) \mathrm dt_1\,,
    \label{eq:iterints}
\end{align}
where $\gamma(t_0)$ is the boundary point, and we defined the pull-back of the forms on the curve $\gamma$ as $\gamma^* \omega_i = f_i(t)\mathrm dt$.
An important special case is constituted by those classes of Feynman integrals for which all $\omega_i$ are logarithmic differential forms  (so-called dlog-forms) of algebraic arguments. 
If the arguments are all rational functions, the ensuing class of iterated integral solutions can be written explicitly
in terms of Multiple Polylogarithms.

Clearly, the problem of finding an $\epsilon$-factorized system of differential equations is intimately related to being able to solve all coupled homogeneous blocks in the original system at $\epsilon=0$. 
As it turns out, if the master integrals can be written in terms of iterated integrals of dlog-forms, 
all homogeneous solutions are, at most, algebraic functions that satisfy simple first-order linear differential equations with rational coefficients. 
On the other hand, Feynman integrals, whose homogeneous solutions are related to more complicated mathematical objects, have been known in the literature for a long time. Most notable are periods of elliptic curves and their higher-genus and higher-dimensional (Calabi-Yau) generalizations.
Studying the homogeneous coupled system of differential equations provides, therefore, important information on the geometry relevant to the problem considered.
This, in turn, can be achieved by considering the independent integration contours associated with the maximal cut of the corresponding Feynman integrals~\cite{Primo:2016ebd, Primo:2017ipr}, which can be most easily evaluated in the so-called Baikov representation~\cite{Baikov:1996iu, Baikov:1996rk, Frellesvig:2017aai}.
In the case under study, it has been known for some time that one encounters at least one elliptic curve (and, in the general case of dijet production, also its crossings) in the two-loop non-planar graph in~\cref{gr:NPA}, see~\cite{Maltoni:2018zvp,Becchetti:2023wev,Calisto:2023vmm}. In particular, an elliptic curve was first recognized in a specific three-point subgraph~\cite{Czakon:2008ii, vonManteuffel:2017hms}, and more recently, it was shown that the very same elliptic curve also appears in the top graph~\cite{Ahmed:2024tsg}. In the same
reference, also an $\epsilon$-factorized basis for the homogeneous differential equations satisfied by the 
top-graph in~\cref{gr:NPA} was identified.

Despite these results, constructing a canonical $\epsilon$-factorized basis for the full problem remained an outstanding problem. In fact, in this case, one encounters a non-trivial mixing of polylogarithmic integrals depending on multiple algebraic letters, together with non-polylogarithmic Feynman integrals 
of elliptic type. 
To construct an $\epsilon$-factorized basis for this problem,
we rely on the algorithm recently proposed in~\cite{Gorges:2023zgv}. 
The properties of our basis also allow us to argue that this is the right generalization of 
a canonical basis to the elliptic case. We will then use this basis, on the one hand, 
to produce analytic results for the amplitudes in terms of independent iterated integrals, 
studying the patterns of cancellations in the final result, and on the other to produce 
arbitrarily precise series expansion results for its fast numerical evaluation.

\subsection{The elliptic Feynman integrals}
We start by describing some basic properties of the elliptic geometry encountered in this problem and how also, in that case, we can rely on an integrand analysis to determine a good starting basis for the application of the algorithm in~\cite{Gorges:2023zgv}. There are two graphs that are individually elliptic on their maximal cut.
They both belong to integral family NPA, see~\cref{fig:toposNP}, and correspond, in particular, to the two sectors identified by the following corner integrals: $\mathcal{I}_{\text{NPA}} (1,1,1,0,0,1,1,1,0)$, $\mathcal{I}_\text{NPA} (1,1,1,1,0,1,1,1,0)$.
As we will demonstrate below, both integrals can be related to the same elliptic curve defined by the algebraic equation
\begin{equation}
    Y^2 = P_4(X)\,, \quad P_4(X) = (\mt^2-X) (\mt^2+s-X) \left(\mt^2 (\mt^2-3 s)-X (2 \mt^2+s)+X^2\right)\,. \label{eq:ellcurve}
\end{equation}
As it is well known, an elliptic curve is characterized by its period matrix. Two out of the four periods are given by the integrals
\begin{align}
    \pi_1 (s,\mt^2) \coloneqq \int_{\mathcal{C}_1} \frac{\mathrm{d}X}{\sqrt{P_4(X)}} \, , \qquad
    \pi_2 (s,\mt^2) \coloneqq \int_{\mathcal{C}_2} \frac{\mathrm{d}X}{\sqrt{P_4(X)}} \,,
\end{align}
where one has integrated the differential form of the first kind, $\mathrm dX/Y$, over the two independent integration contours, $\mathcal{C}_1$ and $\mathcal{C}_2$.
These periods are also the two independent solutions of the associated second-order Picard-Fuchs operator. In what follows, we will not need the specific choice of the contours since we will never use these periods. Instead, both for the construction of a canonical basis and for the numerical evaluation of the resulting integrals, it is more useful to consider the \emph{local} solution to the associated Picard-Fuchs operator, close to each regular singular point of the elliptic curve. Close to each such point, one can define an holomorphic solution and one which contains a single power of a logarithm that diverges as the singularity is approached.\footnote{This is strictly true only close to a point of maximal unipotent monodromy (MUM point). In the elliptic case, each regular singular point is a MUM point and so we will not need this distinction here.} This specific choice of periods is usually referred to as a Frobenius basis. In what follows, we will use the notation $\varpi^{[z]}_0(s,\mt^2)$ for the holomorphic and $\varpi^{[z]}_1(s,\mt^2)$ for the logarithmic solution close to the singular point $s=z \, \mt^2$, respectively.

The curve above has three regular singular points located at
$s=\{-16m^2,0,\infty \}$.
First, let us consider the point $s=0$. The holomorphic solution $\varpi^{[0]}_0(s,\mt^2)$ and the one that diverges logarithmically $\varpi^{[0]}_1(s,\mt^2)$ can be defined as follows 
\begin{align}
    \varpi^{[0]}_0 (s,\mt^2) &= \frac{1}{\sqrt{s\,m^2}} \left( 1 - \frac{s}{64m^2}+\frac{9 s^2}{16384 m^4} -\frac{25 s^3}{1048576 m^6} + \mathcal{O}\left(\frac{s}{m^2}\right)^4\right) \,, \label{eq:varpi0} \\
    \varpi^{[0]}_1 (s,\mt^2) &= \varpi_0 \log{\left(\frac{s}{m^2} \right)}
    \nonumber \\ &+ \frac{1}{\sqrt{s \, m^2}} \left(- \frac{s}{32m^2}
    +\frac{21 s^2}{16384 m^4} 
    -\frac{185 s^3}{3145728 m^6} 
    + \mathcal{O}\left(\frac{s}{m^2}\right)^4\right)\,, \label{eq:varpi1} 
\end{align}
where in writing down all formulas we assumed $s\to 0^+$ and $m^2>0$. Similarly, the holomorphic and the logarithmically divergent solutions for the other two singular points can be chosen as 
\begin{align}
    \varpi^{[\infty]}_0 (s,\mt^2) &= \frac{1}{s} \left( 1 - \frac{4m^2}{s}+\frac{36 m^4}{s^2} -\frac{400 m^6}{s^3} + \mathcal{O}\left(\frac{m^2}{s}\right)^4\right) \,,  \label{eq:varpi0_inf} \\
    \varpi^{[\infty]}_1 (s,\mt^2) &= \varpi^{[\infty]}_0 (s,\mt^2) \log{\left(\frac{m^2}{s} \right)}
    \nonumber \\ &+ \frac{1}{s} \left(- \frac{8m^2}{s}
    +\frac{84 m^4}{s^2} 
    -\frac{2960 m^6}{3 s^3} 
    + \mathcal{O}\left(\frac{m^2}{s}\right)^4\right)\,, \label{eq:varpi1_inf} \\
    \varpi^{[-16]}_0 (s,\mt^2) &= \frac{1}{m^2} \left( 1 + \frac{3v}{64m^2}+\frac{41 v^2}{16384 m^4} +\frac{147 v^3}{1048576 m^6} + \mathcal{O}\left(\frac{v}{m^2}\right)^4\right) \,, \label{eq:varpi0_16} \\
    \varpi^{[-16]}_1 (s,\mt^2) &= \varpi^{[-16]}_0 (s,\mt^2)  \log{\left(\frac{v}{m^2} \right)}
    \nonumber \\ &+ \frac{1}{m^2} \left( \frac{v}{32m^2}
    +\frac{37 v^2}{16384 m^4} 
    +\frac{455 v^3}{3145728 m^6} 
    + \mathcal{O}\left(\frac{v}{m^2}\right)^4\right)\,,\label{eq:varpi1_16}
\end{align}
where $v=s+16m^2$.
After these general considerations on the elliptic curve, let us analyze the two elliptic sectors separately. We will demonstrate their relation to this curve and showcase how to find a good starting basis for master integrals on the maximal cut, starting from an integrand analysis.

\subsubsection*{Six-denominator sector $\mathcal{I}_\text{NPA} (1,1,1,0,0,1,1,1,0)$}
By a simple IBP reduction, one can easily see that the six-denominator subsector requires two master integrals on the maximal cut. 
Using a (loop-by-loop) Baikov representation, one can easily show that its maximal cut in $d=4$ space-time dimensions reads~\cite{vonManteuffel:2017hms}
\begin{align}
    \text{MaxCut} \left[\mathcal{I}_\text{NPA} (1,1,1,0,0,1,1,1,0) \right] &\propto \frac{1}{s} \int \frac{\mathrm{d}z_9}{\sqrt{P_4(z_9)}} \, , \label{eq:cutsub}
\end{align}
where we neglected overall numerical prefactors and $P_4(z_9)$ is the quartic polynomial defined in ~\cref{eq:ellcurve}.
From \cref{eq:cutsub}, we can then see immediately that $\mathcal{I}_\text{NPA} (1,1,1,0,0,1,1,1,0)$ corresponds 
 to a differential form of the first kind on the elliptic curve defined in \cref{eq:ellcurve}.
Following the construction described in~\cite{Gorges:2023zgv}, this is a good initial integral, at least on the maximal cut, and we will keep it as it is. Since this sector requires only two master integrals, the second master must be associated with the form of the second kind, and a convenient choice for it can be made by taking the derivative $\partial_{m^2}$ of the previous integral.\footnote{We stress here than any derivative would work or even a combination of derivatives. Since in our case the mass $\mt$ only appears in the propagators, we choose $\partial_{m^2}$ as it can be more cleanly related to a combination of integrals with squared propagators.} This analysis allows us, therefore, to identify a good start basis on the maximal cut as
\begin{align}
    \varepsilon^4 \, s\,  \mathcal{I}_\text{NPA}(1,1,1,0,0,1,1,1,0) \, , \qquad  \varepsilon^4 \, s\, \partial_{\mt^2} \mathcal{I}_\text{NPA}(1,1,1,0,0,1,1,1,0) \, ,
\end{align}
where the $\epsilon$ normalization was chosen for later convenience.

\subsubsection*{Seven-denominator sector $\mathcal{I}_\text{NPA} (1,1,1,1,0,1,1,1,0)$}
Let us move now to consider the top sector. An IBP reduction 
demonstrates that this sector contains four independent master integrals. 
By analyzing the maximal cut of the corner integral in Baikov representation, we find
\begin{align}
    \text{MaxCut} \left[\mathcal{I}_\text{NPA} (1,1,1,1,0,1,1,1,0) \right] &\propto \frac{1}{s} \int \frac{\mathrm{d}z_5 \mathrm{d}z_9}{P_{2,3}(z_5,z_9)} \, , \label{eq:cuttop} 
\end{align}
where $P_{2,3}(z_5,z_9)$ is a polynomial given as
\begin{align}
    P_{2,3}(z_5,z_9) &= z_5^2 (\mt^2-z_9) (\mt^2+s-z_9) -z_5(\mt^2-z_9) (\mt^2+s-z_9) (3 \mt^2-t-z_9) \nonumber \\
    &+ \mt^2 [\mt^4 (2 \mt^2 + 3 s) - \mt^2 (\mt^2 + 3 s) t + s t^2] \nonumber \\
    &-\mt^2 [5 \mt^2 (\mt^2 + s) - (2 \mt^2 + 3 s) t] z_9 + \mt^2 (4 \mt^2 + 2 s - t) z_9^2 - \mt^2 z_9^3 \, .
\end{align}
This polynomial is cubic in $z_9$ and quadratic in $z_5$. 
While it has single poles in both variables, it is clearly convenient to start the analysis with $z_5$.
Picking the two residues in $z_5$, or alternatively taking a primitive in $z_5$, we can then easily write the maximal cut of the second integral as 
\begin{align}
    \label{eq:maxcutcorner7dens}
    &\text{MaxCut}\left[\mathcal{I}_\text{NPA}(1,1,1,1,0,1,1,1,0) \right] \hspace{-0.5ex}\propto\hspace{-0.5ex} \frac{1}{s} 
    \int \frac{\mathrm{d}z_9}{(\mt^2 -t - z_9)\sqrt{P_{4}(z_9)}} 
    \int \mathrm{d}\log{\left(\frac{1+f(z_5,z_9)}{1-f(z_5,z_9)} \right)}   \, ,
\end{align}
where
\begin{equation}
    f(z_5,z_9) = \frac{(\mt^2-z_9) (\mt^2+s-z_9)(3\mt^2-t-2z_5-z_9)}{(\mt^2 -t - z_9) \sqrt{P_{4}(z_9)}}\,.
\end{equation}
In this form, it becomes evident that, due to the extra pole at $z_9 = \mt^2-t$, this integral can be written as a dlog-integration iterated with a differential form of the third kind on the same elliptic curve given in \cref{eq:ellcurve}.
Indeed, since the reduction to master integrals exposed four master integrals, we expect that, differently from the previous case, we should also be able to identify some forms of the third kind on the maximal cut, which should correspond to the extra two masters.
Taking the extra residue at this pole shows that its leading singularity is
\begin{align}
   \Res_{z_9 = t-\mt^2} \left[ \frac{1}{s} \frac{1}{(\mt^2 -t - z_9) \sqrt{P_{4}(z_9)}} \right] = \frac{1}{s \sqrt{P_4(\mt^2-t)}}\,,
\end{align}
which immediately gives us a first good candidate for this sector as
\begin{equation}
    s\, \sqrt{P_4(\mt^2 - t)}\, \mathcal{I}_\text{NPA}(1,1,1,1,0,1,1,1,0)\,,
\end{equation}
modulo appropriate $\epsilon$ normalization.
From this analysis, we also immediately identify a good candidate for the integral corresponding to the differential of the first kind. This can be obtained by considering an integral with a scalar product constructed exactly to cancel this residue. Using the formulas above, in fact, we see that
\begin{align}
    \text{MaxCut} &\left[ (\mt^2 - t)\mathcal{I}_\text{NPA} (1,1,1,1,0,1,1,1,0)  - \mathcal{I}_\text{NPA} (1,1,1,1,0,1,1,1,-1) \right] \nonumber \\ 
    &\propto \frac{1}{s} \int \mathrm{d}z_5 \mathrm{d}z_9 \frac{ (\mt^2 -t - z_9)}{P_{2,3}(z_5,z_9)} 
    = \frac{1}{s} 
    \int \frac{\mathrm{d}z_9}{\sqrt{P_{4}(z_9)}} 
    \int \mathrm{d}\log{\left(\frac{1+f(z_5,z_9)}{1-f(z_5,z_9)} \right)}
    \, . \label{eq:cuttop1} 
\end{align}
Having identified the integral of the first kind, we can choose its derivative (we choose once more $\partial_{m^2}$) to map the integral of the second kind.
Finally, a similar integrand analysis allows us to identify a second independent integral 
of the third kind as
\begin{align}
    &\text{MaxCut} \left[ (\mt^2 - t)\mathcal{I}_\text{NPA} (1,1,1,1,0,1,1,1,-1)  - \mathcal{I}_\text{NPA} (1,1,1,1,0,1,1,1,-2) \right] \propto \nonumber \\ 
    &\propto \frac{1}{s} \int \mathrm{d}z_5 \mathrm{d}z_9 \frac{ (\mt^2 -t - z_9) z_9 }{P_{2,3}(z_5,z_9)} 
    = \frac{1}{s} 
    \int \mathrm{d}z_9 \frac{z_9}{\sqrt{P_{4}(z_9)}} 
    \int \mathrm{d}\log{\left(\frac{1+f(z_5,z_9)}{1-f(z_5,z_9)} \right)}
    \, , \label{eq:cuttop2} 
\end{align}
which has a single pole at $z_9 = \infty$. 
In this way, we come  to our choice of initial basis for the top sector, which at $\epsilon=0$ explicitly
separates  the integrals of the third kind from the coupled $2\times 2$ block of homogeneous equations satisfied by the first kind and its derivative
\begin{align}
   I_1 &=  \varepsilon^4 \, s\, \left[ (\mt^2 -t) \mathcal{I}_\text{NPA}(1,1,1,1,0,1,1,1,0) - \mathcal{I}_\text{NPA}(1,1,1,1,0,1,1,1,-1) \right] \, ,\\ 
   I_2 &=  \varepsilon^4 \, s\, \partial_{\mt^2} \left[ (\mt^2 -t) \mathcal{I}_\text{NPA}(1,1,1,1,0,1,1,1,0) - \mathcal{I}_\text{NPA}(1,1,1,1,0,1,1,1,-1) \right] \, , \\
   I_3 &= \varepsilon^4 \, s\, \left[ (\mt^2 -t) \mathcal{I}_\text{NPA}(1,1,1,1,0,1,1,1,-1) - \mathcal{I}_\text{NPA}(1,1,1,1,0,1,1,1,-2) \right] \, ,\\ 
   I_4 &= \varepsilon^4 \,  s\, \sqrt{P_4(\mt^2 - t)}\, \mathcal{I}_\text{NPA}(1,1,1,1,0,1,1,1,0) \, . \label{eq:defI4}
\end{align}
As expected, one can easily verify by deriving the homogeneous differential equations satisfied by these four integrals that they take the following form
\begin{align}
    \mathrm d \begin{pmatrix} I_1 \\ I_2 \\ I_3 \\ I_4 \end{pmatrix}
    = \begin{pmatrix}
    \ast & \ast & 0 & 0 \\
    \ast & \ast & 0 & 0 \\
    \ast & \ast & 0 & 0 \\
    \ast & \ast & 0 & 0 \\    
    \end{pmatrix} 
    \begin{pmatrix} I_1 \\ I_2 \\ I_3 \\ I_4 \end{pmatrix}
    + \mathcal{O}(\epsilon) \,.
\end{align}
From this starting basis, we can construct a rotation to a new basis satisfying $\epsilon$-factorized differential equations by going through the following steps identified in~\cite{Gorges:2023zgv}: 
\begin{enumerate}
    \item We first make the homogeneous differential equation of the two elliptic blocks manifestly unipotent by multiplying the initial basis by the inverse of the semi-simple part of the Wronskian matrix. This involves multiplying by a matrix that depends on the periods of the elliptic curve and its derivative with respect to $\mt^2$.
    It is important to realize that the equations so obtained are strictly valid \emph{locally}, i.e., close to each regular singular point, where at each point we must use the different definitions of the periods given in \cref{eq:varpi0,eq:varpi1,eq:varpi0_inf,eq:varpi1_inf,eq:varpi0_16,eq:varpi1_16}\,.
    For definiteness, we consider now the construction of the canonical basis close to the point $s=0$.
    \item This step defines a new set of master integrals close to a regular singular point of the elliptic curve. 
    Since the differential forms in the matrix associated with this
    new basis of integrals turns out to
    have different transcendental weights, we must rescale the various integrals by suitable factors of $\epsilon$ to account for this weight shift. 
    If we start from a basis found following the criteria above, this effectively confines all non $\epsilon$-factorized entries to be below the diagonal of the differential equation matrices.
    \item Finally, we integrate out systematically all remaining non-$\epsilon$-factorized terms by a suitable rotation.
\end{enumerate}

Upon performing the last step, this procedure to obtain a canonical basis can introduce new transcendental functions besides the holomorphic period of the elliptic curve defined in~\cref{eq:varpi0}. 
In our case, we must introduce the following function
\begin{equation}
    G(s,t,\mt^2) = \int^{\mt^2} \mathrm{d}x \, \frac{s (s+2 t)\sqrt{P_4(x - t)}}{(t (s+t)-4 s x)^2} \, \varpi_0 (s,x) \,, \label{eq:Gdef}
\end{equation}
which is related to one of the extra punctures on the elliptic curve. In fact, it also admits a representation in terms of complete elliptic integrals of the first and the third kind, which can be obtained from the maximal cut of $I_4$, see~\cref{eq:maxcutcorner7dens} and~\cref{eq:defI4}, 
\begin{equation}
    G(s,t,\mt^2) \propto \sqrt{P_4(\mt^2 - t)}\, \int \frac{\mathrm{d}z_9}{(\mt^2 -t - z_9)\sqrt{P_{4}(z_9)}} \, .
\end{equation}
This representation is not particularly illuminating and also not needed for what follows, so we decided not to provide it explicitly here.

Following the same logic as for the periods, it is more useful to provide its series representation. 
Being a two-dimensional function,
extra care has to be put in choosing the right variables to define its expansion close to $s=0$, $t=0$.
In view of the construction of series expansion solutions in~\cref{sec:exp}, it is convenient
to introduce the two dimensionless ratios
\begin{equation}
    x_1 = -\frac{t \,\mt^2}{s^2}\,, \qquad x_2 = \frac{s}{4 \mt^2}\,,
\end{equation}
in terms of which we can write
\begin{align}
    G(s,t,\mt^2) =
    -\sqrt{x_1\, x_2^3} \left[
   1-\frac{x_2}{32}-10 x_1 x_2 +\frac{3 x_2^2}{1024} -\frac{43 x_1 x_2^2}{16} + 14 x_1^2 x_2^2 
    +\mathcal{O}\left(x_1^3,x_2^3 \right) \right] \,.
\end{align}   

Following the steps above, supplemented by the usual integrand analysis for polylogarithmic sectors and integrating out by hand some remaining non-$\epsilon$-factorized terms in the differential equations, we can arrive at a fully canonical basis
\begin{align}
    \mathrm d \vec{\mathcal{I}} = \epsilon \left[ \sum_{i=1}^{74} G_i \omega_i \right] \vec{\mathcal{I}}\,. 
    \label{eq:cansys}
\end{align}
We stress that some of the polylogarithmic canonical candidates can be found in~\cite{Becchetti:2023wev,Caron-Huot:2014lda}. Moreover, we provide our canonical basis $\vec{\mathcal{I}}$ in the ancillary files found in a \href{https://doi.org/10.5281/zenodo.14733100}{\tt zenodo.org} repository submission~\cite{zenodo} accompanying this article. 
We find that in order to fully describe all master integrals and their crossings required to compute the diphoton amplitudes, we need a total of $74$ differential forms.
We call the full set of two-loop differential forms $\mathbf{W}^{(2l)}$, and similarly the one-loop ones $\mathbf{W}^{(1l)}$.
In particular, $12$ letters are purely rational, $45$ are algebraic, and $17$ contain kernels of elliptic type.
Among the elliptic letters, $4$ of the latter are modular letters defined on the elliptic curve~\cref{eq:ellcurve}, while the remaining $13$ are more involved, as they mix the period of the elliptic curve, with additional algebraic or rational functions that add extra structure 
related to the polylogarithmic master integrals.
We provide the full set of differential forms in~\cref{app:candiff}.
As we will show explicitly below, a large number of the elliptic letters will eventually drop in the finite part of the scattering amplitude, following a typical pattern already observed in other calculations of QFT correlators related to non-trivial geometries~\cite{Duhr:2024bzt,Forner:2024ojj}.

\subsection{Extra relations among the master integrals}
\label{sec:extrarel}

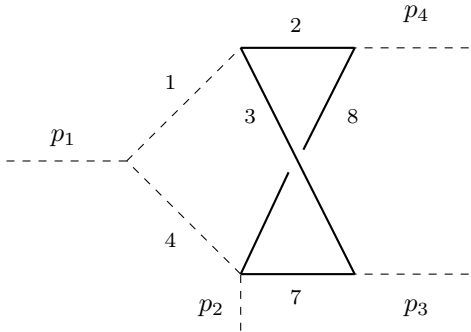
\begin{figure}[t]
\centering
\begin{tikzpicture}[scale=1.5]
\coordinate (links) at (-0.6,0);
\coordinate (mitte) at (0.5,0);
\coordinate (mmitte) at (1.92,-0.1); 
\coordinate (mmmitte) at (2.05,0.1); 
\coordinate (oben) at (1.5,1);
\coordinate (unten) at (1.5,-1);
\coordinate (obenr) at (2.5,1);
\coordinate (untenr) at (2.5,-1);
\coordinate (obenrr) at (3.5,1);
\coordinate (untenrr) at (3.5,-1);
\coordinate (untenl) at (1.5,-1.5);
\begin{scope}
\draw [-, thick,postaction={decorate}] (oben) to [bend right=0]  (obenr);
\draw [-, thick,postaction={decorate}] (untenr) to [bend right=0]  (unten);
\draw [-, thick,postaction={decorate}] (oben) to [bend right=0]  (untenr);
\draw [-, thick,postaction={decorate}] (obenr) to [bend right=0]  (mmmitte);
\draw [-, thick,postaction={decorate}] (unten) to [bend right=0]  (mmitte);
\draw [-, dashed,postaction={decorate}] (obenrr) to [bend right=0]  (obenr);
\draw [-, dashed,postaction={decorate}] (untenrr) to [bend right=0]  (untenr);
\draw [-, dashed,postaction={decorate}] (mitte) to [bend right=0]  (oben);
\draw [-, dashed,postaction={decorate}] (unten) to [bend right=0]  (mitte);
\draw [-, dashed,postaction={decorate}] (mitte) to [bend right=0]  (links);
\draw [-, dashed,postaction={decorate}] (untenl) to [bend right=0]  (unten);
\end{scope}
\node (d1) at (0.9,0.7) [font=\scriptsize, text width=.2 cm]{1};
\node (d2) at (0.9,-0.7) [font=\scriptsize, text width=.2 cm]{4};
\node (d3) at (2,1.2) [font=\scriptsize, text width=.2 cm]{2};
\node (d4) at (2,-1.2) [font=\scriptsize, text width=.2 cm]{7};
\node (d5) at (1.6,0.4) [font=\scriptsize, text width=.2 cm]{3};
\node (d6) at (2.5,0.4) [font=\scriptsize, text width=.2 cm]{8};
\node (d7) at (-0.1,0.2) [font=\small, text width=.2 cm]{$p_{1}$};
\node (d7) at (1.2,-1.3) [font=\small, text width=.2 cm]{$p_{2}$};
\node (d8) at (3,1.3) [font=\small, text width=.2 cm]{$p_4$};
\node (d9) at (3,-1.3) [font=\small, text width=.2 cm]{$p_3$};
\end{tikzpicture}
\caption{ The edge labels correspond to the propagator numbering in family NPA.}
\label{fig:GraphExtraRelations}
\end{figure}

As already hinted at in the previous sections, while a standard IBP reduction exposes 165 independent master integrals for this problem, there exist (at least) 11 extra relations among the latter, which cannot be obtained (with our definitions of integral families) as inhomogeneous ``extra IBPs'' from higher sectors. 
In terms of our canonical integrals, the relations read as follows
\begin{alignat}{4}
    &\mathcal{I}_{80}= \mathcal{I}_{79}-\mathcal{I}_{39} \, , \quad 
    &&  \mathcal{I}_{83}= \mathcal{I}_{82}-\mathcal{I}_{41} \, , 
    \quad &&  \mathcal{I}_{86}= \mathcal{I}_{85}-\mathcal{I}_{43} \, , && \nonumber \\
    &\mathcal{I}_{81}=\mathcal{I}_{40} \, , \quad &&\mathcal{I}_{84}=
   \mathcal{I}_{42} \, , \quad &&\mathcal{I}_{87}=
   \mathcal{I}_{44} \, ,\quad 
   &&\mathcal{I}_{106}= \mathcal{I}_{108} = \mathcal{I}_{110}\,,
\end{alignat}
\begin{align}
    \mathcal{I}_{107}&= -\frac{\mathcal{I}_6}{8}-\frac{5
   \mathcal{I}_8}{8}-\frac{\mathcal{I}_{10}}{8}
   +\frac{\mathcal{I}_{24}}{2}-\frac{3\mathcal{I}_{25}}{2}+\frac{\mathcal{I}_{26}}{2}-\mathcal{I}_{39}+\mathcal{I}_{41}
   -\mathcal{I}_{43} \nonumber \\ 
   &+\frac{\mathcal{I}_{70}}{2}-\frac{\mathcal{I}_{71}}{2}+\frac{\mathcal{I}_{72}}{2}-\frac{\mathcal{I}_{79}}{2}+\frac{\mathcal{I}_{82}}{2}-\frac{\mathcal{I}_{85}}{2} \, , \nonumber \\
   \mathcal{I}_{109}&=
   -\frac{\mathcal{I}_6}{8}-\frac{\mathcal{I}_8}{8}-\frac{5
   \mathcal{I}_{10}}{8}+\frac{\mathcal{I}_{24}}{2}
   +\frac{\mathcal{I}_{25}}{2}-\frac{3\mathcal{I}_{26}}{2}
   -\mathcal{I}_{39}-\mathcal{I}_{41}+\mathcal{I}_{43}
   \nonumber \\ 
   &+\frac{\mathcal{I}_{70}}{2}+\frac{\mathcal{I}_{71}}{2}-\frac{\mathcal{I}_{72}}{2}-\frac{\mathcal{I}_{79}}{2}-\frac{\mathcal{I}_{82}}{2}+\frac{\mathcal{I}_{85}}{2} \, , \nonumber \\
   \mathcal{I}_{111}&= -\frac{5\mathcal{I}_6}{8}-\frac{\mathcal{I}_8}{8}-\frac{\mathcal{I}_{10}}{8}-\frac{3\mathcal{I}_{24}}{2}+\frac{\mathcal{I}_{25}}{2}+\frac{\mathcal{I}_{26}}{2}+\mathcal{I}_{39}-\mathcal{I}_{41}-\mathcal{I}_{43}\nonumber \\ 
   &-\frac{\mathcal{I}_{70}}{2}+\frac{\mathcal{I}_{71}}{2}+\frac{\mathcal{I}_{72}}{2}+\frac{\mathcal{I}_{79}}{2}-\frac{\mathcal{I}_{82}}{2}-\frac{\mathcal{I}_{85
   }}{2} \, .
\end{align}
It is interesting to notice that out of these 11 relations, seven can also be obtained from the remaining four by a suitable crossing of the external momenta.
There are different ways to prove them analytically. The most direct one goes through the canonical system of differential equations in~\cref{eq:cansys}. Indeed, since all differential forms are linearly independent, one can consistently search for linear combinations of the integrals $\mathcal{I}_i$ which fulfill \emph{purely homogeneous} differential equations.
If such integrals exist and if one can use regularity conditions to prove that their boundary conditions are zero to all orders in $\epsilon$,  Cauchy theorem guarantees that the integrals are identically zero.
A similar approach was used in~\cite{Melnikov:2019pdm} to prove extra relations among phase-space master integrals required to compute the triple-real corrections to the  N$^3$LO N-jettines quark beam function.
Extra relations of this type are known to appear in normal amplitudes, even at one loop~\cite{Buccioni:2023okz}. Interestingly, following~\cite{Buccioni:2023okz}, also in the present case we could prove two of these relations by performing non-linear transformations on the Feynman parameter representations of the corresponding integrals. 
Concretely, consider the scalar Feynman integral associated with the graph depicted in \cref{fig:GraphExtraRelations}. 
Its first and second Symanzik polynomials $\mathcal{U}$ and $\mathcal{F}$ read
\begin{align}
    \mathcal{U}(\underline x) =& \, \left(x_3+x_7\right) \left(x_2+x_8\right)+x_1 \left(x_2+x_3+x_7+x_8\right)+x_4 \left(x_2+x_3+x_7+x_8\right) \, , \\
    \mathcal{F}(s,u,\underline x) =& \, s \, x_7 \left(x_2 x_4-x_1 x_8\right)+u \, x_4 \left(x_2 x_7-x_3 x_8\right) +\mt^2 \left[x_1 \left(x_2+x_3+x_7+x_8\right)^2 \right. \nonumber \\ & \ \left.+x_4 \left(x_2+x_3+x_7+x_8\right)^2+\left(x_3+x_7\right) \left(x_2+x_8\right)
   \left(x_2+x_3+x_7+x_8\right)\right] \, ,
\end{align}
where $\underline x =(x_1,x_2,x_3,x_4,x_7,x_8)$ denote the Feynman parameters. The graph has an obvious symmetry under crossing $p_3 \leftrightarrow p_4$ (corresponding to $t \leftrightarrow u$), which can be seen on the level of the graph polynomials by exchanging $x_2 \leftrightarrow x_3$ and $x_7 \leftrightarrow x_8$. 
By studying the differential equations satisfied by the associated Feynman integral as described above, it can be shown that the integral is equal to all its crossings, which implies that it is also invariant under exchanging $p_3$ with any of the other two external momenta, i.e., $p_1$ or $p_2$. 
In terms of our canonical basis, these relations read:
$\mathcal{I}_{106}= \mathcal{I}_{108} = \mathcal{I}_{110}$.

However, there exists no permutation of the Feynman parameters that maps the graph polyomials to themselves with $s$ exchanged with any of the other two Mandelstam variables $t$ or $u$. 
To showcase this symmetry, a quadratic transformation in Feynman parameter space is required. For example, the symmetry under $s \leftrightarrow u$ can be seen from the change of variables
\begin{equation}
\label{eq:quadcov}
\begin{aligned}
    x_1 &\to \frac{\left(x_1+x_4\right)x_3 }{x_3+x_7} \, , \quad &x_2 &\to x_2 \, , \quad &x_3 &\to \frac{
   \left(x_3+x_7\right) x_1}{x_1+x_4} \, , \\
   x_4 &\to \frac{\left(x_1+x_4\right) x_7}{x_3+x_7} \, , \quad &x_7 &\to \frac{\left(x_3+x_7\right)x_4}{x_1+x_4} \, , \quad &x_8&\to x_8 \, ,
\end{aligned}
\end{equation}
whose Jacobian has determinant equal to $-1$ and which also maps the sum of all Feynman parameters to itself. A similar change of variables can be found to prove the symmetry under $s \leftrightarrow t$. 
In contrast to the two relations just discussed, the remaining nine identities we found relate integrals associated to different Feynman graphs, not just crossings of the same integral. It would be very interesting to explore the origin of these equalities and if they can be proven similarly in Feynman parameter space. However, this requires the construction of transformations that map qualitatively different graph polynomials onto each other. There have been attempts in the literature to generalize standard symmetry searching algorithms to find more relations among Feynman integrals~\cite{Wu:2024paw}. It will be interesting in the future to see if also these types of relations can be obtained systematically without having to resort to the underlying system of canonical differential equations.
\section{Helicity amplitudes in terms of iterated integrals}
\label{sec:solution}
With our solutions for the $154$ independent master integrals and the extra relations provided in \cref{sec:extrarel}, it is immediate to obtain analytical solutions for the helicity amplitudes in \cref{eq:helampqq,eq:alphabetagg} in terms of a basis of iterated integrals. We notice here that
the amplitude for the production of two photons in quark-antiquark annihilation is substantially
simpler than the gluon-fusion one, the difference amounting mainly to polylogarithmic
sectors. The discussion below applies to union the sets of iterated integrals appearing in both partonic channels, except the last point, where we discuss the extra simplifications happening in the quark channel.
Upon expanding in $\epsilon$ and substituting the solutions of our master integrals, we can make various observations about the final structure of the bare helicity amplitude, which we summarize here:

\begin{enumerate}
    \item First, we notice that all iterated integrals with kernels of elliptic type identically cancel in the poles of the amplitudes. Also, for both amplitudes, the color factors that are supposed to be finite, 
    are indeed identically finite upon substituting our solutions. Moreover, as expected, a large number of the algebraic letters cancel, leaving an analytic expression for the poles written in terms of iterated integrals up to length three, depending only on the one-loop letters
    \begin{equation}
    \begin{split}
        \mathbf{W}^{(1l)} = \, &\Big\{ \mt^2,s,t,-s-t,s-4 \mt^2,t-4 \mt^2,-4 \mt^2-s-t, 4 s \mt^2+t (-s-t), \\
        &\ 4 t \mt^2+s (-s-t),4 \mt^2 (-s-t)+s t, \frac{s-r_1}{s+r_1},\frac{t-r_2}{t+r_2},\frac{-s-t-r_3}{-s-t+r_3}, \\
        &\ \frac{s t-r_7}{st+r_7},\frac{s (-s-t)-r_8}{s (-s-t)+r_8},\frac{t (-s-t)-r_9}{t(-s-t)+r_9},\frac{s t (s-4 \mt^2)-r_1 r_7}{s t (s-4 \mt^2)+r_1 r_7}, \\
        &\ \frac{s (-s-t) (s-4 \mt^2)-r_1 r_8}{s (-s-t) (s-4 \mt^2)+r_1 r_8},\frac{s t(t-4 \mt^2)-r_2 r_7}{s t (t-4 \mt^2)+r_2 r_7}, \\
        &\ \frac{t (-s-t)(t-4 \mt^2)-r_2 r_9}{t (-s-t) (t-4 \mt^2)+r_2 r_9},\frac{s (-s-t) (-s-t-4\mt^2)-r_3 r_8}{s (-s-t) (-s-t-4 \mt^2)+r_3 r_8}, \\
        &\ \frac{t (-s-t) (-s-t-4 \mt^2)-r_3 r_9}{t (-s-t) (-s-t-4 \mt^2)+r_3 r_9} \Big\}\,,
    \end{split}
    \end{equation}
    where the square roots $r_i$ are defined in \cref{app:candiff}.
    \item All algebraic polylogarithmic letters appear in the finite part of the bare amplitude, while two of the rational ones drop, in particular, $s+4 m^2$ and $s+ 16 m^2$.
    \item Finally, out of the 17 elliptic letters, 6 drop and 11 appear instead in the finite part of the bare amplitude. We stress here that the pattern of the disappearing letters follows exactly what was expected from previous calculations of two-point correlators~\cite{Duhr:2024bzt, Forner:2024ojj}, namely all letters which contain the holomorphic period squared $\varpi_0^2$ at the numerator drop from the
    amplitude. We stress here that this also includes letters that depend \emph{indirectly} 
    on two positive powers of $\varpi_0$, for example through the new function $G$ defined in \cref{eq:Gdef}.
    Following the nomenclature for the elliptic letters introduced in~\cref{app:candiff}, the elliptic letters that drop are $\{M_{4},E_4,E_6,E_{11},E_{12},E_{13}\}$.

    \item As we already discussed, the quark-antiquark channel is substnatially simpler. In particular, all dependence from the massive quarks cancels up to one loop and, therefore, the poles of the two-loop amplitudes only depend on a small subset of the one-loop letters
    \begin{equation}
    \begin{split}
        \mathbf{W}^{(1l)} = \, &\Big\{ \mt^2,s,t,-s-t\Big\}\,,
    \end{split}
    \end{equation}
    Moreover, for what concerns the finite part of these amplitudes, an additional elliptic letter drops compared to the gluon channel, which is $E_2$. From the alphabet, 30 letters (4 rational, 26 with square roots) contribute and are a subset of the gg alphabet.
\end{enumerate}

\section{UV renormalization and IR structure}
\label{sec:ren}
The bare helicity amplitudes contain both divergences of UV and IR origin.
The UV divergences can easily be removed by renormalizing the strong coupling constant, the quark and gluon wave functions, and the top mass.
We work in the $\overline{ \rm MS}$ scheme for the massless quarks and in the on-shell scheme for the top quark. In practice, this means that the relations between the bare and the renormalized parameters to the $\epsilon$-order needed for this calculation read
\begin{align}
 \mu_0^{2 \epsilon} S_{\epsilon}\alpha_{s,b} &= \mu^{2\epsilon} \alpha_s
\left[ 1- \left( \frac{\beta_0 }{\epsilon}  + \delta_w \right) \left(\frac{\alpha_s}{2\pi}\right)  + \mathcal O(\alpha_s^2) \right]\,,  \\ 
m_0 &= m \left[ 1+ \delta_m\, \left(\frac{\alpha_s}{2\pi}\right)+ \mathcal O(\alpha_s^2) \right]\,,
\end{align}
with $\alpha_s = \alpha_s(\mu)$ the renormalized coupling evaluated at the renormalization scale $\mu$. 
In the formulas above, we have put
\begin{equation}
    S_{\epsilon} = (4\pi)^{\epsilon}e^{-\gamma_E \epsilon} \, ,
\end{equation}
and
\begin{align}
    \beta_0 &= \frac{11}{6}C_A - \frac{2}{3}T_R\, n_f\,, \\ 
    \delta_w &=  T_R\, n_h\, \left( \frac{\mt^2}{\mu^2}\right)^{-\epsilon} \left(-\frac{2}{3 \epsilon} - \frac{\pi^2}{18} \epsilon + \frac{2}{9} \zeta_3 \epsilon^2 + \mathcal{O}(\epsilon^3)\right)\,, \label{eq:renconst} \\ 
    \delta_m &= - C_F \left( \frac{\mt^2}{\mu^2}\right)^{-\epsilon} \left( \frac{3}{2 \epsilon} + 2 + \mathcal{O}(\epsilon)\right) 
\end{align}
with $n_f$ the number of massless active fermions, $n_h$ the number of massive flavours (which we assume to be $n_h = 1$ throughout the calculation)
$C_A = N_c$  the Casimir of the adjoint representation, $C_F =(N_c^2-1)/(2N_c)$ the Casimir of the fundamental representation and $T_R = 1/2$ (see also~\cite{Barnreuther:2013qvf}).
Clearly, $N_c = 3$ is the number of colors and from
here on we fix $\mu_0 = \mu$\,.
Finally, we also need the wave-function renormalization constants for gluons and massless quarks, which, to this order, read
\begin{align}
    &Z_A = 1 + \left(\frac{\alpha_s}{2\pi}\right) \delta_w + \mathcal O(\alpha_s^2)\,, \nonumber \\ 
    &Z_q = 1 + \left(\frac{\alpha_s}{2\pi}\right)^2 C_F\, T_R\, n_h\, \left[\frac{1}{ 4\epsilon} - \frac{5}{24} \right] 
    \left(\frac{\mu^2}{m^2}\right)^{2\epsilon} + \mathcal O(\alpha_s^3)\,,
\end{align}
respectively, where $\delta_w$ was defined in \cref{eq:renconst}.
Putting everything together, and focussing only on the part of the two-loop amplitude which contains
at least one massive loop, 
the renormalized helicity coefficients for the two channels can be written as
\begin{equation}
\Omega_{qq}^{\rm UV} = \delta_{kl} (4 \pi \alpha)\, \sum_{\ell=0}^{2} \left(\frac{\alpha_{s}}{2\pi}\right)^\ell \Omega_{qq}^{(\ell, \rm UV)}\,,
\end{equation}
\begin{equation}
\Omega_{gg}^{\rm UV} = 
\delta_{a_1 a_2} (4 \pi \alpha)\, \sum_{\ell=1}^{2} \left(\frac{\alpha_{s}}{2\pi}\right)^\ell\Omega_{gg}^{(\ell, \rm UV)}\,,
\end{equation}
where explicitly we find\footnote{We stress that we assume here that the bare amplitude is defined using the normalization for the integration measure provided in~\cref{eq:norm}.} 
\begin{align}
    \Omega_{qq}^{(2, \rm UV)} &= \left(\frac{C_{\epsilon}}{S_\epsilon}\right)^{2} 
    \Omega_{qq}^{(2, b)} - \left(\frac{C_{\epsilon}}{S_\epsilon}\right) 
    \delta_{w} \, \Omega_{qq}^{(1, b)} + 
    C_F T_R n_h \left[\frac{1}{ 4\epsilon} - \frac{5}{24} \right] \left(\frac{\mu^2}{m^2}\right)^{2\epsilon}  \Omega_{qq}^{(0, b)} \,,
\end{align}
and 
\begin{align}
    \Omega_{gg}^{(1, \rm UV)} &= \Omega_{gg}^{(1, b)}  \,, \\
    \Omega_{gg}^{(2, \rm UV)} &= \left(\frac{C_{\epsilon}}{S_\epsilon}\right)^2 
    \Omega_{gg}^{(2, b)} - \left(\frac{C_{\epsilon}}{S_\epsilon}\right)  
    \frac{\beta_0}{\epsilon} \Omega_{gg}^{(1, b)} 
    + \left(\frac{C_{\epsilon}}{S_\epsilon}\right)  (2 m^2) \, \delta_{m} \frac{\partial \Omega_{gg}^{(1, b)}}{\partial m^2}  \,.
\end{align}
We notice that, to this perturbative order, the renormalization of the top quark mass does 
not enter in $\Omega_{qq}^{\ell,\rm UV}$.

After UV renormalization, the massive helicity coefficients for the quark-antiquark channel are finite, while the ones for the gluon-gluon channel
still contain divergences of IR origin. As it was first shown by Catani up to two loops~\cite{Catani:1998bh}, at every order in perturbation theory in the strong coupling these can be reorganized in terms of lower loop amplitudes~\cite{Becher:2009cu,Becher:2009kw,Becher:2009qa,Gardi:2009qi,Ferroglia:2009ii}. In our case, we can write
\begin{align}
&\Omega_{qq}^{(2, \rm fin)} = \Omega_{qq}^{(2, \rm UV)} \,, \nonumber \\
        &\Omega_{gg}^{(1, \rm fin)} = \Omega_{gg}^{(1, \rm UV)}  \,, \qquad
    \Omega_{gg}^{(2, \rm fin)} = \Omega_{gg}^{(2, \rm UV)} - \mathcal{I}_{gg}^{(1)} \Omega_{gg}^{(1, \rm UV)} \,,
\end{align}
where the operator $\mathcal{I}_{gg}^{(1)}$ reads
\begin{equation}
    \mathcal{I}_{gg}^{(1)} = - \frac{e^{\gamma_{E} \epsilon}}{\Gamma(1-\epsilon)} \left(\frac{C_A}{\epsilon^2} + \frac{\beta_0}{\epsilon}\right) \left(\frac{-s-i0^+}{\mu^2}\right)^{-\epsilon}\,.
\end{equation}

While we performed UV renormalization and the subtraction of IR poles as a consistency check, we prefer to provide the results for the bare helicity amplitudes, which can be found in a \href{https://doi.org/10.5281/zenodo.14733100}{\tt zenodo.org} repository submission~\cite{zenodo} accompanying this article. 
These allow the implementation of any UV renormalization and IR subtraction in an arbitrary scheme.
\section{Series expansions and numerical evaluation}
\label{sec:exp}
Our discussion until now has been limited to expressions of the amplitudes in terms of linearly independent iterated integrals. This has been beneficial in studying some of their analytic properties, in particular, the cancellation of large numbers of integration kernels that could 
have otherwise naively been expected to appear in the final result.

In this section, we describe a possible strategy to efficiently evaluate our amplitudes numerically.
At variance with other approaches that aim at evaluating all functions numerically on the fly producing series expansion solutions for them for each individual phase-space point (see, for example~\cite{Moriello:2019yhu,Hidding:2020ytt,Prisco:2025wqs}), our approach will consist of constructing series expansion representations for the whole amplitude and patching them together in order to cover large portions of the relevant phase space. While our construction is not new, it is important to note a subtle point. Following the standard theory of differential equations, we would expect that a larger number of different series expansions should be necessary to cover the whole phase space due to the intricated structure of singular lines in the differential equations. 
Instead, as already observed in different contexts~\cite{Remiddi:1999ew, Niggetiedt:2023uyk, Duhr:2024uid, Forner:2024ojj}, 
the fact that most of these singularities are pseudo-thresholds makes it such that the series obtained around the various singular 
points often converge beyond their naive radius of convergence. 
This typically allows one to cover large regions of the phase space with a very small number of expansions. 
Additionally, we can utilize the symmetry properties of the amplitude (see~\cref{eq:bosesymmqq,eq:bosesymmgg}) 
to derive further series expansions valid around different points from a single expansion point. 
This reduces the actual number of series expansions, which have to be derived by explicitly expanding the individual master integrals. 

As a proof of concept of this approach, we provide series expansions for our amplitudes close to $\mt^2 = \infty$, which corresponds to the large mass expansion of the amplitude, and $s= 4\mt^2$, which is one of the physical thresholds, and showcase how they are sufficient to cover a large patch of the physical phase-space. We recall here that $s=4m^2$ is a very non-trivial point. In particular, the elliptic curve in~\cref{eq:ellcurve} does not degenerate at this point, which means that our series expansions contain new numbers related to the periods of the curve evaluated on this kinematical point.
The results presented here can be easily supplemented by adding more expansion points. 
In particular, it is natural to
consider the value of the amplitude close to $\mt^2 = 0$. Here, one can easily prove from direct inspection of the system of differential equations, that 
the amplitude can be expanded to arbitrary high orders in $(\mt^2)^n \log^m{(\mt^2)}$ in terms of harmonic polylogarithms~\cite{Remiddi:1997ny} of $x = -t/s$, see for example the recent calculation~\cite{Lee:2024dbp}.
Below, we briefly discuss how we obtained our series of solutions in the regions discussed above.

\subsection{Large-mass expansion}
The large-mass expansion of an arbitrary scattering amplitude can typically be obtained by a straightforward application of standard methods of asymptotic expansions of Feynman integrals~\cite{Smirnov:2002pj}.
This approach has been used effectively to obtain results in this limit for many non-trivial scattering amplitudes, see for example~\cite{Melnikov:2015laa, Davies:2018qvx, Davies:2020drs, Davies:2020lpf} and many others. Here, we follow a different approach, deriving the large-mass expansion directly from our system of canonical differential equations. 

\begin{figure}[h]
    \centering
    \subfloat[Singularities gluon fusion]{\includegraphics[width=0.45\textwidth]{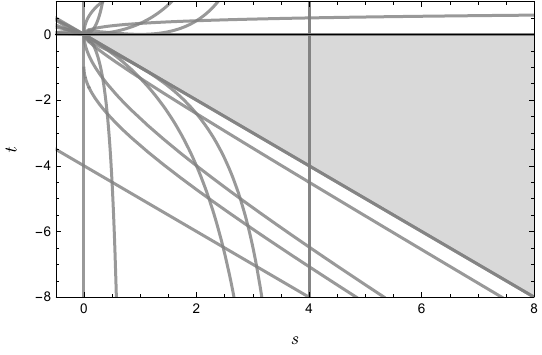}\label{subfig1}}
    \hspace{0.08\textwidth}
    \subfloat[Singularities quark-antiquark annihilation]{\includegraphics[width=0.45\textwidth]{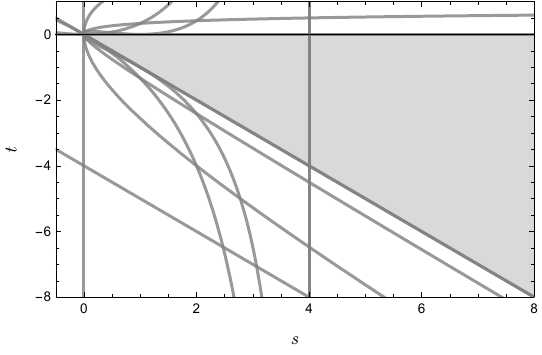}\label{subfig2}} \\
    
    \caption{Plot of all singular lines in the variables $s,t$ for gluon fusion and quark-antiquark annihilation, respectively. We have highlighted the physically relevant parameter space in light gray. Notice that the physical threshold is at $s=4m^2$.}
    \label{fig:plotsingularities}
\end{figure}
To consistently expand around any regular singular point of a system of differential equations, it is crucial to consider a good set of variables. 
These variables have to be chosen such that all possible singular lines intersecting at the considered singular point are disentangled. 
In detail, this means that we must separate the singular lines so that at most two singular lines intersect. In addition, we must make sure that all intersections happen non tangentially. This condition is referred to as \emph{normal crossing divisors} in the mathematical literature.
One can achieve this by performing a so-called \emph{blow-up}~\cite{MR1288523}.
The number of blow-ups required depends on the strength of the intersection of the singular lines. 
Reference~\cite{MR0199184} demonstrates that, after performing a sufficient number of blow-ups, 
one can always locally disentangle all singular lines yielding normal crossing divisors.

\begin{figure}[h]
    \centering
    \subfloat[Singularities gluon fusion]{\includegraphics[width=0.45\textwidth]{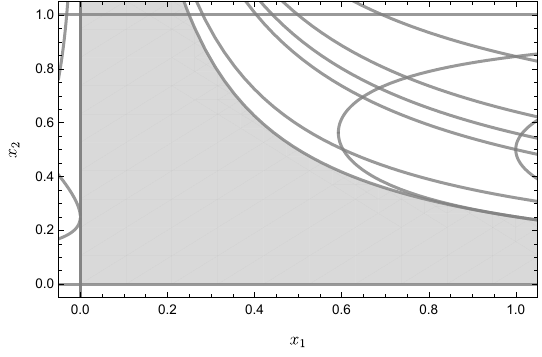}\label{subfig1}}
    \hspace{0.08\textwidth}
    \subfloat[Singularities quark-antiquark annihilation]{\includegraphics[width=0.45\textwidth]{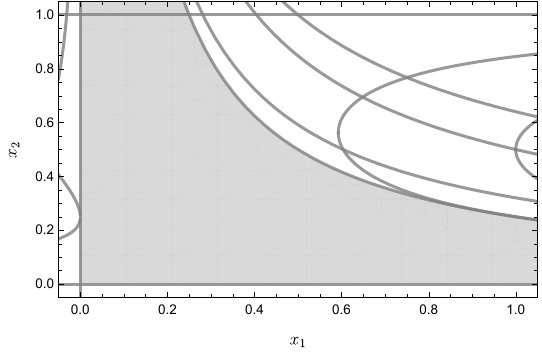}\label{subfig2}} \\
    
    \caption{Plot of all singular lines after blow-up in the variables $x_1,x_2$ for gluon fusion and quark-antiquark annihilation, respectively. The threshold is now given at $x_2=1$. We have again highlighted the physically relevant parameter space in light gray.}
    \label{fig:plotsingularitiesblowup}
\end{figure}

In the specific case of the large-mass expansion, the point $\mt^2 \to \infty$ corresponds to 
$s=0$, $t=0$, and $u=0 \Leftrightarrow s=-t$ in the kinematical plane.  
In ~\cref{fig:plotsingularities}, we draw all singular lines relevant for gluon fusion and quark-antiquark annihilation, 
respectively, with a particular focus on the ones passing through $(s,t)=(0,0)$. As it can be seen, many lines intersect at that point and most
of them cross tangentially to each other.
Concretely, these singular lines correspond to the following set
\begin{equation}
\begin{aligned}
    &\left\{s+t,-4 s-4 t+s t,s+2 s^2+s^3+t-2 s t+s^2 t,-4 s+s t+t^2, s-2 s t-4 t^2+s t^2,  \right. \\
    &\qquad \left. s-2 s^2+s^3-6 s t+2 s^2 t-4 t^2+s t^2,s+t-2 s t+2 t^2+s t^2+t^3, s^2-4 t+s t, \right. \\
    &\qquad \left. -4 s^2+t-2 s t+s^2 t,-4 s^2+t-6 s t+s^2 t-2 t^2+2 s t^2+t^3\right\} \, 
\end{aligned}
\label{eq:singdiphoton}
\end{equation}
for gluon fusion and the subset
\begin{equation}
\begin{aligned}
    &\left\{s+t,-4 s-4 t+s t,-4 s+s t+t^2,s-2 s t-4 t^2+s t^2, \right. \\
    &\qquad \left. s-2 s^2+s^3-6 s t+2 s^2 t-4 t^2+s t^2,s^2-4 t+s t\right\} \, 
\end{aligned}
\label{eq:singqq}
\end{equation}
for quark-antiquark annihilation.
Following the blow-up procedure, in our case, a good set of variables that resolves all singularities is given by
\begin{equation}
\label{eq:blowupvariables}
    x_1 = -\frac{t \,\mt^2}{s^2}\,, \qquad x_2 = \frac{s}{4 \mt^2}\,.
\end{equation}
In \cref{fig:plotsingularitiesblowup}, we report the same singularities in the new set of coordinates $x_1,x_2$. As it is easy to see, 
all singular lines are appropriately disentangled in the large mass limit, i.e., for $x_1=x_2=0$. 
Moreover, in the blow-up variables $x_1,x_2$, the threshold $s=4m^2$ is at $x_2=1$. 
With these new variables, we can now expand all singular denominators in series consistently. 
We stress here that, in terms of the original kinematic variables, this choice of expansion variables corresponds to having taken the 
directional limit $-t/\mt^2 < s^2/\mt^4$. 

Before integrating the differential equations, we have to determine the necessary boundary constants. We follow the standard
approach and first use regularity properties of the individual master integrals to minimize the number of independent boundary integrals that must be computed explicitly. The remaining boundaries can then all be related to simple products of tadpole and bubble integrals, 
which can be easily computed analytically. Alternatively, one can also recover the boundaries from high-precision 
numerical evaluations of the master integrals close to $s=t=0$ via the method of auxiliary mass flow, relying on the \textsc{Mathematica} package \texttt{AMFlow}~\cite{Liu:2022chg} .

Having all boundary conditions at our disposal, we can expand the system of differential equations and integrate them iteratively up to arbitrary orders.
For definiteness, we produce solutions for the canonical master integrals expanding up to $40$ orders, 
which allows us to obtain a corresponding expansion for the helicity amplitudes up to $34$ orders. 
The order of expansion for the helicity amplitudes is reduced due to rational functions in the amplitudes, which have poles of up to power six. 
We stress here that using a canonical basis is beneficial, as the loss of terms due to divergences in the numerators is typically less severe.

\subsection{Expansion at $s=4\mt^2$}
As a second, non-trivial expansion point, we consider the threshold $s=4 \mt^2$ \footnote{As $s=4\mt^2$ is a regular point of the elliptic curve, both periods are holomorphic in a neighborhood of this point. For concreteness, we used the specific linear combination corresponding to $\varpi_0^{[0]}$ (see~\cref{eq:varpi0}) to function as $\varpi_0$ in the definition of our local canonical basis, while any other choice would have also been allowed.}. Having good control of the amplitudes 
close to this point is important since  we can expect that the effects induced by the massive loops will be most visible 
in this region of the phase space.

The periods of the elliptic curve do not depend on $t$. Nevertheless, an expansion of the helicity amplitudes in the single variable 
$s$ around the threshold value $4\mt^2$ remains non-trivial. This is due to the function $G(s,t,\mt^2)$, for which $s=4\mt^2$ is a regular point. 
As a consequence, after expanding in the variable $s$, the remaining integration of the canonical differential equations in the variable $t$ 
still requires integration over elliptic kernels. To handle them, we perform a double series and expand the helicity amplitudes simultaneously in both variables around the singular point $(s,t)=(4\mt^2,0)$, rendering all integrations trivial. In contrast to the large-mass expansion, there are only two singular lines $s=4\mt^2$ and $t=0$ crossing at this point, such that no blow-up is necessary. As expansion variables, we therefore use
\begin{equation}
    y_1= 4 - \frac{s}{\mt^2} \, , \qquad y_2 = - \frac{t}{\mt^2} \, .
\end{equation}

To find the values of the master integrals at the expansion point, we make use of the fact that the boundary conditions are trivial in the large mass limit considered previously. 
We can then use the canonical differential equations to transport them from the point $(s,t)=(0,0)$ to $(s,t)=(4\mt^2,0)$ along the path given by the straight line connecting the two points.\footnote{The master integrals are finite as $t \rightarrow 0$, with the exception of $\mathcal{J}_{162}$ and $\mathcal{J}_{165}$, which correspond to massless one-loop integrals with a tadpole attached, that can be computed exactly instead.} This can be supplemented by regularity conditions, as well as information on the branching structure of the master integrals to reduce the number of single-scale integrals that have to be computed explicitly. In practice, this approach proves convenient for most of our master integrals \footnote{To this end, the functionalities of the \textsc{Mathematica} package \texttt{PolyLogTools} proved highly useful~\cite{Duhr:2019tlz}.} except those that require integration over elliptic kernels. For these few integrals, we rely on numerical evaluations obtained with \texttt{AMFlow}.

We performed the expansion of the master integrals in this limit up to $35$ orders, from which we can obtain at least $30$ orders in the corresponding expansion of any of the helicity coefficients. As for the previous case, 
the reduction in the number of orders obtained stems from rational functions in the amplitude, similar to the large-mass case.

\subsection{Numerical Evaluations}
By merging together the expansion for large mass and the one around the point $(s,t)=(4\mt^2,0)$, and exploiting the symmetry
properties of the helicity amplitudes under $t \leftrightarrow u$ (see~\cref{eq:bosesymmqq} and~\cref{eq:bosesymmgg}), 
we can obtain a stable and reliable numerical evaluation of all helicity 
coefficients on a large portion of the phase space. 
In particular, we use the symmetry under $t \leftrightarrow u$ to obtain two more expansions, one around the additional expansion point $(s,u)=(4\mt^2,0) \ \Leftrightarrow (s,t)=(4\mt^2,-4\mt^2)$ and another one in the large-mass limit. Of course, in this case, the overlapping singularities 
are disentangled by different blow-up variables, obtained from those given in~\cref{eq:blowupvariables} by $t\rightarrow u$. 
With these four expansions put together, we can cover the entire phase space region around the 
massive-particle production threshold $s=4\mt^2$, down to the region covered by the large-mass expansion. 

The series have to be produced once and are extremely efficient to evaluate numerically. For instance, without any further optimization, 
the evaluation of our results on a single phase-space point in \textsc{Mathematica} using a single core on a standard laptop takes 
around $0.02s$ to $0.07s$ per helicity coefficient.

To double-check our results, we compared the numerical value of the expanded helicity coefficients in multiple different phase space points to evaluations of the unexpanded coefficients where the numerical values of the master integrals have been obtained via \texttt{AMFlow}. In~\cref{tb:AMFlowComparison}, we show the absolute value of the difference for the finite part of the bare helicity coefficients, normalized by the unexpanded result, for several phase space points. 
\begin{table}[h!]
\begin{center}
\begin{tabular}{ |p{1cm}||p{1.75cm}|p{1.75cm}|p{1.75cm}|p{1.75cm}|p{1.75cm}|p{1.75cm}| }
 \hline
  $s/\mt^2$ & 13/10 & 23/10 & 28/9 & 11/3 & 22/5 & 51/10 \\
 \hline
  $t/\mt^2$ & -3/5 & -1 & -1/10 & -5/2& -3/5 & -11/10 \\
 \hline
 \hline
  $f_{++++}$ & $ 2.3 \times 10^{-8}$ & $ 2.5 \times 10^{-6}$ & $ 2.7 \times 10^{-18}$ & $ 3.8 \times 10^{-18}$ & $ 5.2 \times 10^{-21}$ & $ 9.2 \times 10^{-9}$ \\
  \hline
  $f_{-+++}$ & $ 1.2 \times 10^{-9}$ & $ 6.4 \times 10^{-7}$ & $ 1.9 \times 10^{-16}$ & $ 1.6 \times 10^{-18}$ & $ 1.3 \times 10^{-18}$ & $ 7.4 \times 10^{-7}$ \\
  \hline
  $f_{+-++}$ & $ 1.2 \times 10^{-9}$ & $ 6.4 \times 10^{-7}$ & $ 1.9 \times 10^{-16}$ & $ 1.6 \times 10^{-18}$ & $ 1.3 \times 10^{-18}$ & $ 7.4 \times 10^{-7}$ \\
  \hline
  $f_{++-+}$ & $ 1.8 \times 10^{-9}$ & $ 7.9 \times 10^{-7}$ & $ 2.1 \times 10^{-16}$ & $ 2.0 \times 10^{-18}$ & $ 1.4 \times 10^{-18}$ & $ 6.4 \times 10^{-7}$ \\
  \hline
  $f_{+++-}$ & $ 1.8 \times 10^{-9}$ & $ 7.9 \times 10^{-7}$ & $ 2.1 \times 10^{-16}$ & $ 2.0 \times 10^{-18}$ & $ 1.4 \times 10^{-18}$ & $ 6.4 \times 10^{-7}$ \\
  \hline
  $f_{--++}$ & $ 5.1 \times 10^{-10}$ & $ 3.7 \times 10^{-7}$ & $ 4.0 \times 10^{-18}$& $ 3.5 \times 10^{-19}$ & $ 1.9 \times 10^{-22}$ &  $ 5.2 \times 10^{-10}$ \\
  \hline
  $f_{-+-+}$ & $ 1.1 \times 10^{-9}$ & $ 1.6 \times 10^{-6}$& $ 2.1 \times 10^{-15}$ & $ 4.9 \times 10^{-18}$ & $ 1.5 \times 10^{-18}$ & $ 5.6 \times 10^{-7}$ \\
  \hline
  $f_{+--+}$& $ 2.3 \times 10^{-8}$ & $ 1.0 \times 10^{-6}$ & $ 5.1 \times 10^{-19}$ & $ 7.0 \times 10^{-17}$ & $ 6.7 \times 10^{-20}$  & $ 7.7 \times 10^{-8}$ \\
  \hline
  $\alpha_{h}$ & $ 3.0 \times 10^{-9}$ & $ 4.1 \times 10^{-7}$ & $ 2.3 \times 10^{-17}$& $ 1.0 \times 10^{-21}$ & $ 5.2 \times 10^{-21}$ & $ 3.5 \times 10^{-8}$ \\
  \hline
  $\beta_{h}$ & $ 8.4 \times 10^{-12}$ & $ 7.7 \times 10^{-9}$ & $ 2.1 \times 10^{-18}$ & $ 6.7 \times 10^{-23}$ & $ 7.0 \times 10^{-22}$ & $ 1.5 \times 10^{-9}$ \\
  \hline
  $\gamma_{h}$ & $ 5.7 \times 10^{-11}$ & $ 2.9 \times 10^{-8}$ & $ 3.3 \times 10^{-19}$ & $ 6.8 \times 10^{-21}$ & $ 3.3 \times 10^{-22}$ & $ 2.5 \times 10^{-9}$ \\
  \hline
  $\delta_{h}$ & $ 3.0 \times 10^{-9}$ & $ 4.1 \times 10^{-7}$ & $ 2.3 \times 10^{-17}$ & $ 1.0 \times 10^{-21}$ & $ 5.2 \times 10^{-21}$ &  $ 3.5 \times 10^{-8}$ \\
 \hline
\end{tabular}
\end{center}
\caption{Absolute value of the difference of the helicity coefficients evaluated via the method of auxiliary mass flow and the series expansion representation, normalized by the former. For every phase space point, we use the series that performed best at that point. }
\label{tb:AMFlowComparison}
\end{table}

For the quark-antiquark channel, the helicity amplitudes can be split into a part proportional to the square of the electric charge of the heavy quark in units of the electron charge, $e_h$, and a part proportional to the number of heavy quark flavors $n_h$ times the electric charge squared of the light quark $e_q$ instead,
\begin{equation}
\label{eq:partsqqamp}
     \Omega_{qq} = n_h\, e_q^2 \, \Omega_{l} + e_h^2 \, \Omega_{h} \, .
\end{equation}
Note that the coefficients $\Omega_{l}$ are substantially simpler, as the only dependence on the heavy flavour is contained in bubble subdiagrams
and all integrals can, in principle, be expressed in terms of multiple polylogarithms.
For this reason, in~\cref{tb:AMFlowComparison}, we have shown this difference only for $\Omega_{h}$, which receives contributions from those diagrams where the two photons attach directly to a heavy-quark loop.

\begin{figure}[htbp]
    \centering
    \subfloat[$f_{++++}^{(2,b)}$]{\includegraphics[width=0.3\textwidth]{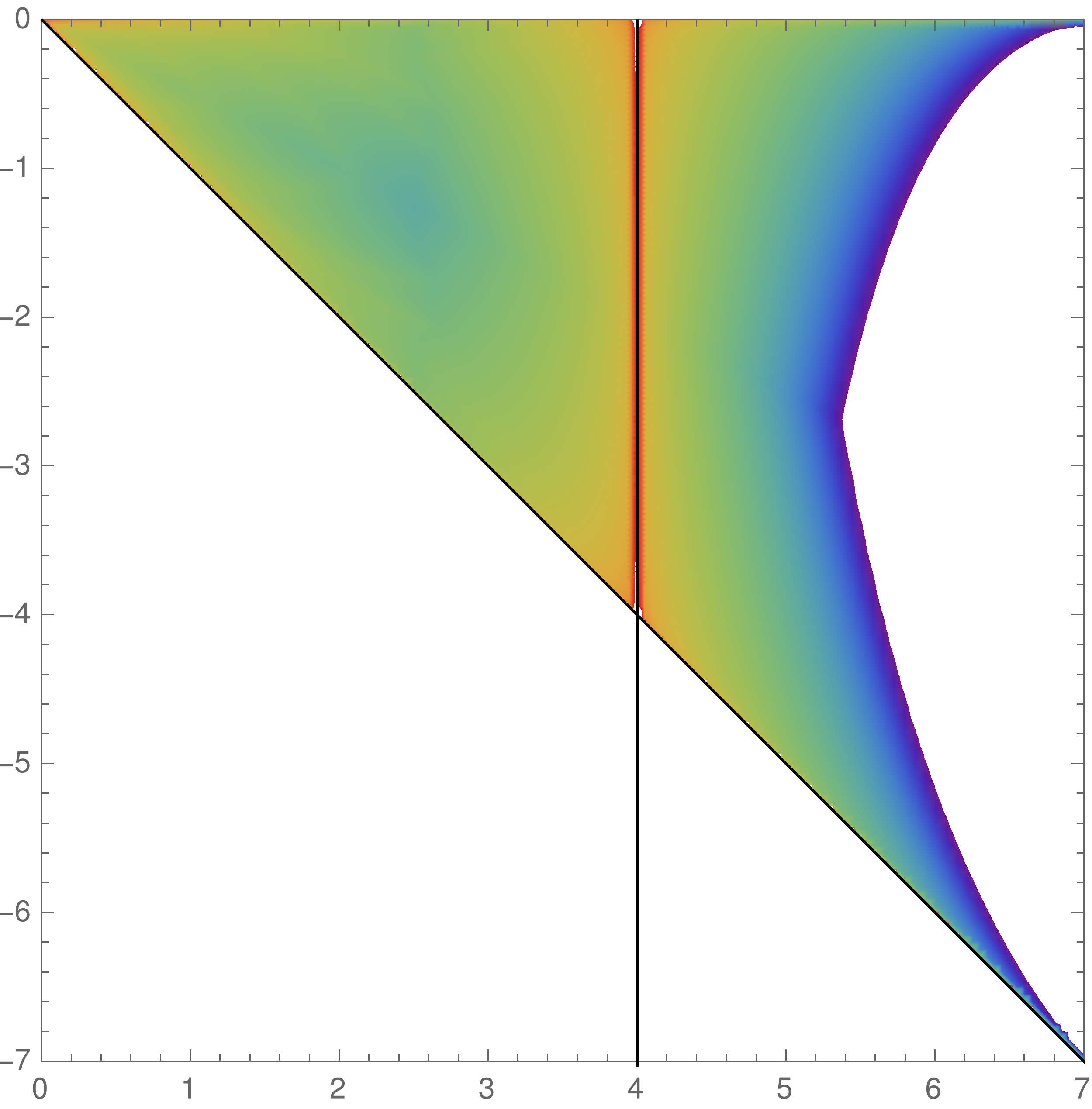}\label{subfig1}}
    \hspace{0.03\textwidth}
    \subfloat[$f_{--++}^{(2,b)}$]{\includegraphics[width=0.3\textwidth]{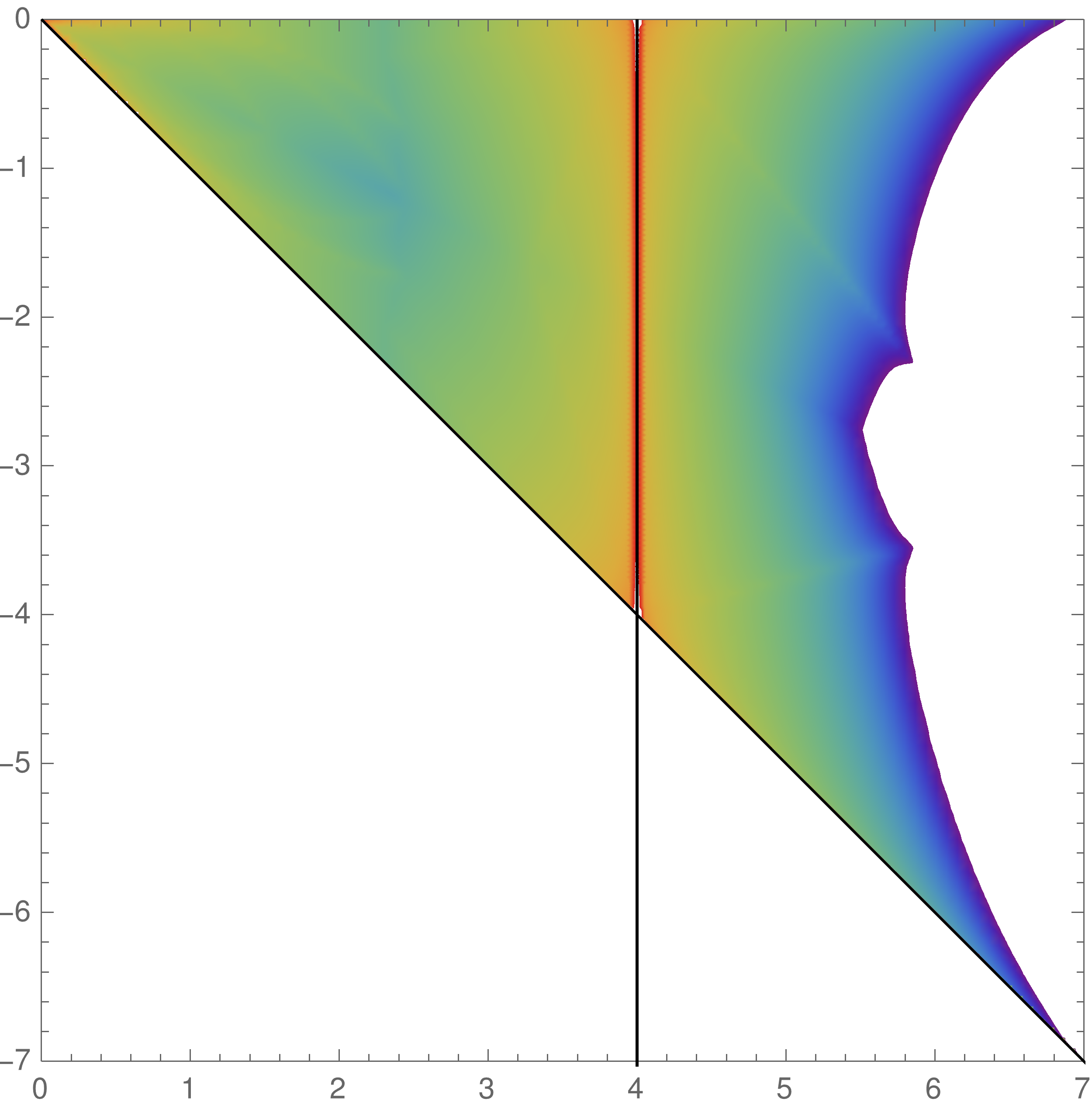}\label{subfig2}} 
    \hspace{0.03\textwidth}
    \subfloat[$f_{-+-+}^{(2,b)}$]{\includegraphics[width=0.3\textwidth]{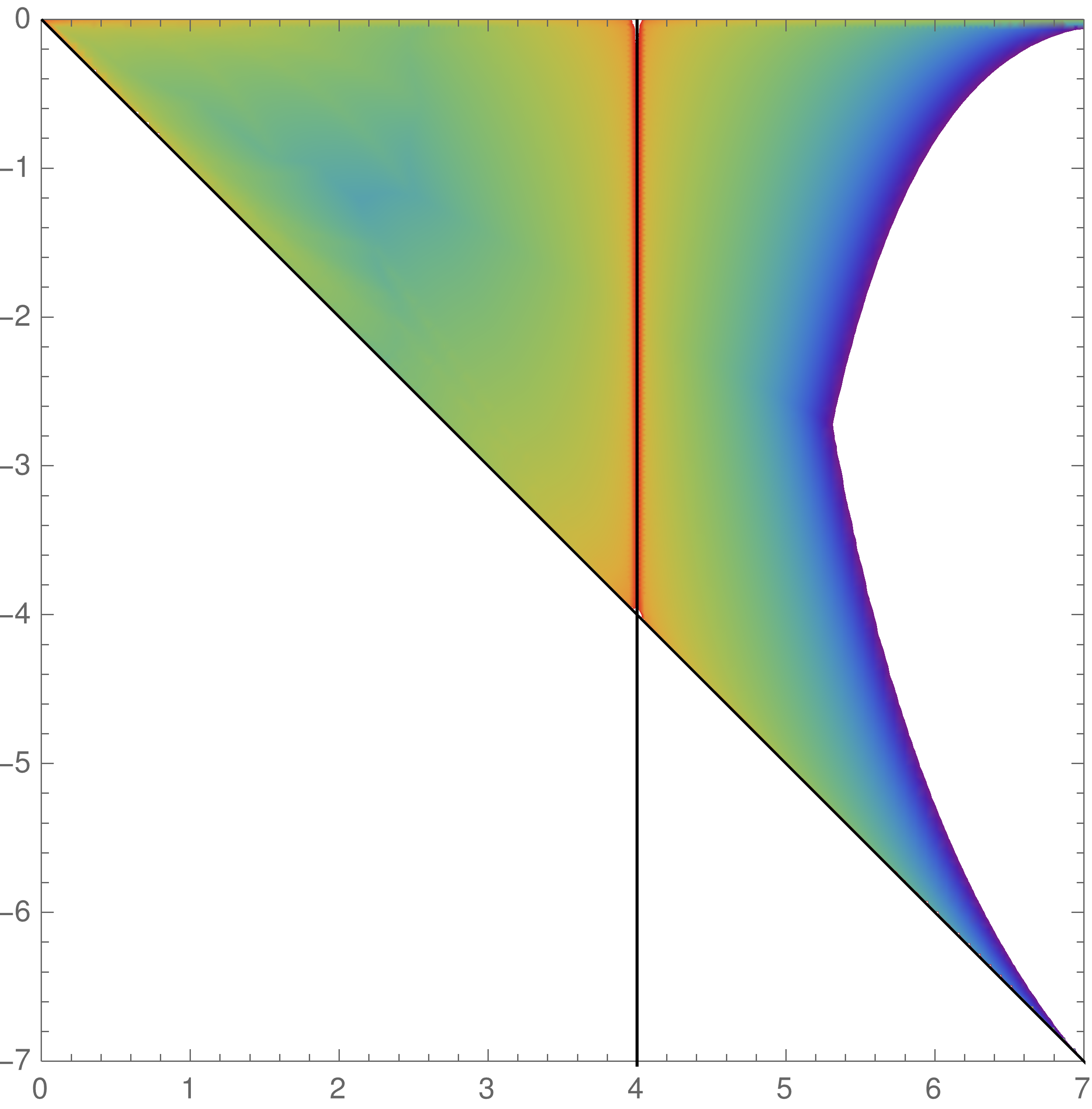}\label{subfig3}}
    \\
    
    \subfloat[$f_{+--+}^{(2,b)}$]{\includegraphics[width=0.3\textwidth]{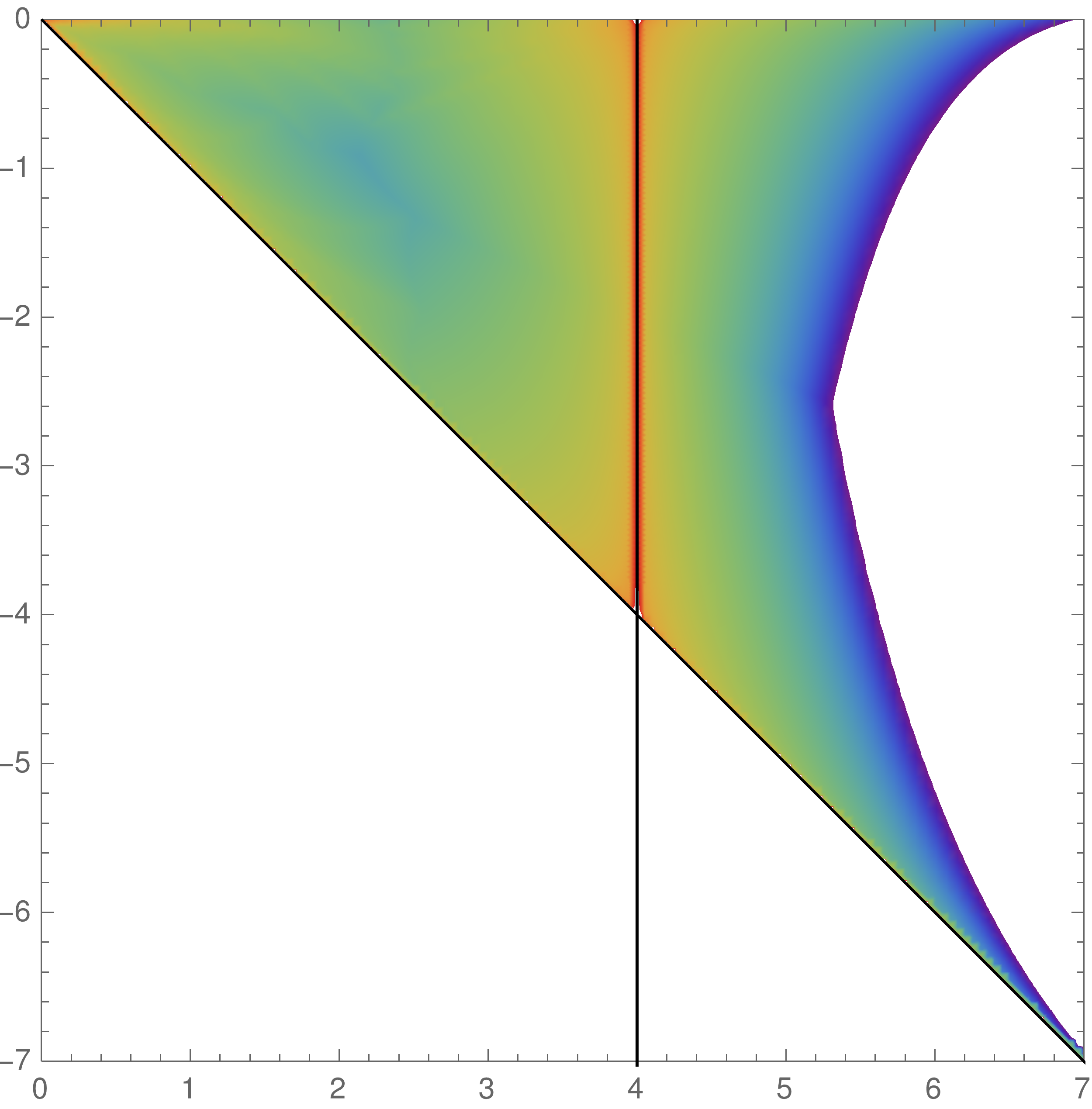}\label{subfig4}}
    \hspace{0.03\textwidth}
    \subfloat[$f_{-+++}^{(2,b)}$]{\includegraphics[width=0.3\textwidth]{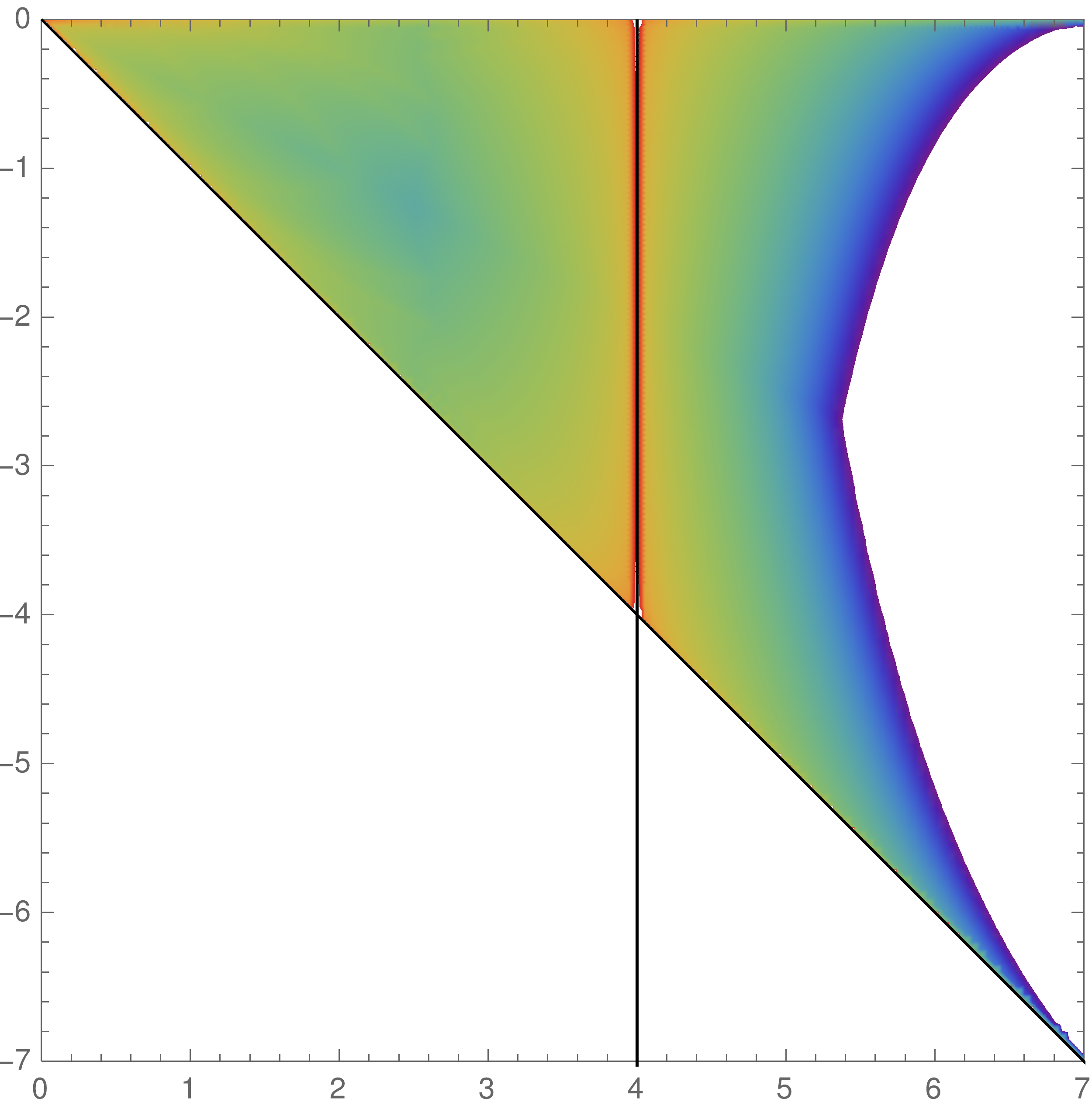}\label{subfig5}}
    \hspace{0.03\textwidth}
    \subfloat[$f_{+-++}^{(2,b)}$]{\includegraphics[width=0.3\textwidth]{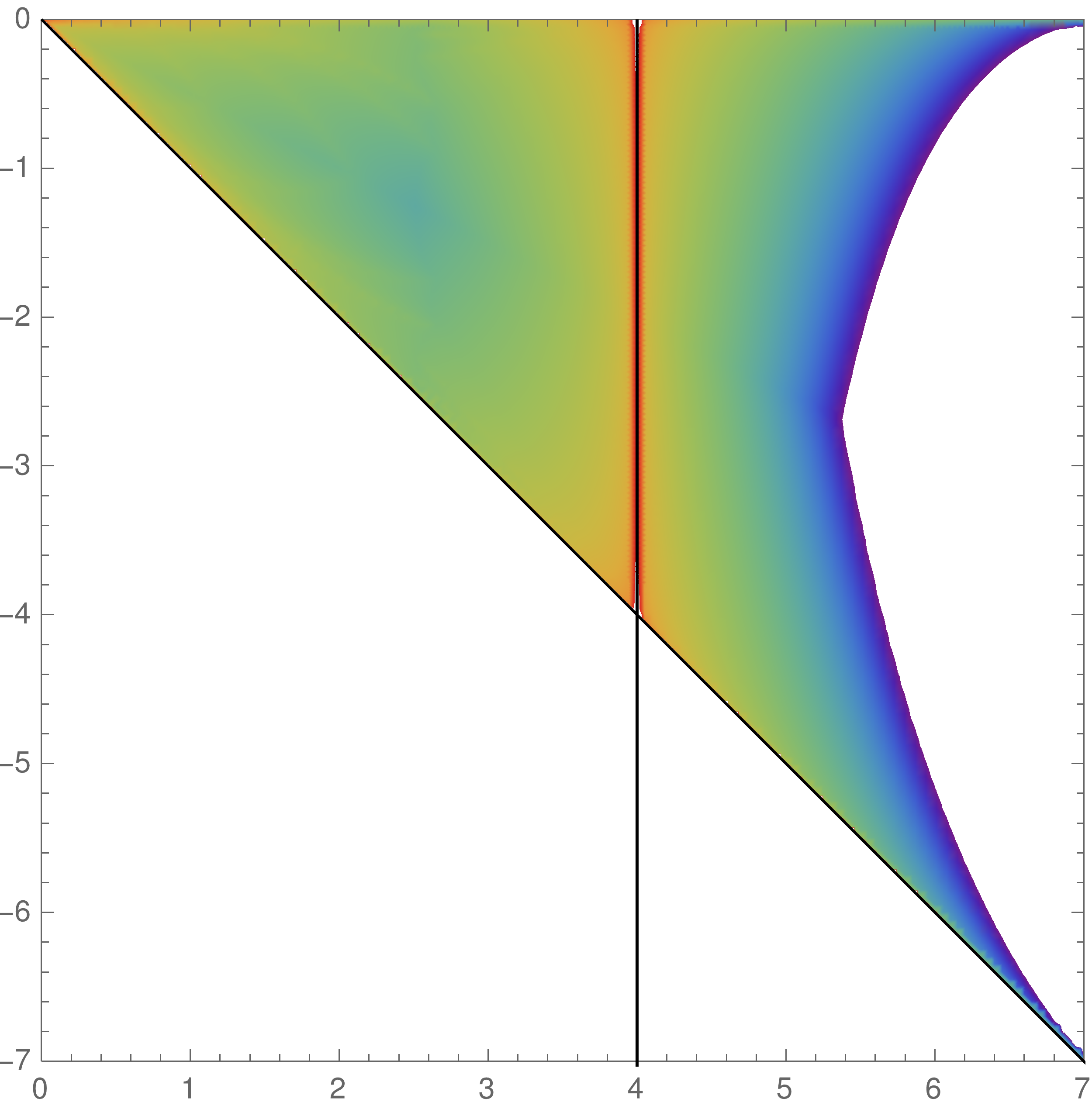}\label{subfig6}} \\
    
    \subfloat[$f_{++-+}^{(2,b)}$]{\includegraphics[width=0.3\textwidth]{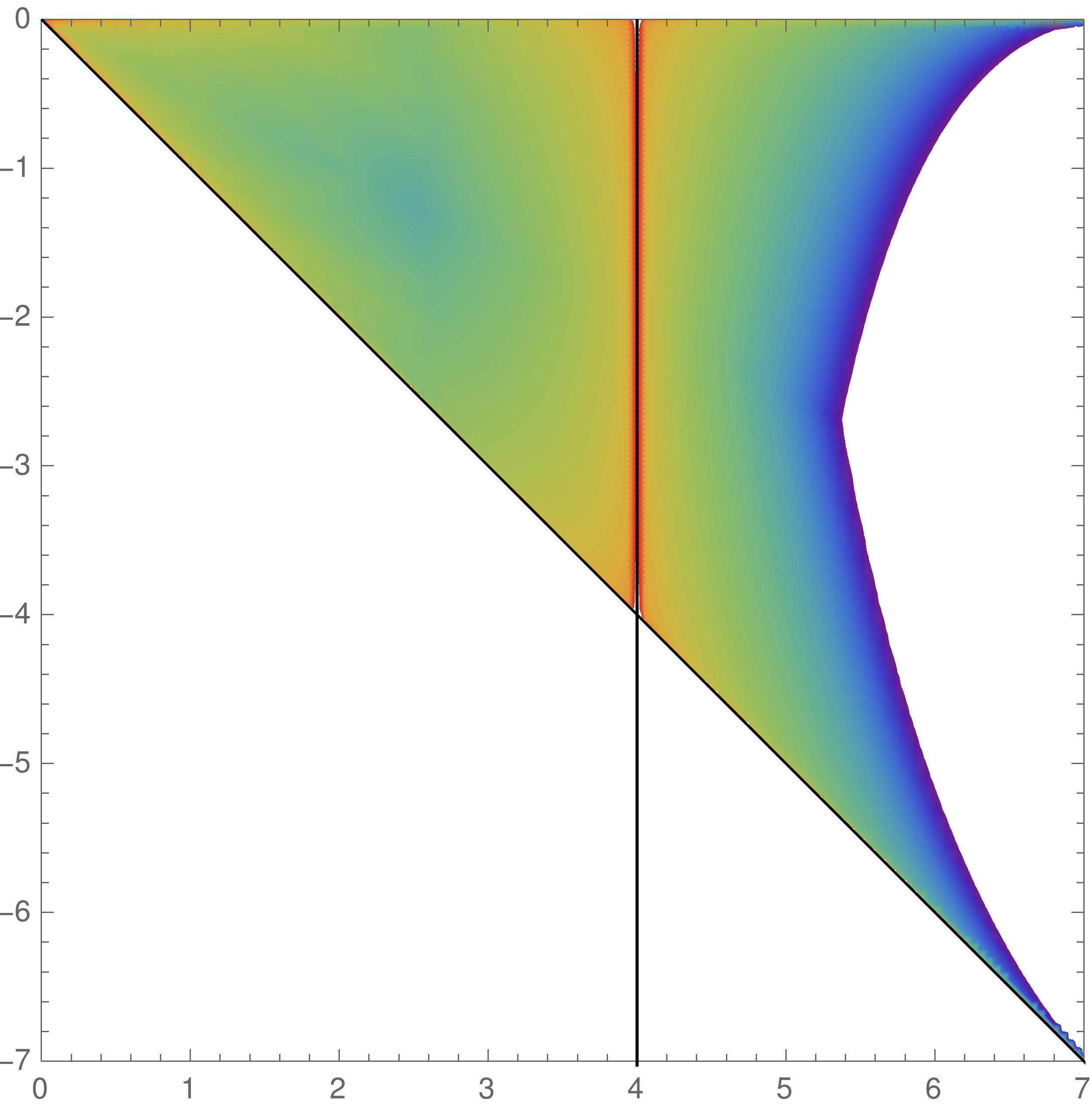}\label{subfig7}}
    \hspace{0.03\textwidth}
    \subfloat[$f_{+++-}^{(2,b)}$]{\includegraphics[width=0.3\textwidth]{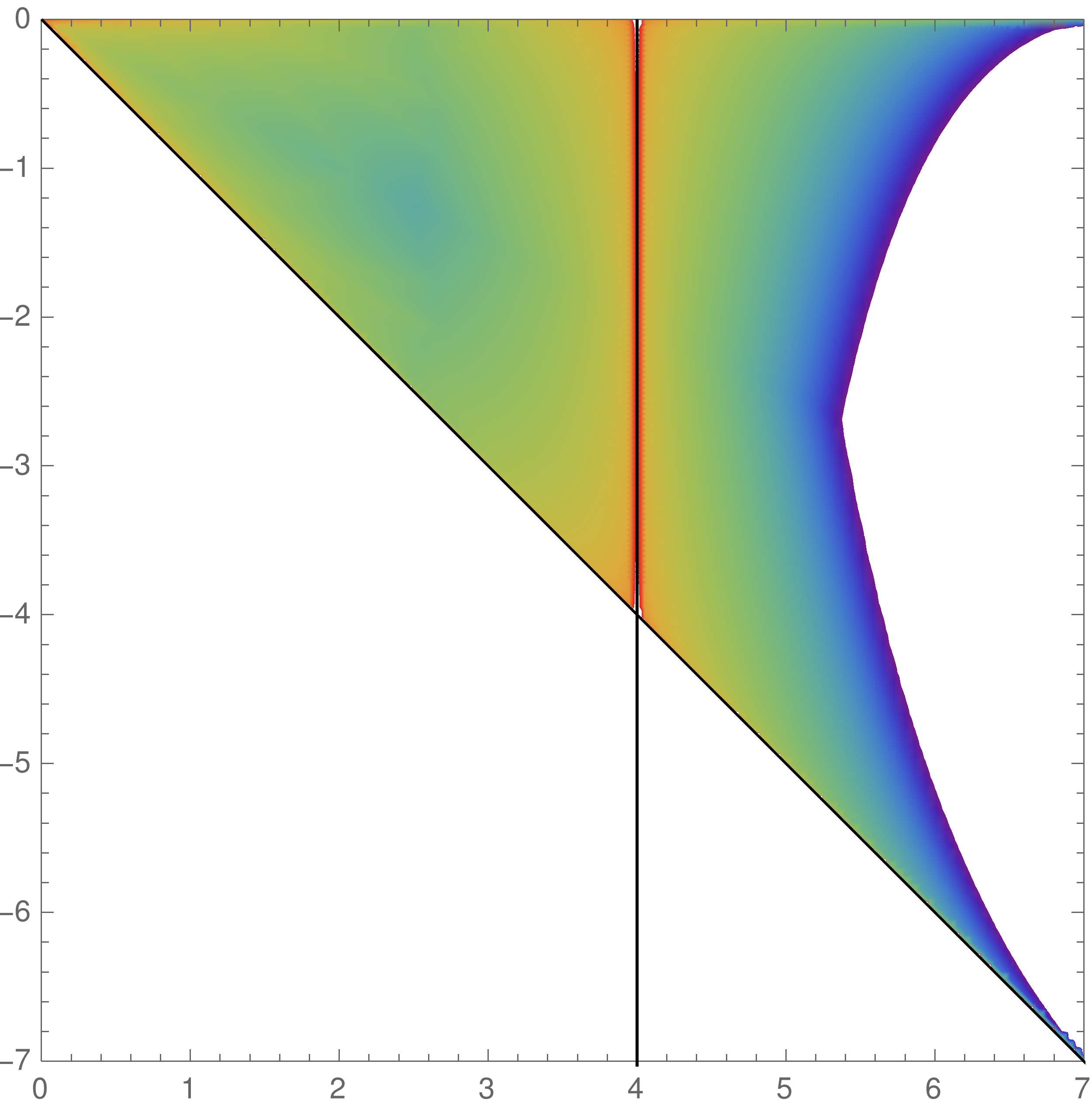}\label{subfig8}}
    \hspace{0.1\textwidth}
    {\includegraphics[width=0.065\textwidth]{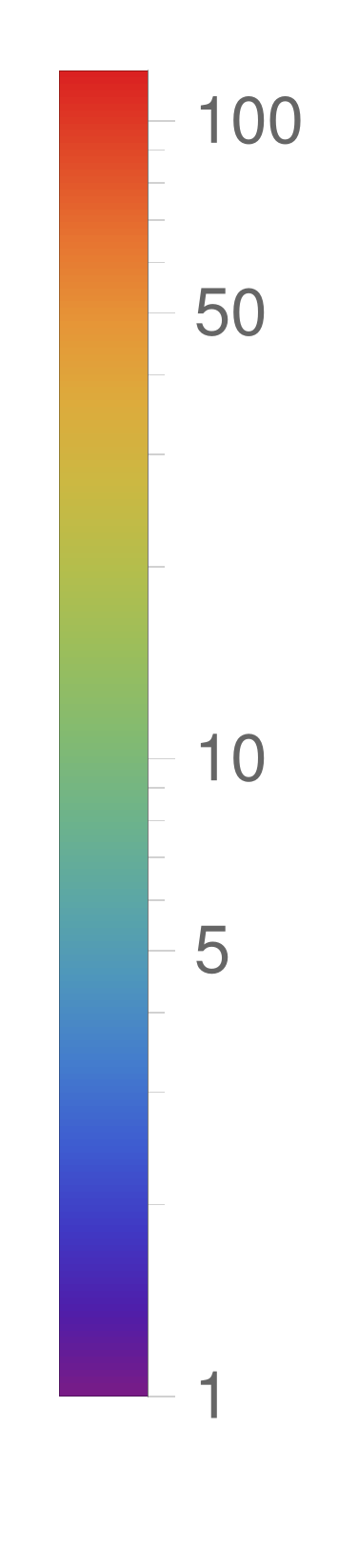}\label{subfig9}}
    \hspace{0.165\textwidth}
    
    \caption{Number of digits of precision obtained for the finite part of the bare helicity amplitudes in the gluon-fusion channel. The horizontal axis corresponds to $s/\mt^2$, the vertical axis to $t/\mt^2$.}
    \label{fig:ggaa}
\end{figure}

\begin{figure}[htbp]
    \centering
    \subfloat[$\alpha_{l}$]{\includegraphics[width=0.3\textwidth]{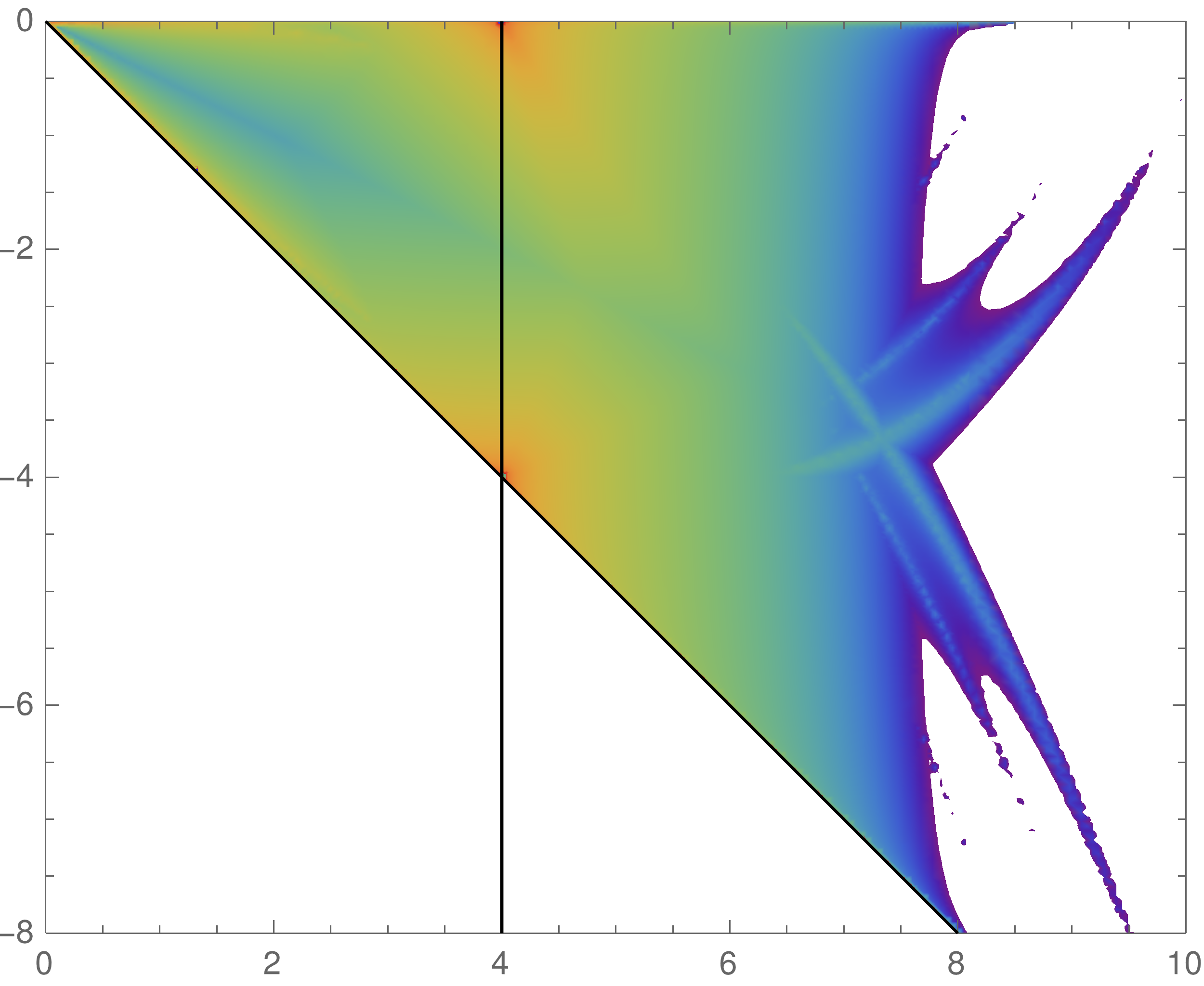}\label{subfig10}}
    \hspace{0.03\textwidth}
    \subfloat[$\beta_{l}$]{\includegraphics[width=0.3\textwidth]{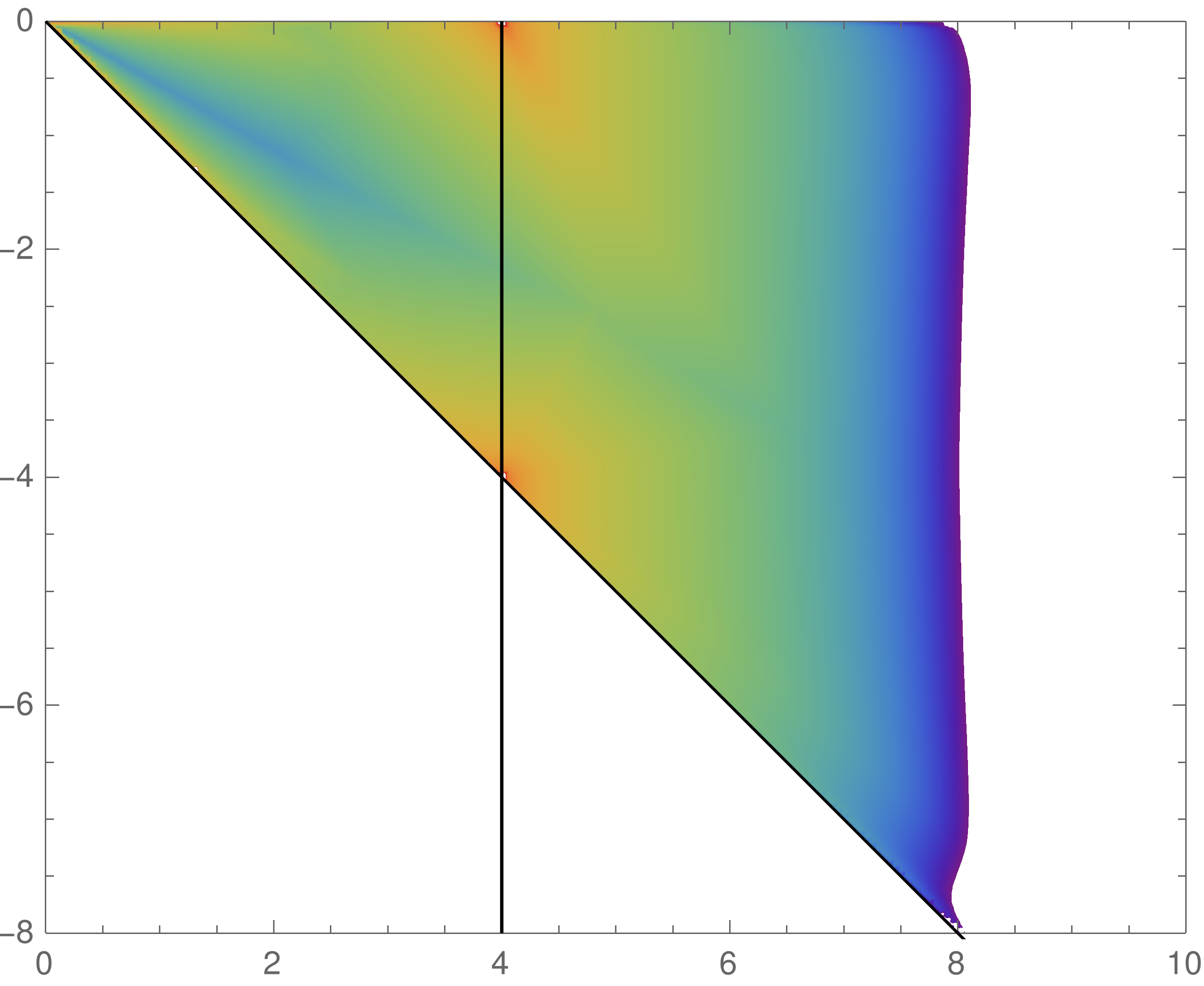}\label{subfig11}} 
    \hspace{0.03\textwidth}
    \subfloat[$\gamma_{l}$]{\includegraphics[width=0.3\textwidth]{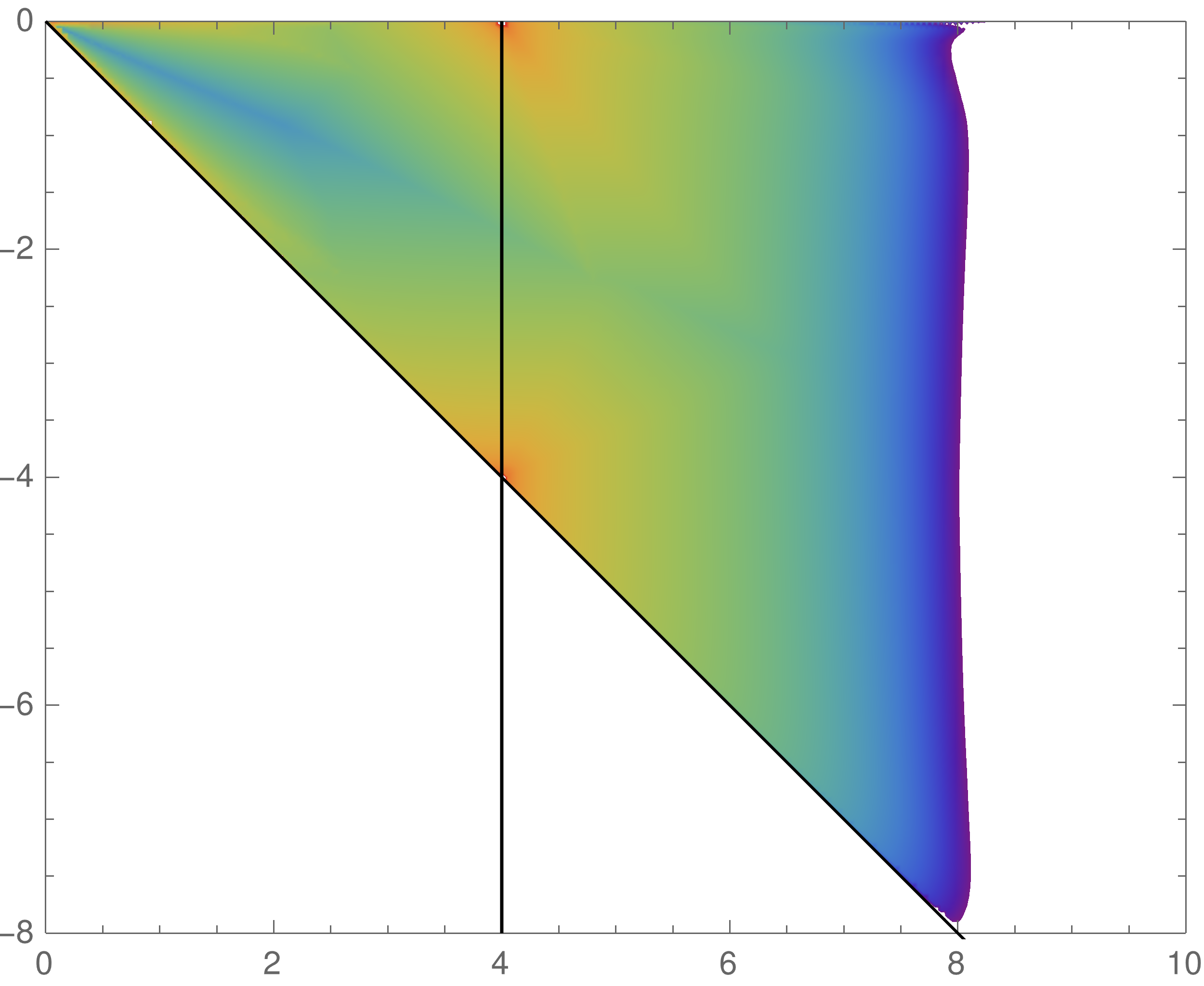}\label{subfig12}}
    \\
    
    \subfloat[$\delta_{l}$]{\includegraphics[width=0.3\textwidth]{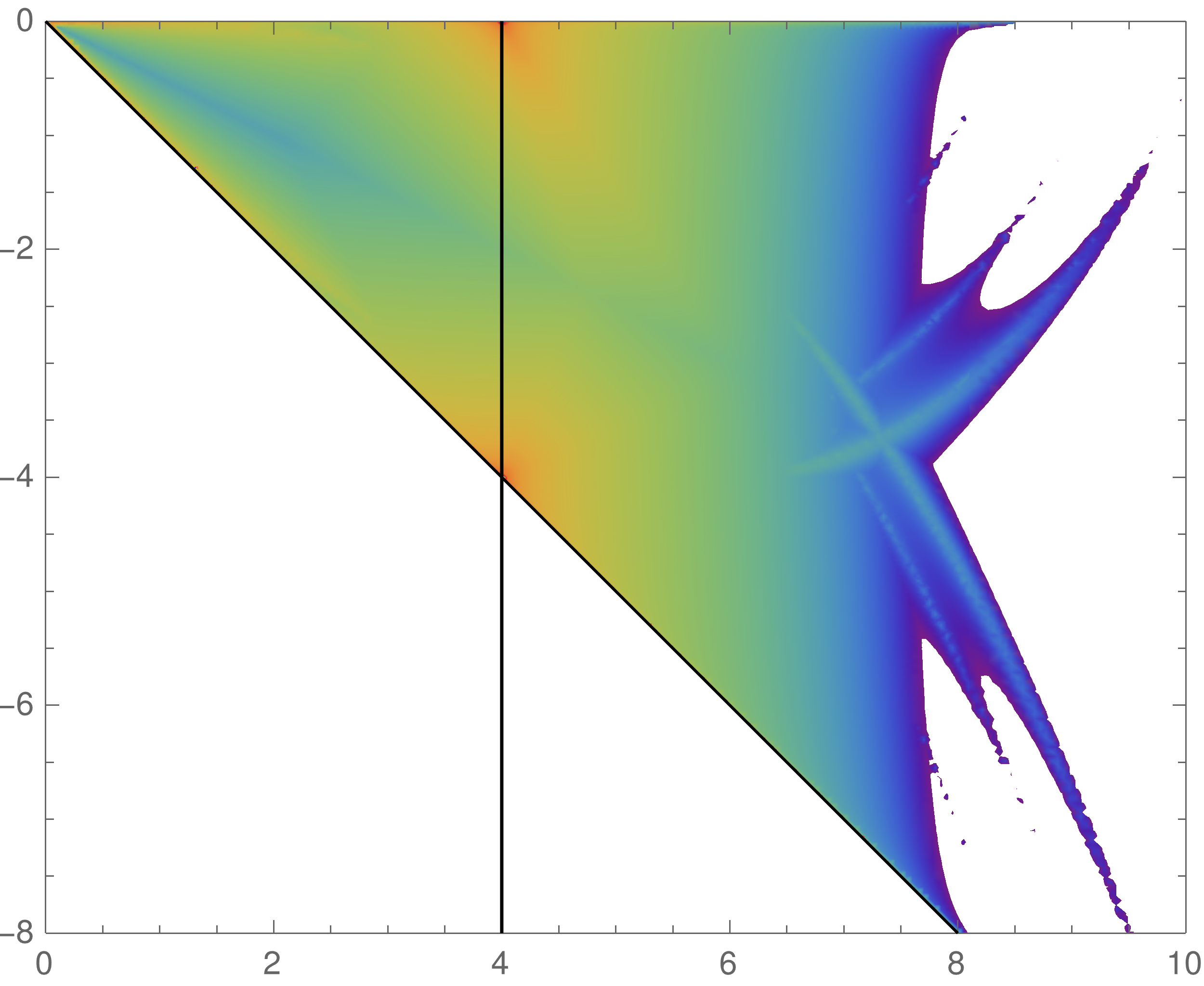}\label{subfig13}}
    \hspace{0.03\textwidth}
    \subfloat[$\alpha_{h}$]{\includegraphics[width=0.3\textwidth]{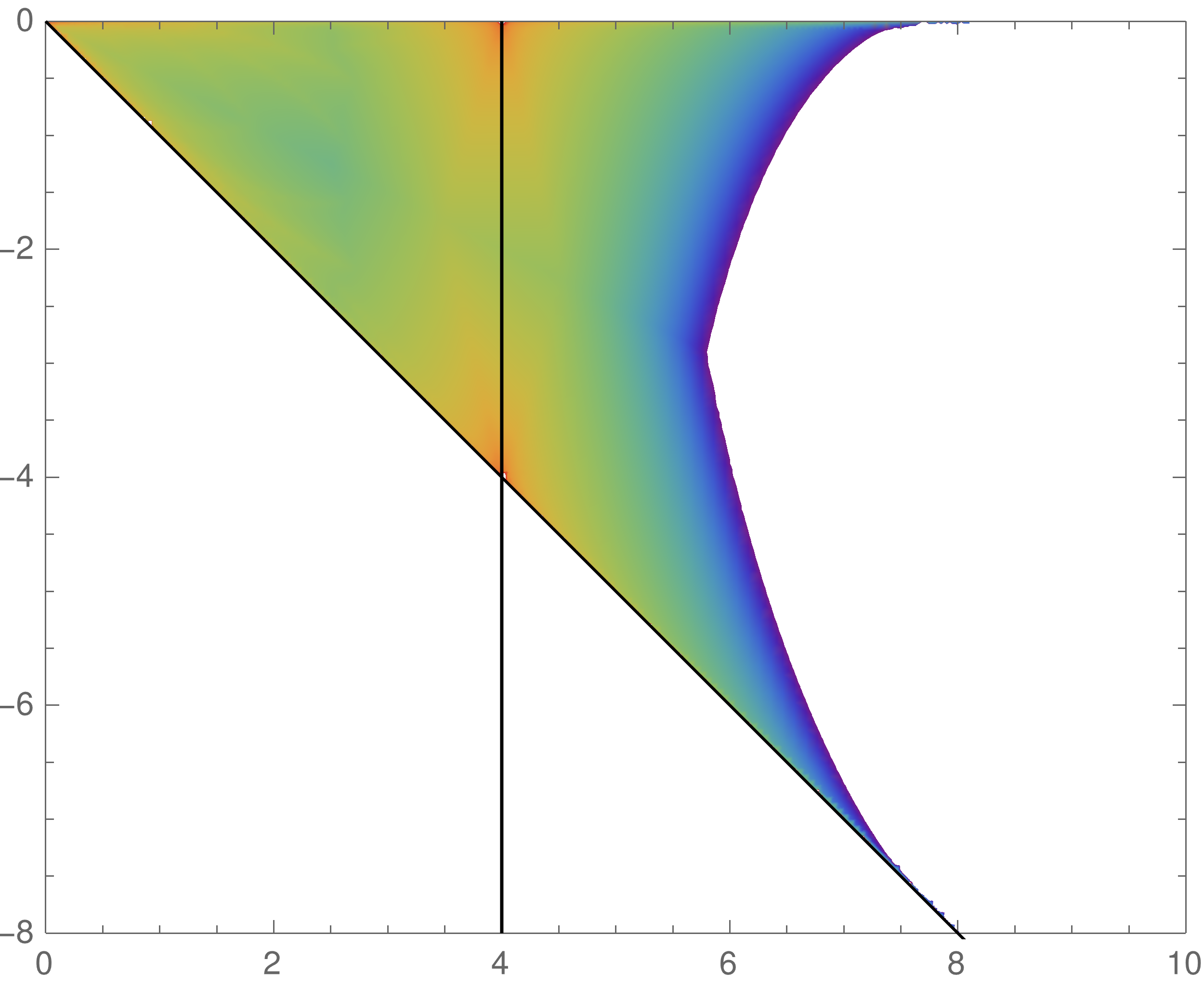}\label{subfig14}}
    \hspace{0.03\textwidth}
    \subfloat[$\beta_{h}$]{\includegraphics[width=0.3\textwidth]{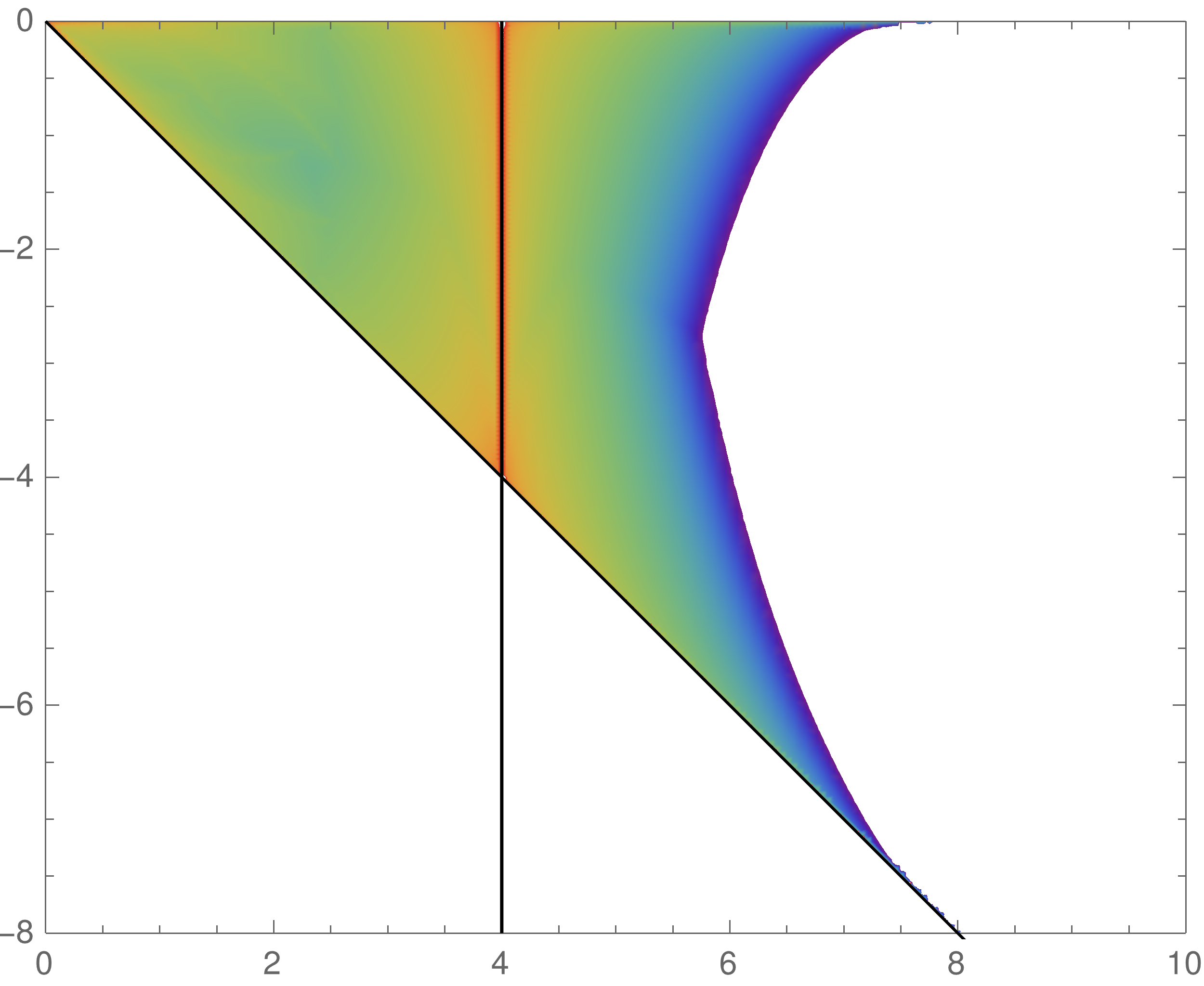}\label{subfig15}} \\
    
    \subfloat[$\gamma_{h}$]{\includegraphics[width=0.3\textwidth]{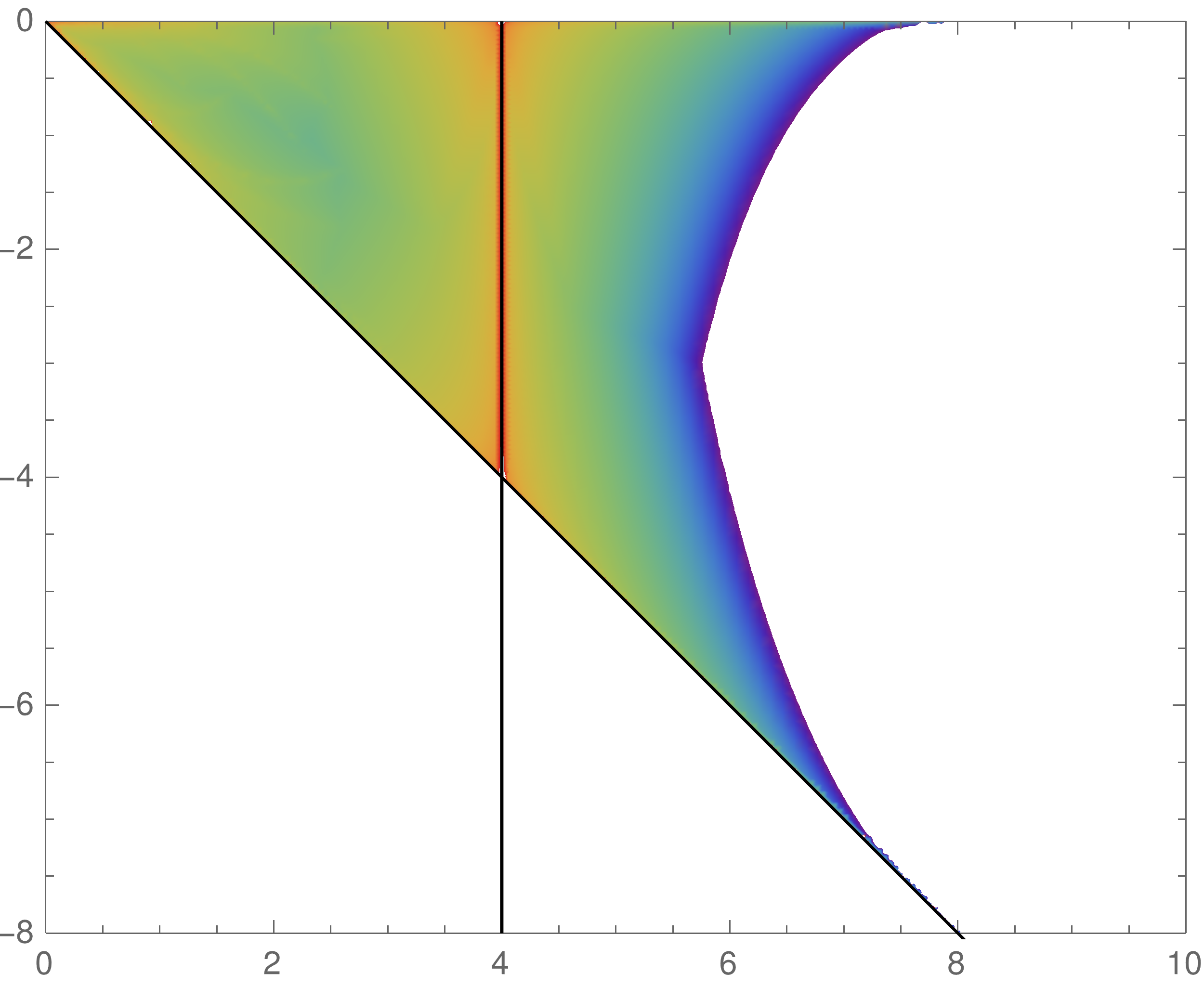}\label{subfig16}}
    \hspace{0.03\textwidth}
    \subfloat[$\delta_{h}$]{\includegraphics[width=0.3\textwidth]{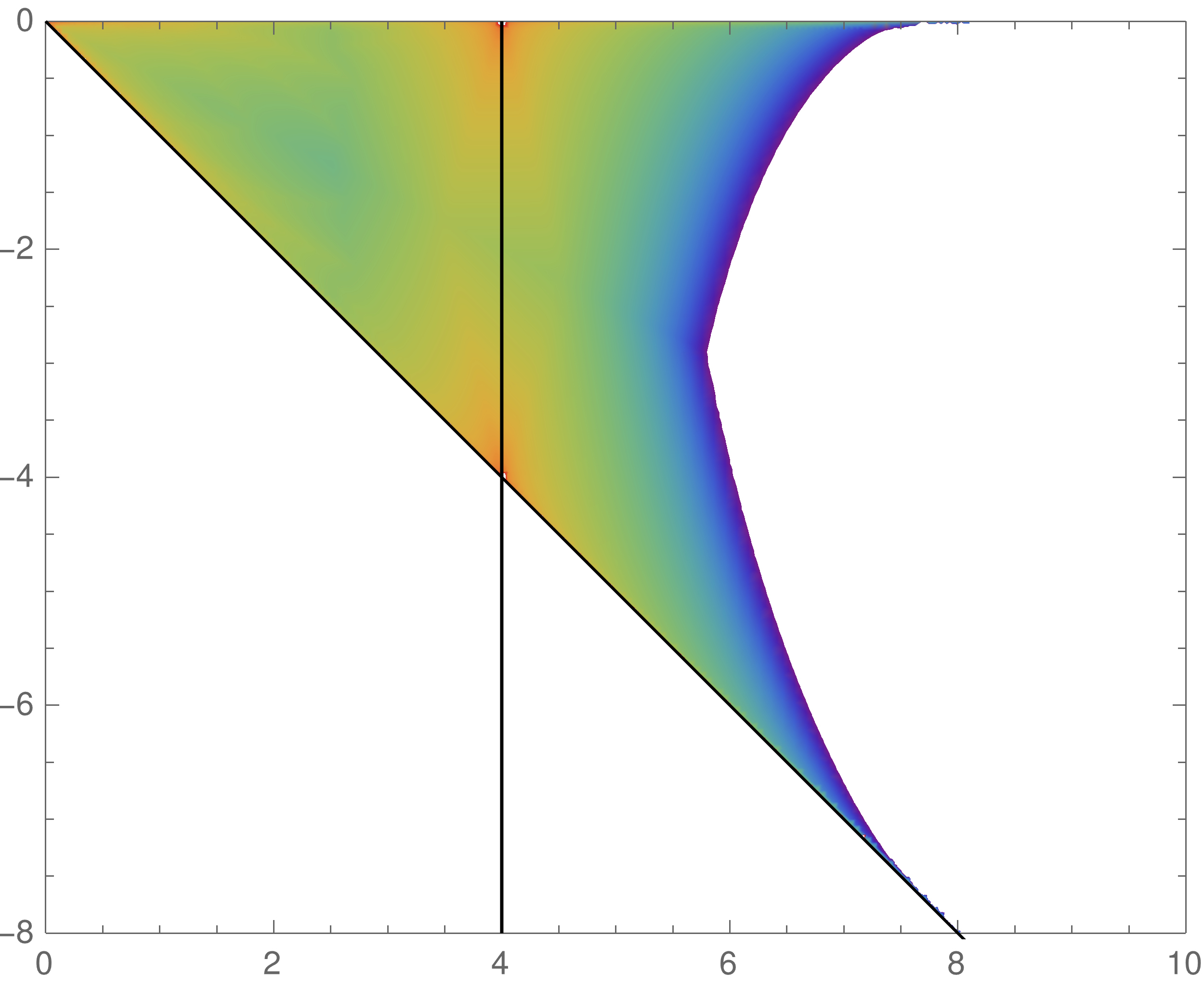}\label{subfig17}}
    \hspace{0.1\textwidth}
    {\includegraphics[width=0.05\textwidth]{Plots/PrecisionPlotScale.png}\label{subfig18}}
    \hspace{0.18\textwidth}
    
    \caption{Number of digits of precision obtained for the finite part of the bare helicity amplitudes in the quark-antiquark annihilation channel. The horizontal axis corresponds to $s/\mt^2$, the vertical axis to $t/\mt^2$.}
    \label{fig:qqaa}
\end{figure}

To get an estimate of the convergence of our series expansion representation across the whole phase space region and identify where our representation is not reliable anymore, a possible approach is to take differences of the partial sums to check how many digits are preserved when adding one more term to the expansion. This is based on the fact that, as long as the series converges, adding one more term to the series should improve on the previous result.
The best estimate of whether our series converges or not can then be obtained by the highest order term in the series, which is also the difference between the partial sums with the largest number of terms and the partial sum with one term less. 
Calling this term $\Omega_{X,\text{max}}$ for the helicity amplitude $\Omega_{X}$, we turn it into an estimate for the number of digits of precision as follows: We take its absolute value, compute its logarithm with base $10$, and multiply the result by minus one,
\begin{equation}
    - \log_{10} \left( \left\vert \Omega_{X,\text{max}} \right\vert \right) \, .
\end{equation}
The result for the finite part of all the bare two-loop helicity amplitudes is shown in~\cref{fig:ggaa} and~\cref{fig:qqaa}, where we have taken in every point the best value of the four expansions available. Note that we treated $\Omega_{h}$ and $\Omega_{l}$ as defined in~\cref{eq:partsqqamp} separately. 

Close to the expansion points, the convergence is very rapid, and $\Omega_{X,\text{max}}$ is of order $10^{-100}$, corresponding to the red-colored regions in the figures. As the distance to the expansion points increases, so does $\Omega_{X,\text{max}}$, as the speed of convergence slows down, resulting in the colors becoming colder. Finally, at the dark blue- to purple-colored edges bordering the white regions, $\Omega_{X,\text{max}}$ is of order $10^{-1}$, and we expect that the series expansion representation up to the given order will cease to provide an accurate numerical value for the helicity coefficients.

On general grounds, each of the expansions is expected to converge in a region around the expansion point, whose size is essentially determined by the distance to the next physical singularity, with the convergence improving as additional orders in the expansions are calculated. While most expansions appear to converge at most for $s = 8\mt^2$, convergence beyond such bounds can, however, not be excluded, as it happens, for instance, for $\alpha_{l}$ and $\delta_{l}$, see~\cref{fig:qqaa}. To fully cover phase space regions with $s \gtrsim 8\mt^2$, the expansions performed can be supplemented by additional expansions, for instance, an expansion in the limit of small quark mass~\cite{Lee:2024dbp}.
We will consider this problem in a subsequent publication, where also all amplitudes for the production of up to two massless partons will
be computed.

Finally, we show in~\cref{tb:HelAmpsValues} the numerical results for all the bare $2$-loop helicity coefficients in a selected phase-space point for reference\footnote{We thank Taushif Ahmed, Amlan Chakraborty, Ekta Chaubey, and Mandeep Kaur for enabling a comparison of results prior to publication of this article~\cite{AEMT:2025}. We found complete agreement in the point used for~\cref{tb:HelAmpsValues}, as well as two further phase-space points, $(s/\mt^2,t/\mt^2)=(13/10,-3/5)$ and $(s/\mt^2,t/\mt^2)=(51/10,-11/10)$.}.

\begin{table}[h!]
\begin{center}
\begin{tabular}{ |p{1cm}||p{0.8cm}|p{3.3cm}|p{3.3cm}|p{4.1cm}| }
 \hline
   & $1/\epsilon^3$ & $1/\epsilon^2$ & $1/\epsilon$ & $\epsilon^0$ \\
 \hline
 \hline
  $f_{++++}$ & 0 & $-2.6098$ & $-19.546 - 8.1990 i $& $-32.044 - 3.5038 i$ \\
  \hline
  $f_{-+++}$ & 0 & $-0.032704$ & $-0.17702 - 0.10274 i$ & $-0.36482 - 0.22727 i$\\
  \hline
  $f_{+-++}$ & 0 & $-0.032704$ & $-0.17702 - 0.10274 i$ & $-0.36482 - 0.22727 i$\\
  \hline
  $f_{++-+}$ & 0 & $-0.032704$ & $-0.17702 - 0.10274 i$ & $-0.08140 - 0.34717 i$ \\
  \hline
  $f_{+++-}$ & 0 & $-0.032704$ & $-0.17702 - 0.10274 i$ & $-0.08140 - 0.34717 i$ \\
  \hline
  $f_{--++}$ & 0 & $4.8124$ & $29.196 + 15.119 i$ & $36.108 + 4.2217 i$ \\
  \hline
  $f_{-+-+}$ & 0 & $1.2813$ & $4.4086 + 4.0254 i$ & $4.4725 + 6.9322 i$ \\
  \hline
  $f_{+--+}$ & 0 & $0.29001$ & $1.00104 + 0.91108 i$ & $1.0263 + 1.6390 i$ \\
  \hline
  $\alpha_{l}$ & 0 & 0 & $80/99$ & $0.19228 + 0.05399 i$ \\
  \hline
  $\beta_{l}$ & -4/9 & $-0.08921 - 1.39626 i$ & $1.6850 - 1.5085 i$& $0.75751 + 0.83552 i$ \\
  \hline 
  $\gamma_{l}$ & -4/9 & $-0.08921 - 1.39626 i$ & $2.0198 - 2.0749 i$ & $2.6238 - 1.1535 i$ \\
  \hline
  $\delta_{l}$ & 0 & 0 & $-80/99$ & $-0.19228 - 0.05399 i$ \\
 \hline
  $\alpha_{h}$ & 0 & 0 & 0 & $-0.0088466 + 0.0055248 i$   \\
  \hline
  $\beta_{h}$ & 0 & 0 & 0 & $0.19945 - 0.55326 i$ \\
  \hline 
  $\gamma_{h}$ & 0 & 0 & 0 & $0.19886 - 0.55796 i$ \\
  \hline
  $\delta_{h}$ & 0 & 0 & 0 & $0.0088466 - 0.0055248 i$ \\
 \hline
\end{tabular}
\end{center}
\caption{Numerical value of the 2-loop bare helicity coefficients in the point $(s/\mt^2,t/\mt^2)=(11/3,-5/2)$ with $N_c=3$ and the electric charge of the respective quark coupling to the photon in units of the electron charge set to one. Further, we extracted a factor $(\mu_0^2/\mt^2)^{2\epsilon}$. We stress that the bare helicity amplitudes are computed with the normalization of the integrals defined in~\cref{eq:measure}.} 
\label{tb:HelAmpsValues}
\end{table}

\section{Conclusions}
\label{sec:conc}
In this paper, we have addressed the analytic computation of the two-loop scattering amplitudes for the production of a pair of photons in quark-antiquark annihilation and gluon fusion, mediated by a loop of massive quarks. While these amplitudes had already been computed using semi-numerical methods, new developments in the theory of Feynman integrals defined on non-trivial geometries, including canonical differential equations and their solutions as iterated integrals, allowed us to reconsider this problem from a different perspective.
In particular, we have derived a system of canonical differential equations from which we could obtain analytic results in terms of independent iterated integrals defined over mixtures of logarithmic and elliptic differential forms. 
With these expressions, we could verify recent conjectures on the cancellation of some entire classes of elliptic differential forms from the finite part in $\epsilon=0$ of the scattering amplitudes. 
Moreover, these equations can be used as the starting point to obtain fast converging series representations for the result close to any regular or singular point from the differential equations. In particular, here we produced double series expansion solutions close to the point $s=4\mt^2$, $t=0$, and the point $\mt^2 = \infty$, corresponding to what is usually referred to as large-mass expansion.
While in the large mass limit, the calculation becomes trivial, at the point $s=4\mt^2$, $t=0$, the elliptic curve that characterizes the problem does not degenerate, and one obtains series expansion solutions in two variables with coefficients that are defined as iterated integrals over elliptic kernels. All these numerical coefficients can be evaluated with very large precision and allow us to produce a compact series expansion representation for the entire amplitude close to the threshold, which can be evaluated numerically in fractions of a second. This result can be used to supplement the much simpler series expansions that can be obtained at  $\mt^2 = \infty$ to provide reliable solutions across large portions of the phase space where mass effects are expected to be sizable.

We believe that the results provided here constitute a proof of concept of how amplitudes, which require Feynman integrals defined on non-trivial geometries, can be computed analytically and numerically efficiently, leveraging new developments in the theory of differential equations for Feynman integrals. It is to be expected that a minor generalization of these results will allow us to obtain similar results for the production of two jets or a jet and photon, mediated by a top quark. It will be interesting to apply the same techniques to more involved problems in the future.

\acknowledgments
We thank Federico Ripani for collaboration during the initial stages of this project and Felix Forner and Nikolaos Syrrakos for insightful discussions. 
We are grateful to the Munich Institute for Astro-, Particle and BioPhysics (MIAPbP), funded by the Deutsche Forschungsgemeinschaft (DFG, German Research Foundation) under Germany´s Excellence Strategy – EXC-2094 – 390783311, where part of these results have been obtained.
CN, LT and FW were supported in part by the Excellence Cluster ORIGINS 
funded by the Deutsche Forschungsgemeinschaft (DFG, German Research Foundation) under Germany’s 
Excellence Strategy – EXC-2094-390783311 and in part by the European Research Council (ERC) under the European Union’s research and innovation program grant agreements 949279 (ERC Starting Grant HighPHun). 
The work of MB was supported by the European Research Council (ERC) under the European Union’s Horizon Europe research and innovation program grant agreement 101040760, (ERC Starting Grant FFHiggsTop).
The work of FC was supported by the European Research Council (ERC) under the European Union’s Horizon Europe research and innovation program grant agreement 101078449 (ERC Starting Grant MultiScaleAmp) and by Generalitat Valenciana GenT Excellence Programme (CIDEGENT/2020/011), ILINK22045 and Generalitat Valenciana Grants No. PROMETEO/2021/071.
Views and opinions expressed are however those of the authors only and do not necessarily reflect those of the European Union or the European Research Council Executive Agency. Neither the European Union nor the granting authority can be held responsible for them.

\appendix

\section{Helicity Projectors}
\label{app:helproj}
In this appendix, we provide the explicit analytic form of the helicity projectors that we use to compute directly the helicity amplitudes in~\cref{eq:helampqq,eq:alphabetagg}.
We use the notation $\mathcal{P}_{\omega}$ with $\omega=\{\alpha,\beta,\gamma,\delta\}$ for the projectors onto the quark helicity amplitudes, i.e.,
\begin{equation*}
    \mathcal{P}_{\omega} \cdot \mathcal{A}_{qq} = \omega(x,y)\,,
\end{equation*}
where $x$ and $y$ were defined in~\cref{eq:variabledefs}. Similarly, we define $\mathcal{P}_{\lambda_1\lambda_2\lambda_3\lambda_4}$ 
the projectors on the gluon helicity amplitudes
\begin{equation*}
\mathcal{P}_{\lambda_1\lambda_2\lambda_3\lambda_4} \cdot \mathcal{A}_{gg} = f_{\lambda_1\lambda_2\lambda_3\lambda_4}(x,y)\,.
\end{equation*}
We decompose the quark helicity projectors onto the basis of tensors as
\begin{align*}
    \mathcal{P}_{\alpha}& = \frac{1}{(d-3)} \left[ \frac{u}{8 s^2} \left( \tau_1^{\dag} + \tau_2^{\dag}\right) + \frac{2t + (4-d)u}{8s^2t} \tau_3^{\dag} + \frac{t^2 + t u}{8s^2t} \tau_4^{\dag} \right]\, , \nonumber\\
    \mathcal{P}_{\beta}& = \frac{1}{(d-3)} \left[ \frac{u}{8st} \left(\tau_2^{\dag} - \tau_1^{\dag}\right) + \frac{2t + du}{8s^2t} \tau_3^{\dag} - \frac{1}{8t} \tau_4^{\dag}\right]\, , \nonumber\\
    \mathcal{P}_{\gamma}& = \frac{1}{(d-3)} \left[ \frac{1}{8s} \left(\tau_2^{\dag} - \tau_1^{\dag}\right) + \frac{1}{8u} \tau_4^{\dag} - \frac{2t + (d-4)u}{8stu} \tau_3^{\dag}\right]\, , \nonumber\\
    \mathcal{P}_{\delta}& = \frac{1}{(d-3)} \left[ \frac{u}{8s^2} \left(\tau_1^{\dag} + \tau_2^{\dag}\right) + \frac{1}{8s} \tau_4^{\dag} + \frac{(d-4)u - 2t}{8s^2t} \tau_3^{\dag}\right] \, , \nonumber
\end{align*}
where 
\begin{equation*}
    \tau_i^\dagger = \bar{u}(p_1) \Gamma_i^{\mu \nu} u(p_2) \epsilon_{3,\mu}^*(p_3) \epsilon_{4,\nu}^*(p_4)\,,
\end{equation*}
and the $\Gamma_i^{\mu\nu}$ have been defined in~\cref{eq:tensqq}.

In the same way, we decompose the gluon helicity projectors as
\begin{align*}
\mathcal{P}_{++++} &= \frac{1}{3(d-1)(d-3)t^2}\biggl[
    \frac{3 \bigl((4 - d)(d - 2) s^2 + (4 - d)(d - 2) s t - 2 (d - 2) t^2\bigr)}{4 s (s+t)} \mathcal{T}_1^{\dag}\\
    &\quad+ \frac{3 \bigl((d - 2) s^2 t + (d - 2) s t^2\bigr)}{4 s (s+t)} \mathcal{T}_2^{\dag} 
    + \frac{3\bigl(-d s^2 t - 2 s t^2\bigr)}{4 s (s+t)}  \mathcal{T}_3^{\dag}\\
    &\quad- \frac{3 \bigl(d s^2 t + d s t^2 - 2 s^2 t - 2 s t^2\bigr)}{4 s (s+t)} \left( \mathcal{T}_4^{\dag} - \mathcal{T}_5^{\dag} + \mathcal{T}_7^{\dag}\right) \\
    &\quad+ \frac{3 \bigl(d s^2 t + 2 d s t^2 + (d - 2) t^3 - 2 s t^2\bigr)}{4 s (s+t)} \mathcal{T}_6^{\dag}
    + \frac{3 \bigl(s^2 t^2 + s t^3\bigr)}{4 s (s+t)}\mathcal{T}_8^{\dag}
\biggr], \\[6pt]
\mathcal{P}_{-+++} &= \frac{1}{3(d-1)(d-3)t^2}\biggl[
    \frac{3 \bigl(d^2 s (s+t) - 4 d s^2 - 8 d s t - 2 d t^2 + 4 t (s+t)\bigr)}{4 s (s+t)} \mathcal{T}_1^{\dag} \\
    &\quad- \frac{3 \bigl(d s^2 t + d s t^2\bigr)}{4 s (s+t)} \left( \mathcal{T}_2^{\dag} - \mathcal{T}_4^{\dag}\right) 
    + \frac{3 \bigl(d s^2 t + d s t^2 - 2 s t (s+t)\bigr)}{4 s (s+t)}\left( \mathcal{T}_5^{\dag} - \mathcal{T}_7^{\dag} \right) \\
    &\quad+ \frac{3 \bigl(d s^2 t + 2 d s t^2 + d t^3 - 2 t^2 (s+t)\bigr)}{4 s (s+t)}\mathcal{T}_6^{\dag} + \frac{3 \bigl(d s^2 t - 2 s t (s+t)\bigr)}{4 s (s+t)}\mathcal{T}_3^{\dag} + \frac{3 t^2 }{4}\mathcal{T}_8^{\dag}
\biggr], \\[6pt]
\mathcal{P}_{+-++} &= \frac{1}{3(d-1)(d-3)t^2}\biggl[
    \frac{3 \bigl(d^2 s (s+t) - 4 d s^2 + 2 d t^2 - 4 s t\bigr)}{4 s (s+t)} \mathcal{T}_1^{\dag}\\
    &\quad- \frac{3 \bigl(d s^2 t + d s t^2\bigr)}{4 s (s+t)}
    \left( \mathcal{T}_2^{\dag} + \mathcal{T}_5^{\dag} \right) + \frac{3 \bigl(-d s^2 t - 2 s t^2\bigr)}{4 s (s+t)} \mathcal{T}_3^{\dag}\\
    &\quad+ \frac{3 \bigl(-d s^2 t - d s t^2 + 2 s^2 t + 2 s t^2\bigr)}{4 s (s+t)} \left( \mathcal{T}_4^{\dag} + \mathcal{T}_7^{\dag} \right)\\
    &\quad+ \frac{3 \bigl(-d s^2 t - 2 d s t^2 - d t^3 + 2 s^2 t + 2 s t^2\bigr)}{4 s (s+t)} \mathcal{T}_6^{\dag} + \frac{3 t^2}{4} \mathcal{T}_8^{\dag}
\biggr], \\[6pt]
\mathcal{P}_{++-+} &= \frac{1}{3(d-1)(d-3)t^2}\biggl[
    \frac{3 \mathcal{T}_1^{\dag}\bigl(d^2 s (s+t) - 4 d s^2 - 8 d s t - 2 d t^2 + 4 t (s+t)\bigr)}{4 s (s+t)} \\
    &\quad+ \frac{3 \bigl(d s^2 t + d s t^2 - 2 s t (s+t)\bigr)}{4 s (s+t)} \left(  \mathcal{T}_2^{\dag} -  \mathcal{T}_4^{\dag}\right) 
    + \frac{3 \bigl(d s^2 t - 2 s t (s+t)\bigr)}{4 s (s+t)} \mathcal{T}_3^{\dag}\\
    &\quad- \frac{3 \bigl(d s^2 t + d s t^2\bigr)}{4 s (s+t)} \left( \mathcal{T}_5^{\dag} - \mathcal{T}_7^{\dag}\right) + \frac{3 t^2 }{4}\mathcal{T}_8^{\dag} \\
    &\quad+ \frac{3 \bigl(d s^2 t + 2 d s t^2 + d t^3 - 2 t^2 (s+t)\bigr)}{4 s (s+t)}  \mathcal{T}_6^{\dag}   
\biggr], \\[6pt]
\mathcal{P}_{+++-} &= \frac{1}{3(d-1)(d-3)t^2}\biggl[
    \frac{3 \bigl(d^2 s (s+t) - 4 d s^2 + 2 d t^2 - 4 s t\bigr)}{4 s (s+t)} \mathcal{T}_1^{\dag}\\
    &\quad+ \frac{3 \bigl(d s^2 t + d s t^2 - 2 s^2 t - 2 s t^2\bigr)}{4 s (s+t)} \left( \mathcal{T}_2^{\dag} + \mathcal{T}_5^{\dag}\right) \\
    &\quad+ \frac{3 \bigl(-d s^2 t - 2 s t^2\bigr)}{4 s (s+t)}\mathcal{T}_3^{\dag} 
    + \frac{3 \bigl(d s^2 t + d s t^2\bigr)}{4 s (s+t)} \left( \mathcal{T}_4^{\dag} + \mathcal{T}_7^{\dag}\right)\\
    &\quad+ \frac{3 \bigl(-d s^2 t - 2 d s t^2 - d t^3 + 2 s t (s+t)\bigr)}{4 s (s+t)}\mathcal{T}_6^{\dag} + \frac{3 t^2 }{4}\mathcal{T}_8^{\dag}
\biggr], \\[6pt]
\mathcal{P}_{--++} &= \frac{1}{3(d-1)(d-3)t^2}\biggl[
    -\frac{3 \bigl((d - 4)(d - 2) s^2 + (d - 4)(d - 2) s t - 2 d t^2\bigr)}{4 s (s+t)}\mathcal{T}_1^{\dag} \\
    &\quad- \frac{3 \bigl(-d s^2 t + (2 - d) s t^2 + 2 s^2 t\bigr)}{4 s (s+t)} \mathcal{T}_2^{\dag} - \frac{3 \bigl(-d s^2 t + 2 s^2 t + 2 s t^2\bigr)}{4 s (s+t)}\mathcal{T}_3^{\dag} \\
    &\quad+ \frac{3 \bigl(d s^2 t + d s t^2\bigr)}{4 s (s+t)}\left( \mathcal{T}_4^{\dag} - \mathcal{T}_5^{\dag}\right) 
    - \frac{3 \bigl(d s^2 t + d s t^2 - 2 s^2 t - 2 s t^2\bigr)}{4 s (s+t)} \mathcal{T}_7^{\dag}\\
    &\quad- \frac{3\bigl(d s^2 t + 2 d s t^2 + d t^3 - 2 s^2 t - 2 s t^2\bigr)}{4 s (s+t)}  \mathcal{T}_6^{\dag}- \frac{3 \bigl(-s^2 t^2 - s t^3\bigr)}{4 s (s+t)} \mathcal{T}_8^{\dag}
\biggr], \\[6pt]
\mathcal{P}_{-+-+} &= \frac{1}{3(d-1)(d-3)t^2}\biggl[
    -\frac{3 \bigl(d^2 s (s+t) + 2 d s^2 + 2 d s t + 2 d t^2 - 4 t^2\bigr)}{4 s (s+t)} \mathcal{T}_1^{\dag} \\
    &\quad- \frac{3 \bigl(d s^2 t + d s t^2\bigr)}{4 s (s+t)} \left( \mathcal{T}_2^{\dag} - \mathcal{T}_4^{\dag} + \mathcal{T}_5^{\dag} - \mathcal{T}_7^{\dag}\right)
    - \frac{3 \bigl(d s^2 t + 2 s t^2\bigr)}{4 s (s+t)} \mathcal{T}_3^{\dag} \\
    &\quad- \frac{3 \bigl(-d s^2 t - 2 d s t^2 - d t^3 + 2 t^2 (s+t)\bigr)}{4 s (s+t)} \mathcal{T}_6^{\dag} + \frac{3 t^2 }{4}\mathcal{T}_8^{\dag} 
\biggr], \\[6pt]
\mathcal{P}_{+--+} &= \frac{1}{3(d-1)(d-3)t^2}\biggl[
    -\frac{3 \bigl((d - 4)(d - 2) s^2 + (d - 4)(d - 2) s t - 2 d t^2\bigr)}{4 s (s+t)}\mathcal{T}_1^{\dag} \\
    &\quad- \frac{3 \bigl(d s^2 t + d s t^2\bigr)}{4 s (s+t)} 
    \left(\mathcal{T}_2^{\dag} - \mathcal{T}_7^{\dag} \right) - \frac{3 \bigl(-d s^2 t + 2 s^2 t + 2 s t^2\bigr)}{4 s (s+t)} \mathcal{T}_3^{\dag}\\
    &\quad- \frac{3 \bigl(d s^2 t + d s t^2 - 2 s^2 t - 2 s t^2\bigr)}{4 s (s+t)} \left( \mathcal{T}_4^{\dag} - \mathcal{T}_5^{\dag}\right) \\
    &\quad- \frac{3 \bigl(d s^2 t + 2 d s t^2 + d t^3 - 2 s^2 t - 2 s t^2\bigr)}{4 s (s+t)} \mathcal{T}_6^{\dag}
    + \frac{3\bigl(s^2 t^2 + s t^3\bigr)}{4 s (s+t)} \mathcal{T}_8^{\dag}
\biggr],
\end{align*}
where
$$\mathcal{T}_i^\dagger = T_i^{\mu \nu \rho \sigma} 
\epsilon_{1,\mu}^*(p_1) \epsilon_{2,\nu}^*(p_2)
\epsilon_{3,\rho}^*(p_3) \epsilon_{4,\sigma}^*(p_4)\,,$$
and the $T_i^{\mu \nu \rho \sigma}$ have been defined in \cref{eq:tensgg}.



\section{Master Integrals}
\label{app:masters}
Our choice of initial basis of master integrals for the amplitude is given by
\begin{align*}
&\mathcal{J}_1 = \mathcal{I}_{\text{PLA}}(0,2,2,0,0,0,0,0,0)\, , \quad
&&\mathcal{J}_2 = \mathcal{I}_{\text{PLA}}(2,2,0,0,0,1,0,0,0)\, , \\
&\mathcal{J}_3 = \mathcal{I}_{\text{PLA}}(0,2,2,0,0,0,1,0,0)\, , \quad
&&\mathcal{J}_4 = \mathcal{I}_{\text{PLAx124}}(0,2,2,0,0,0,1,0,0)\, , \\
&\mathcal{J}_5 = \mathcal{I}_{\text{PLAx123}}(0,2,2,0,0,0,1,0,0)\, , \quad
&&\mathcal{J}_6 = \mathcal{I}_{\text{PLA}}(0,2,2,0,0,1,0,0,0)\, , \\
&\mathcal{J}_7 = \mathcal{I}_{\text{PLA}}(0,2,1,0,0,2,0,0,0)\, , \quad
&&\mathcal{J}_{8}  = \mathcal{I}_{\text{PLAx12}}(0,0,2,1,0,0,0,0,2)\, , \\ 
&\mathcal{J}_{9}  = \mathcal{I}_{\text{PLAx12}}(0,0,2,2,0,0,0,0,1)\, , \quad
&&\mathcal{J}_{10}  = \mathcal{I}_{\text{PLA}}(0,0,2,1,0,0,0,0,2)\, , \\ 
&\mathcal{J}_{11}  = \mathcal{I}_{\text{PLA}}(0,0,2,2,0,0,0,0,1)\, , \quad
&&\mathcal{J}_{12} = \mathcal{I}_{\text{PLA}}(0,1,2,1,0,0,1,0,0)\, , \\
&\mathcal{J}_{13} = \mathcal{I}_{\text{PLA}}(0,1,1,2,0,0,1,0,0)\, , \quad
&&\mathcal{J}_{14} = \mathcal{I}_{\text{PLA}}(0,2,2,1,0,0,1,0,0)\, , \\
&\mathcal{J}_{15} = \mathcal{I}_{\text{PLAx124}}(0,1,2,1,0,0,1,0,0)\, , \quad
&&\mathcal{J}_{16} = \mathcal{I}_{\text{PLAx124}}(0,1,1,2,0,0,1,0,0)\, , \\ 
&\mathcal{J}_{17} = \mathcal{I}_{\text{PLAx124}}(0,2,2,1,0,0,1,0,0)\, , \quad
&&\mathcal{J}_{18} = \mathcal{I}_{\text{PLAx123}}(0,1,2,1,0,0,1,0,0)\, , \\
&\mathcal{J}_{19} = \mathcal{I}_{\text{PLAx123}}(0,1,1,2,0,0,1,0,0)\, , \quad
&&\mathcal{J}_{20} = \mathcal{I}_{\text{PLAx123}}(0,2,2,1,0,0,1,0,0)\, , \\
&\mathcal{J}_{21} = \mathcal{I}_{\text{PLA}}(2,2,0,0,0,1,1,0,0)\, , \quad
&&\mathcal{J}_{22} = \mathcal{I}_{\text{PLA}}(1,0,2,0,0,1,0,0,1)\, , \\
&\mathcal{J}_{23} = \mathcal{I}_{\text{PLA}}(2,0,2,0,0,1,0,0,1)\, , \quad
&&\mathcal{J}_{24} = \mathcal{I}_{\text{PLA}}(0,1,3,0,0,1,0,0,1)\, , \\
&\mathcal{J}_{25} = \mathcal{I}_{\text{PLAx12}}(0,1,3,1,0,0,0,0,1)\, , \quad
&&\mathcal{J}_{26} = \mathcal{I}_{\text{PLA}}(0,1,3,1,0,0,0,0,1)\, , \\
&\mathcal{J}_{27} = \mathcal{I}_{\text{PLA}}(0,1,2,0,0,0,1,0,1)\, , \quad
&&\mathcal{J}_{28} = \mathcal{I}_{\text{PLAx124}}(0,1,2,0,0,0,1,0,1)\, , \\
&\mathcal{J}_{29} = \mathcal{I}_{\text{PLAx123}}(0,1,2,0,0,0,1,0,1)\, , \quad
&&\mathcal{J}_{30} = \mathcal{I}_{\text{PLC}}(2,2,0,0,0,1,1,0,0)\, , \\
&\mathcal{J}_{31} = \mathcal{I}_{\text{PLCx12}}(0,0,0,2,2,0,0,1,1)\, , \quad
&&\mathcal{J}_{32} = \mathcal{I}_{\text{PLC}}(0,0,0,2,2,0,0,1,1)\, , \\
&\mathcal{J}_{33} = \mathcal{I}_{\text{PLA}}(1,1,1,0,0,1,1,0,0)\, , \quad
&& \mathcal{J}_{34} = \mathcal{I}_{\text{PLA}}(1,0,2,1,0,1,0,0,1)\, , \\
&\mathcal{J}_{35} = \mathcal{I}_{\text{PLA}}(1,0,2,1,0,1,0,-1,1)\, , \quad
&&\mathcal{J}_{36} = \mathcal{I}_{\text{PLAx12}}(1,0,2,1,0,1,0,0,1)\, , \\
&\mathcal{J}_{37} = \mathcal{I}_{\text{PLAx12}}(1,0,2,1,0,1,0,-1,1)\, , \quad
&&\mathcal{J}_{38} = \mathcal{I}_{\text{PLA}}(1,1,1,0,0,1,0,0,1)\, , \\
&\mathcal{J}_{39} = \mathcal{I}_{\text{PLAx123}}(0,1,1,1,0,1,0,0,1)\, , \quad
&&\mathcal{J}_{40} = \mathcal{I}_{\text{PLAx123}}(0,1,2,1,0,1,0,0,1)\, , \\
&\mathcal{J}_{41} = \mathcal{I}_{\text{PLA}}(0,1,1,1,0,1,0,0,1)\, , \quad
&&\mathcal{J}_{42} = \mathcal{I}_{\text{PLA}}(0,1,2,1,0,1,0,0,1)\, , \\ 
&\mathcal{J}_{43} = \mathcal{I}_{\text{PLAx12}}(0,1,1,1,0,1,0,0,1)\, , \quad
&&\mathcal{J}_{44} = \mathcal{I}_{\text{PLAx12}}(0,1,2,1,0,1,0,0,1)\, , \\
&\mathcal{J}_{45} = \mathcal{I}_{\text{PLA}}(0,1,2,1,0,0,1,0,1)\, , \quad
&&\mathcal{J}_{46} = \mathcal{I}_{\text{PLA}}(0,1,3,1,0,0,1,0,1)\, , \\
&\mathcal{J}_{47} = \mathcal{I}_{\text{PLA}}(0,1,2,1,-1,0,1,0,1)\, , \quad
&&\mathcal{J}_{48} = \mathcal{I}_{\text{PLAx12}}(0,1,2,1,0,0,1,0,1)\, , \\
&\mathcal{J}_{49} = \mathcal{I}_{\text{PLAx12}}(0,1,3,1,0,0,1,0,1)\, , \quad
&&\mathcal{J}_{50} = \mathcal{I}_{\text{PLAx12}}(0,1,2,1,-1,0,1,0,1)\, ,\\
&\mathcal{J}_{51} = \mathcal{I}_{\text{PLAx123}}(0,1,2,1,0,0,1,0,1)\, , \quad
&&\mathcal{J}_{52} = \mathcal{I}_{\text{PLAx123}}(0,1,3,1,0,0,1,0,1)\, , \\
&\mathcal{J}_{53} = \mathcal{I}_{\text{PLAx123}}(0,1,2,1,-1,0,1,0,1)\, , \quad
&&\mathcal{J}_{54} = \mathcal{I}_{\text{PLAx124}}(0,1,2,1,0,0,1,0,1)\, , \\
&\mathcal{J}_{55} = \mathcal{I}_{\text{PLAx124}}(0,1,3,1,0,0,1,0,1)\, , \quad
&&\mathcal{J}_{56} = \mathcal{I}_{\text{PLAx124}}(0,1,2,1,-1,0,1,0,1)\, , \\
&\mathcal{J}_{57} = \mathcal{I}_{\text{PLAx1234}}(0,1,2,1,0,0,1,0,1)\, , \quad
&&\mathcal{J}_{58} = \mathcal{I}_{\text{PLAx1234}}(0,1,3,1,0,0,1,0,1)\, , \\
&\mathcal{J}_{59} = \mathcal{I}_{\text{PLAx1234}}(0,1,2,1,-1,0,1,0,1)\, , \quad
&&\mathcal{J}_{60} = \mathcal{I}_{\text{PLAx1243}}(0,1,2,1,0,0,1,0,1)\, , \\
&\mathcal{J}_{61} = \mathcal{I}_{\text{PLAx1243}}(0,1,3,1,0,0,1,0,1)\, , \quad
&&\mathcal{J}_{62} = \mathcal{I}_{\text{PLAx1243}}(0,1,2,1,-1,0,1,0,1)\, , \\
&\mathcal{J}_{63} = \mathcal{I}_{\text{PLA}}(2,1,0,0,0,1,1,0,1)\, , \quad 
&&\mathcal{J}_{64} = \mathcal{I}_{\text{PLC}}(1,1,1,0,1,1,0,0,0)\, , \\
&\mathcal{J}_{65} = \mathcal{I}_{\text{PLC}}(1,1,1,0,1,2,0,0,0)\, , \quad
&&\mathcal{J}_{66} = \mathcal{I}_{\text{PLCx12}}(0,1,1,1,1,0,0,1,0)\, , \\
&\mathcal{J}_{67} = \mathcal{I}_{\text{PLCx12}}(0,1,1,1,1,0,0,2,0)\, , \quad 
&&\mathcal{J}_{68} = \mathcal{I}_{\text{PLC}}(0,1,1,1,1,0,0,1,0)\, , \\
&\mathcal{J}_{69} = \mathcal{I}_{\text{PLC}}(0,1,1,1,1,0,0,2,0)\, , \quad 
&&\mathcal{J}_{70} = \mathcal{I}_{\text{PLC}}(0,1,1,1,1,1,0,0,0)\, , \\
&\mathcal{J}_{71} = \mathcal{I}_{\text{PLCx12}}(1,1,1,0,1,0,0,1,0)\, , \quad
&&\mathcal{J}_{72} = \mathcal{I}_{\text{PLC}}(1,1,1,0,1,0,0,1,0)\, , \\
&\mathcal{J}_{73} = \mathcal{I}_{\text{PLC}}(1,2,0,1,0,1,1,0,0)\, , \quad 
&&\mathcal{J}_{74} = \mathcal{I}_{\text{PLCx12}}(1,0,0,1,2,0,0,1,1)\, , \\
&\mathcal{J}_{75} = \mathcal{I}_{\text{PLC}}(1,0,0,1,2,0,0,1,1)\, , \quad 
&&\mathcal{J}_{76} = \mathcal{I}_{\text{PLCx123}}(1,2,0,1,0,1,0,1,0)\, , \\
&\mathcal{J}_{77} = \mathcal{I}_{\text{PLC}}(1,2,0,1,0,1,0,1,0)\, , \quad
&&\mathcal{J}_{78} = \mathcal{I}_{\text{PLCx12}}(1,2,0,1,0,1,0,1,0)\, , \\
&\mathcal{J}_{79} = \mathcal{I}_{\text{PLCx123}}(0,1,1,0,1,1,0,1,0)\, , \quad 
&&\mathcal{J}_{80} = \mathcal{I}_{\text{PLCx123}}(0,2,1,0,1,1,0,1,0)\, , \\
&\mathcal{J}_{81} = \mathcal{I}_{\text{PLCx123}}(0,1,2,0,1,1,0,1,0)\, , \quad 
&&\mathcal{J}_{82} = \mathcal{I}_{\text{PLC}}(0,1,1,0,1,1,0,1,0)\, , \\ 
&\mathcal{J}_{83} = \mathcal{I}_{\text{PLC}}(0,2,1,0,1,1,0,1,0)\, , \quad
&&\mathcal{J}_{84} = \mathcal{I}_{\text{PLC}}(0,1,2,0,1,1,0,1,0)\, ,\\
&\mathcal{J}_{85} = \mathcal{I}_{\text{PLCx12}}(0,1,1,0,1,1,0,1,0)\, , \quad
&&\mathcal{J}_{86} = \mathcal{I}_{\text{PLCx12}}(0,2,1,0,1,1,0,1,0)\, , \\
&\mathcal{J}_{87} = \mathcal{I}_{\text{PLCx12}}(0,1,2,0,1,1,0,1,0)\, , \quad 
&&\mathcal{J}_{88} = \mathcal{I}_{\text{PLA}}(1,1,1,0,0,1,1,0,1)\, , \\
&\mathcal{J}_{89} = \mathcal{I}_{\text{PLA}}(1,1,1,1,0,1,0,0,1)\, , \quad 
&&\mathcal{J}_{90} = \mathcal{I}_{\text{PLAx12}}(1,1,1,1,0,1,0,0,1)\, , \\
&\mathcal{J}_{91} = \mathcal{I}_{\text{PLC}}(1,1,0,1,1,1,1,0,0)\, , \quad
&&\mathcal{J}_{92} = \mathcal{I}_{\text{PLCx12}}(1,1,0,1,1,0,0,1,1)\, , \\
&\mathcal{J}_{93} = \mathcal{I}_{\text{PLC}}(1,1,0,1,1,0,0,1,1)\, , \quad
&&\mathcal{J}_{94} = \mathcal{I}_{\text{PLC}}(1,1,1,0,1,1,0,1,0)\, , \\
&\mathcal{J}_{95} = \mathcal{I}_{\text{PLC}}(1,1,1,-1,1,1,0,1,0)\, , \quad
&&\mathcal{J}_{96} = \mathcal{I}_{\text{PLCx12}}(1,1,1,0,1,1,0,1,0)\, , \\
&\mathcal{J}_{97} = \mathcal{I}_{\text{PLCx12}}(1,1,1,-1,1,1,0,1,0)\, , \quad
&&\mathcal{J}_{98} = \mathcal{I}_{\text{PLCx123}}(1,1,1,0,1,1,0,1,0)\, , \\
&\mathcal{J}_{99} = \mathcal{I}_{\text{PLCx123}}(1,1,1,-1,1,1,0,1,0)\, , \quad
&&\mathcal{J}_{100} = \mathcal{I}_{\text{PLCx12}}(0,1,1,1,1,1,0,1,0)\, , \\
&\mathcal{J}_{101} = \mathcal{I}_{\text{PLCx12}}(-1,1,1,1,1,1,0,1,0)\, , \quad
&&\mathcal{J}_{102} = \mathcal{I}_{\text{PLC}}(0,1,1,1,1,1,0,1,0)\, ,\\
&\mathcal{J}_{103} = \mathcal{I}_{\text{PLC}}(-1,1,1,1,1,1,0,1,0)\, , \quad
&&\mathcal{J}_{104} = \mathcal{I}_{\text{PLCx123}}(0,1,1,1,1,1,0,1,0)\, , \\
&\mathcal{J}_{105} = \mathcal{I}_{\text{PLCx123}}(-1,1,1,1,1,1,0,1,0)\, , \quad
&&\mathcal{J}_{106} = \mathcal{I}_{\text{NPA}}(1,1,1,1,0,0,1,1,0)\, , \\
&\mathcal{J}_{107} = \mathcal{I}_{\text{NPA}}(1,2,1,1,0,0,1,1,0) \quad
&&\mathcal{J}_{108} = \mathcal{I}_{\text{NPAx124}}(1,1,1,1,0,0,1,1,0)\, , \\ 
&\mathcal{J}_{109} = \mathcal{I}_{\text{NPAx124}}(1,2,1,1,0,0,1,1,0) \quad
&&\mathcal{J}_{110} = \mathcal{I}_{\text{NPAx123}}(1,1,1,1,0,0,1,1,0)\, , \\
&\mathcal{J}_{111} = \mathcal{I}_{\text{NPAx123}}(1,2,1,1,0,0,1,1,0) \quad
&&\mathcal{J}_{112} = \mathcal{I}_{\text{NPA}}(1,1,1,0,0,1,1,1,0)\, , \\
&\mathcal{J}_{113} = \mathcal{I}_{\text{NPA}}(1,2,1,0,0,1,1,1,0)\, , \quad 
&&\mathcal{J}_{114} = \mathcal{I}_{\text{NPB}}(1,1,1,0,0,1,1,1,0)\, , \\
&\mathcal{J}_{115} = \mathcal{I}_{\text{NPBx124}}(1,1,1,0,0,1,1,1,0)\, , \quad
&&\mathcal{J}_{116} = \mathcal{I}_{\text{NPBx123}}(1,1,1,0,0,1,1,1,0)\, , \\
&\mathcal{J}_{117} = \mathcal{I}_{\text{PLA}}(1,1,1,1,0,1,1,0,1)\, , \quad
&&\mathcal{J}_{118} = \mathcal{I}_{\text{PLA}}(1,1,1,1,-1,1,1,0,1)\, , \\
&\mathcal{J}_{119} = \mathcal{I}_{\text{PLA}}(1,1,1,1,0,1,1,-1,1)\, , \quad 
&&\mathcal{J}_{120} = \mathcal{I}_{\text{PLA}}(1,1,1,1,-1,1,1,-1,1)\, , \\
&\mathcal{J}_{121} = \mathcal{I}_{\text{PLAx12}}(1,1,1,1,0,1,1,0,1)\, , \quad 
&&\mathcal{J}_{122} = \mathcal{I}_{\text{PLAx12}}(1,1,1,1,-1,1,1,0,1)\, , \\
&\mathcal{J}_{123} = \mathcal{I}_{\text{PLAx12}}(1,1,1,1,0,1,1,-1,1)\, , \quad 
&&\mathcal{J}_{124} = \mathcal{I}_{\text{PLAx12}}(1,1,1,1,-1,1,1,-1,1)\, , \\
&\mathcal{J}_{125} = \mathcal{I}_{\text{PLC}}(1,1,1,0,1,1,1,1,0)\, , \quad
&&\mathcal{J}_{126} = \mathcal{I}_{\text{PLC}}(1,1,1,-1,1,1,1,1,0)\, , \\
&\mathcal{J}_{127} = \mathcal{I}_{\text{PLC}}(1,1,1,-1,1,1,1,1,-1)\, , \quad 
&&\mathcal{J}_{128} = \mathcal{I}_{\text{PLCx12}}(1,1,1,0,1,1,1,1,0)\, , \\
&\mathcal{J}_{129} = \mathcal{I}_{\text{PLCx12}}(1,1,1,-1,1,1,1,1,0)\, , \quad
&&\mathcal{J}_{130} = \mathcal{I}_{\text{PLCx12}}(1,1,1,-1,1,1,1,1,-1)\, ,\\
&\mathcal{J}_{131} = \mathcal{I}_{\text{PLCx123}}(1,1,1,0,1,1,1,1,0)\, , \quad
&&\mathcal{J}_{132} = \mathcal{I}_{\text{PLCx123}}(1,1,1,-1,1,1,1,1,0)\, , \\
&\mathcal{J}_{133} = \mathcal{I}_{\text{PLCx123}}(1,1,1,-1,1,1,1,1,-1)\, , \quad 
&&\mathcal{J}_{134} = \mathcal{I}_{\text{PLCx12}}(0,1,1,1,1,1,0,1,1)\, , \\
&\mathcal{J}_{135} = \mathcal{I}_{\text{PLCx12}}(-1,1,1,1,1,1,0,1,1)\, , \quad
&&\mathcal{J}_{136} = \mathcal{I}_{\text{PLCx12}}(-1,1,1,1,1,1,-1,1,1)\, , \\
&\mathcal{J}_{137} = \mathcal{I}_{\text{PLC}}(0,1,1,1,1,1,0,1,1)\, , \quad
&&\mathcal{J}_{138} = \mathcal{I}_{\text{PLC}}(-1,1,1,1,1,1,0,1,1)\, , \\
&\mathcal{J}_{139} = \mathcal{I}_{\text{PLC}}(-1,1,1,1,1,1,-1,1,1)\, , \quad
&&\mathcal{J}_{140} = \mathcal{I}_{\text{PLCx123}}(0,1,1,1,1,1,0,1,1)\, , \\
&\mathcal{J}_{141} = \mathcal{I}_{\text{PLCx123}}(-1,1,1,1,1,1,0,1,1)\, , \quad
&&\mathcal{J}_{142} = \mathcal{I}_{\text{PLCx123}}(-1,1,1,1,1,1,-1,1,1)\, ,\\
&\mathcal{J}_{143} = \mathcal{I}_{\text{NPA}}(1,1,1,1,0,1,1,1,-1)\, , \quad
&&\mathcal{J}_{144} = \mathcal{I}_{\text{NPA}}(1,2,1,1,0,1,1,1,-1)\, , \\
&\mathcal{J}_{145} = \mathcal{I}_{\text{NPA}}(1,1,1,1,0,1,1,1,-2) \quad
&&\mathcal{J}_{146} = \mathcal{I}_{\text{NPA}}(1,1,1,1,0,1,1,1,0)\, , \\
&\mathcal{J}_{147} = \mathcal{I}_{\text{NPB}}(1,1,1,1,0,1,1,1,0)\, , \quad 
&&\mathcal{J}_{148} = \mathcal{I}_{\text{NPB}}(1,1,1,1,0,1,1,1,-1)\, , \\
&\mathcal{J}_{149} = \mathcal{I}_{\text{NPB}}(1,1,1,1,-1,1,1,1,0)\, , \quad
&&\mathcal{J}_{150} = \mathcal{I}_{\text{NPB}}(1,1,1,1,0,1,1,1,-2)\, , \\
&\mathcal{J}_{151} = \mathcal{I}_{\text{NPB}}(1,1,1,1,-1,1,1,1,-1) \quad
&&\mathcal{J}_{152} = \mathcal{I}_{\text{NPBx124}}(1,1,1,1,0,1,1,1,0)\, , \\
&\mathcal{J}_{153} = \mathcal{I}_{\text{NPBx124}}(1,1,1,1,0,1,1,1,-1)\, , \quad
&&\mathcal{J}_{154} = \mathcal{I}_{\text{NPBx124}}(1,1,1,1,-1,1,1,1,0)\, , \\
&\mathcal{J}_{155} = \mathcal{I}_{\text{NPBx124}}(1,1,1,1,0,1,1,1,-2)\, , \quad
&&\mathcal{J}_{156} = \mathcal{I}_{\text{NPBx124}}(1,1,1,1,-1,1,1,1,-1) \\
&\mathcal{J}_{157} = \mathcal{I}_{\text{NPBx123}}(1,1,1,1,0,1,1,1,0)\, , \quad 
&&\mathcal{J}_{158} = \mathcal{I}_{\text{NPBx123}}(1,1,1,1,0,1,1,1,-1)\, , \\
&\mathcal{J}_{159} = \mathcal{I}_{\text{NPBx123}}(1,1,1,1,-1,1,1,1,0)\, , \quad
&&\mathcal{J}_{160} = \mathcal{I}_{\text{NPBx123}}(1,1,1,1,0,1,1,1,-2)\, , \\
&\mathcal{J}_{161} = \mathcal{I}_{\text{NPBx123}}(1,1,1,1,-1,1,1,1,-1)\, , \quad
&&\mathcal{J}_{162} = \mathcal{I}_{\text{PLAx12}}(0,2,0,2,0,0,0,1,0)\, , \\ 
&\mathcal{J}_{163} = \mathcal{I}_{\text{PLA}}(0,2,0,2,0,0,0,1,0)\, , \quad
&&\mathcal{J}_{164} = \mathcal{I}_{\text{PLA}}(1,2,0,1,0,1,0,1,0)\, , \\ 
&\mathcal{J}_{165} = \mathcal{I}_{\text{PLAx12}}(1,2,0,1,0,1,0,1,0)\, , \quad
&&
\end{align*}
where we include also crossings of the original integral families. For the crossings,
we follow the notation introduced by \texttt{Reduze2}~\cite{vonManteuffel:2012np} such that $x_{i_1...i_n}$ indicates a cyclic permutation of the $n$ momenta $p_{i_1}, \cdots, p_{i_n}$. More explicitly, in our case we need crossings of up to four external momenta, which are defined as
\begin{equation}
    \left\{ 
    \begin{array}{ll} 
    x_{ij}: & \qquad p_i \leftrightarrow p_j   \\
    x_{ijk}: & \qquad p_i \to p_j \to p_k \to p_i  \\
    x_{ijkl}: & \qquad p_i \to p_j \to p_k \to p_l \to p_i \,.
    \end{array}
    \right.
\end{equation}
The integrals $\mathcal{J}_{1},\dots,\mathcal{J}_{161}$ are relevant to the helicity amplitudes in the gluon-fusion channel, while for the amplitudes in the quark annihilation channel, only the following $63$ integrals are needed:
\begin{equation}
\begin{aligned}
    &\left\{ \mathcal{J}_{1}, \mathcal{J}_{2} ,\mathcal{J}_{3} ,\mathcal{J}_{6},\mathcal{J}_{7},\mathcal{J}_{8}, \mathcal{J}_{9} ,\mathcal{J}_{10} ,\mathcal{J}_{11},\mathcal{J}_{12},\mathcal{J}_{13}, \mathcal{J}_{14} ,\mathcal{J}_{21} ,\mathcal{J}_{22},\mathcal{J}_{23},\mathcal{J}_{24}, \mathcal{J}_{25} ,\mathcal{J}_{26} ,\mathcal{J}_{27}, \right. \\
    & \ \, \mathcal{J}_{33},\mathcal{J}_{34}, \mathcal{J}_{35} , \mathcal{J}_{36} ,\mathcal{J}_{37}, \mathcal{J}_{38},\mathcal{J}_{41}, \mathcal{J}_{42} ,\mathcal{J}_{43} ,\mathcal{J}_{44},\mathcal{J}_{45},\mathcal{J}_{46}, \mathcal{J}_{47} ,\mathcal{J}_{48} ,\mathcal{J}_{49},\mathcal{J}_{50},\mathcal{J}_{63}, \mathcal{J}_{70} , \\
    & \ \, \mathcal{J}_{79} ,\mathcal{J}_{80},\mathcal{J}_{81},\mathcal{J}_{88}, \mathcal{J}_{89} ,\mathcal{J}_{90} ,\mathcal{J}_{106},\mathcal{J}_{107},\mathcal{J}_{112}, \mathcal{J}_{113} ,\mathcal{J}_{117} ,\mathcal{J}_{118},\mathcal{J}_{119},\mathcal{J}_{120}, \mathcal{J}_{121} ,\mathcal{J}_{122} , \\ 
    & \ \left. \mathcal{J}_{123},\mathcal{J}_{124},\mathcal{J}_{143}, \mathcal{J}_{144} ,\mathcal{J}_{145} ,\mathcal{J}_{146},\mathcal{J}_{162},\mathcal{J}_{163}, \mathcal{J}_{164} ,\mathcal{J}_{165} \right\} \, .
\end{aligned}
\end{equation}


\section{Canonical Differential Equations}
\label{app:candiff}

\allowdisplaybreaks

In this appendix, we list all the kernels appearing in our $\epsilon$-factorised differential equations. They depend on the following 16 square roots of the kinematic invariants and mass
\begin{alignat*}{3}
r_1 &= \sqrt{s(s-4\mt^2)}\,, \quad &&r_2 &&= \sqrt{t(t-4\mt^2)}\,, \\ 
r_3 &= \sqrt{(s + t)(4\mt^2 + s + t)}\,, \quad &&r_4 &&=\sqrt{s(s+4\mt^2)}\,, \\
r_5 &= \sqrt{t+t(4\mt^2)}\,, \quad &&r_6 &&= \sqrt{(s + t)( s + t - 4\mt^2)}\,, \\ 
r_7 &= \sqrt{st(st - 4\mt^2(s + t))}\,, \quad &&r_8 &&= \sqrt{s(s + t)(s(s + t)-4\mt^2t )}\,, \\
r_9 &= \sqrt{t(s + t)(t(s + t)-4\mt^2s )}\,, \quad &&r_{10} &&= \sqrt{s(-4\mt^2(s + t)^2 + s(\mt^2 + s + t)^2)}\,, \\
r_{11} &= \sqrt{s(-4\mt^2t^2 + s(-\mt^2 + t)^2)}\,, \quad &&r_{12} &&= \sqrt{(s + t)(4\mt^2t^2 + (-\mt^2 + t)^2(s + t))}\,, \\
r_{13} &= \sqrt{t(-4\mt^2s^2 + (-\mt^2 + s)^2t)}\,, \quad &&r_{14} &&= \sqrt{(s + t)(4\mt^2s^2 + (-\mt^2 + s)^2(s + t))}\,, \\
r_{15} &= \sqrt{t(-4\mt^2(s + t)^2 + t(\mt^2 + s + t)^2)}\,, \quad &&r_{16} &&= \sqrt{-(\mt^2st(s + t))} \, .
\end{alignat*}
Strictly speaking, the roots $r_5,r_6$ are not required. We include them anyway to render the set of roots closed under crossings of the external momenta. Next, we list all dlog-kernels,
\begin{equation}
    \omega_i = \mathrm{d}\log \alpha_i
\end{equation}
with letters $\alpha_i$. We split the alphabet (the set of letters) into sets of rational (even) and algebraic (odd) letters
\begin{align*}
    \mathbf{W}_R \, = \, &\{\mt^2,s,t,-s-t,s-4 \mt^2,4 \mt^2+s,t-4 \mt^2,-4 \mt^2-s-t, \\
    &\ 4 s \mt^2+t (-s-t),4 t \mt^2+s (-s-t),4 \mt^2 (-s-t)+s t,16 \mt^2+s \} \, , \\
    \mathbf{W}_A \, = \, &\Bigl\{\frac{s-r_1}{s+r_1},\frac{s-r_4}{s+r_4},\frac{t-r_2}{t+r_2},\frac{-s-t-r_3}{-s-t+r_3},\frac{s \left(-4 \mt^2-s-t\right)-r_3 r_4}{s \left(-4\mt^2-s-t\right)+r_3 r_4}, \\
    &\ \frac{s \left(t-4 \mt^2\right)-r_2 r_4}{s \left(t-4 \mt^2\right)+r_2 r_4},\frac{s t-r_7}{s t+r_7},\frac{s (-s-t)-r_8}{s
   (-s-t)+r_8},\frac{(-s-t) t-r_9}{(-s-t) t+r_9},\frac{s t \left(s-4 \mt^2\right)-r_1 r_7}{s t \left(s-4 \mt^2\right)+r_1 r_7}, \\
   &\ \frac{s (-s-t) \left(s-4
   \mt^2\right)-r_1 r_8}{s (-s-t) \left(s-4 \mt^2\right)+r_1 r_8},\frac{s t \left(t-4 \mt^2\right)-r_2 r_7}{s t \left(t-4 \mt^2\right)+r_2 r_7},\frac{(-s-t) t
   \left(t-4 \mt^2\right)-r_2 r_9}{(-s-t) t \left(t-4 \mt^2\right)+r_2 r_9}, \\
   &\ \frac{s (-s-t) \left(-4 \mt^2-s-t\right)-r_3 r_8}{s (-s-t) \left(-4 \mt^2-s-t\right)+r_3
   r_8},\frac{(-s-t) t \left(-4 \mt^2-s-t\right)-r_3 r_9}{(-s-t) t \left(-4 \mt^2-s-t\right)+r_3 r_9}, \\
   &\ \frac{-s (-s-t) t \left(s-4 \mt^2\right)-r_7 r_8}{r_7 r_8-s
   (-s-t) t \left(s-4 \mt^2\right)},\frac{-s (-s-t) t \left(-4 \mt^2-s-t\right)-r_8 r_9}{r_8 r_9-s (-s-t) t \left(-4 \mt^2-s-t\right)}, \\
   &\ \frac{-s (-s-t) t \left(t-4
   \mt^2\right)-r_7 r_9}{r_7 r_9-s (-s-t) t \left(t-4 \mt^2\right)},\frac{s \mt^2+(-s-t) (s+2 t)-r_{10}}{s \mt^2+(-s-t) (s+2 t)+r_{10}}, \\
   &\ \frac{s \mt^2+(-s-2 t)
   t-r_{11}}{s \mt^2+(-s-2 t) t+r_{11}},\frac{(-s-t) \mt^2+(s-t) t-r_{12}}{(-s-t) \mt^2+(s-t) t+r_{12}},\frac{t \mt^2+s (-2 s-t)-r_{13}}{t \mt^2+s (-2
   s-t)+r_{13}}, \\
   &\ \frac{(-s-t) \mt^2+s (t-s)-r_{14}}{(-s-t) \mt^2+s (t-s)+r_{14}},\frac{t \mt^2+(-s-t) (2 s+t)-r_{15}}{t \mt^2+(-s-t) (2 s+t)+r_{15}}, \\
   &\ \frac{(s+2 t)
   \mt^2+s (-s-t)-r_{10}}{(s+2 t) \mt^2+s (-s-t)+r_{10}},\frac{(-s-2 t) \mt^2+s t-r_{11}}{(-s-2 t) \mt^2+s t+r_{11}}, \\
   &\ \frac{(s-t) \mt^2+(-s-t) t-r_{12}}{(s-t)
   \mt^2+(-s-t) t+r_{12}}, \frac{(-2 s-t) \mt^2+s t-r_{13}}{(-2 s-t) \mt^2+s t+r_{13}}, \\
   &\ \frac{(t-s) \mt^2+s (-s-t)-r_{14}}{(t-s) \mt^2+s (-s-t)+r_{14}},\frac{(2 s+t)
   \mt^2+(-s-t) t-r_{15}}{(2 s+t) \mt^2+(-s-t) t+r_{15}}, \\
   &\ \frac{s \left(s (-s-t)-(-3 s-4 t) \mt^2\right)-r_1 r_{10}}{s \left(s (-s-t)-(-3 s-4 t) \mt^2\right)+r_1
   r_{10}},\frac{s \left(s t-(s+4 t) \mt^2\right)-r_1 r_{11}}{s \left(s t-(s+4 t) \mt^2\right)+r_1 r_{11}}, \\
   &\ \frac{t \left(s t-(4 s+t) \mt^2\right)-r_2 r_{13}}{t
   \left(s t-(4 s+t) \mt^2\right)+r_2 r_{13}},\frac{t \left((-s-t) t-(-4 s-3 t) \mt^2\right)-r_2 r_{15}}{t \left((-s-t) t-(-4 s-3 t) \mt^2\right)+r_2
   r_{15}}, \\
   &\ \frac{(-s-t) \left((-s-t) t-(3 t-s) \mt^2\right)-r_3 r_{12}}{(-s-t) \left((-s-t) t-(3 t-s) \mt^2\right)+r_3 r_{12}}, \\
   &\ \frac{(-s-t) \left(s (-s-t)-(3 s-t)
   \mt^2\right)-r_3 r_{14}}{(-s-t) \left(s (-s-t)-(3 s-t) \mt^2\right)+r_3 r_{14}}, \\
   &\ \frac{s (-s-t) \left(s (-s-t)-(-s-4 t) \mt^2\right)-r_8 r_{10}}{s (-s-t) \left(s
   (-s-t)-(-s-4 t) \mt^2\right)+r_8 r_{10}},\frac{s t \left(s t-(3 s+4 t) \mt^2\right)-r_7 r_{11}}{s t \left(s t-(3 s+4 t) \mt^2\right)+r_7 r_{11}}, \\
   &\ \frac{(-s-t) t
   \left((-s-t) t-(t-3 s) \mt^2\right)-r_9 r_{12}}{(-s-t) t \left((-s-t) t-(t-3 s) \mt^2\right)+r_9 r_{12}},\frac{s t \left(s t-(4 s+3 t) \mt^2\right)-r_7 r_{13}}{s
   t \left(s t-(4 s+3 t) \mt^2\right)+r_7 r_{13}}, \\
   &\ \frac{s (-s-t) \left(s (-s-t)-(s-3 t) \mt^2\right)-r_8 r_{14}}{s (-s-t) \left(s (-s-t)-(s-3 t) \mt^2\right)+r_8
   r_{14}}, \\
   &\ \frac{(-s-t) t \left((-s-t) t-(-4 s-t) \mt^2\right)-r_9 r_{15}}{(-s-t) t \left((-s-t) t-(-4 s-t) \mt^2\right)+r_9 r_{15}}, \\
   &\ \frac{s t \left(4 (-s-t)
   \mt^2+s t\right)-2 r_7 r_{16}}{s t \left(4 (-s-t) \mt^2+s t\right)+2 r_7 r_{16}},\frac{s (-s-t) \left(4 t \mt^2+s (-s-t)\right)-2 r_8 r_{16}}{s (-s-t) \left(4 t
   \mt^2+s (-s-t)\right)+2 r_8 r_{16}}, \\
   &\ \frac{(-s-t) t \left(4 s \mt^2+(-s-t) t\right)-2 r_9 r_{16}}{(-s-t) t \left(4 s \mt^2+(-s-t) t\right)+2 r_9 r_{16}}\Bigl\} \, .
\end{align*}
Notice that all algebraic letters appearing in the differential equations also appear in the amplitude.

We can split the elliptic letters into two groups. The first group contains 4 elliptic letters, which are easily related to modular forms. We call them $M_i$ and give here their defining differential relations
\begin{equation*}
    \mathrm{d} M_i = M_i^s \, \mathrm{d}s + M_i^t \, \mathrm{d}t + M_i^{\mt^2} \, \mathrm{d}\mt^2\,,
\end{equation*}
with
\begin{alignat*}{5}
    M_1^s &= -\frac{1}{s^2  (16 \mt^2+s)\varpi_0^2} \, , \quad
    &M_1^t &= 0 \, , \quad
    &M_1^{\mt^2} &= \frac{1}{s  \mt^2 \left(16 \mt^2+s\right)\varpi_0^2}\, , \\
    M_2^s &= \varpi_0 \, , \quad
    &M_2^t &= 0 \, ,\quad
    &M_2^{\mt^2} &= -\frac{s \varpi_0}{\mt^2}\, ,\\
    M_3^s &= -\frac{r_4 \varpi_0}{4 \mt^2+s} \, , \quad
    &M_3^t &= 0 \, ,\quad
    &M_3^{\mt^2} &= \frac{r_4 s \varpi_0}{\mt^2 \left(4 \mt^2+s\right)}\, , \\
    M_{4}^s &= -\frac{ (8\mt^2+s)^2\varpi_0^2}{16 \mt^2+s} \, , \quad
    &M_{4}^t &= 0 \, ,\quad
    &M_{4}^{\mt^2} &= \frac{s  \left(8 \mt^2+s\right)^2\varpi_0^2}{\mt^2 \left(16 \mt^2+s\right)} \, .
\end{alignat*}
The other 13 elliptic letters contain the function $G$ defined in \cref{eq:Gdef}. We denote them with $E_i$, and they are similarly defined by
\begin{equation*}
    \mathrm{d} E_i = E_i^s \, \mathrm{d}s + E_i^t \, \mathrm{d}t + E_i^{\mt^2} \, \mathrm{d}\mt^2\,,
\end{equation*}
with
\begin{align*}
    E_1^s =& \frac{4 G}{s^2  (16 \mt^2+s)\varpi_0^2}+\frac{2 r_9}{s  (s+t) (4 \mt^2 s-t (s+t))\varpi_0} \, , \\
    E_1^t =& \frac{2 r_9}{t  (s+t) (t (s+t)-4 \mt^2 s)\varpi_0} \, , \\
    E_1^{\mt^2} =& -\frac{4 G}{s  \mt^2 \left(16 \mt^2+s\right)\varpi_0^2} \, , \\
    E_2^s =& -\frac{2 G^2}{s^2  (16 \mt^2+s)\varpi_0^2}+\frac{2  r_9 G}{s  (s+t) (t (s+t)-4 \mt^2 s)\varpi_0}+\frac{8 \mt^2-t}{8 \mt^2 s-2 t
   (s+t)} \, , \\
    E_2^t =& -\frac{2  r_9 G}{t \varpi_0 (s+t) (t (s+t)-4 \mt^2 s)}-\frac{s+2 t}{8 \mt^2 s-2 t (s+t)} \, ,\\
    E_2^{\mt^2} =& \frac{2 G^2}{s  \mt^2 \left(16 \mt^2+s\right)\varpi_0^2} \, , \\
    E_3^s =& \frac{4  (s+3 t) G}{s (s+t)}+\frac{2  (16 \mt^2+s)r_9 \varpi_0}{(s+t) (4 \mt^2 s-t (s+t))} \, , \\
    E_3^t =&  \left(-\frac{8}{s+t}-\frac{8}{t}\right) G+\frac{2  s  (16 \mt^2+s)r_9\varpi_0}{t (s+t) (t (s+t)-4 \mt^2 s)} \, ,\\
    E_3^{\mt^2} =& \frac{4 G}{\mt^2} \, , \\
    E_4^s =& \frac{r_9 \varpi_0}{2 (t (s+t)-4 \mt^2 s)}-\frac{G}{s+t} \, , \\
    E_4^t =& \frac{ s G}{t (s+t)} \, ,\\
    E_4^{\mt^2} =& \frac{s r_9  \varpi_0}{2 \mt^2 \left(4 s \mt^2-t (s+t)\right)} \, , \\
    E_5^s =& \frac{ (4 \mt^2 s-t (8 \mt^2+s))\varpi_0}{8 \mt^2 s-2 t (s+t)}-\frac{  t r_9 G}{s (s+t) (t (s+t)-4 \mt^2 s)} \, , \\
    E_5^t =& \frac{(s+2 t) r_9 G}{t (s+t) (t (s+t)-4 \mt^2 s)}+\frac{s  (16 \mt^2+s)\varpi_0}{8 \mt^2 s-2 t (s+t)} \, , \\
    E_5^{\mt^2} =& \frac{ r_9G}{4 s \mt^4-t \mt^2 (s+t)}+\frac{2 s  (s+2 t)\varpi_0}{t (s+t)-4 s \mt^2} \, , \\
    E_6^s =& -\frac{2 G^3}{s^2  (16 \mt^2+s)\varpi_0^2}+\frac{3  r_9G^2}{s  (s+t) (t (s+t)-4 \mt^2 s)\varpi_0} \\
    &+  \left(\frac{6 \mt^2}{4\mt^2 s-t (s+t)}+\frac{1}{32 \mt^2+2 s}-\frac{2}{s+t}+\frac{2}{s}\right)G \\
    &+\frac{ \left(64 \mt^4 s+16 \mt^2 (s-2 t)(s+t)-3 s t (s+t)\right)r_9 \varpi_0}{4 (s+t) (t (s+t)-4 \mt^2 s)^2} \, , \\
    E_6^t =& -\frac{3 r_9G^2 }{t  (s+t) (t (s+t)-4 \mt^2 s)\varpi_0}+ \left(-\frac{2}{s+t}-\frac{2}{t}\right)G \\
    &+\frac{ s  \left(-64 \mt^4 s+64 \mt^2 t (s+t)+3 s t (s+t)\right)r_9\varpi_0}{4 t (s+t) (t (s+t)-4 \mt^2 s)^2} \, ,\\
    E_6^{\mt^2} =& \frac{2 G^3}{ \left(s^2 \mt^2+16 s \mt^4\right)\varpi_0^2}+\frac{ \left(64 s \mt^4-64 t \mt^2 (s+t)-3 s t (s+t)\right)G}{2 \mt^2 \left(16 \mt^2+s\right) \left(4 s
    \mt^2-t (s+t)\right)}- \\
    &-\frac{4  s  (s+2 t)r_9\varpi_0}{\left(t (s+t)-4 s \mt^2\right)^2} \, , \\
    E_7^s =& \frac{4   (s+2 t)r_2\varpi_0}{t (s+t)-4 \mt^2 s}-\frac{8  r_2 r_9G}{t (s+t) (t (s+t)-4 \mt^2 s)} \, , \\
    E_7^t =& \frac{8  \left(4 \mt^2 s+t^2\right)r_2 r_9 G}{t^2 (4 \mt^2-t) (s+t) (t (s+t)-4 \mt^2 s)}+\frac{4  s  (16 \mt^2+s)r_2\varpi_0}{(4\mt^2-t) (4 \mt^2 s-t (s+t))} \, ,\\
    E_7^{\mt^2} =& \frac{8  r_2 r_9G}{\mt^2 \left(t-4 \mt^2\right) \left(t (s+t)-4 s \mt^2\right)}-\frac{8  s  \left(t (s+t)-2 \mt^2 (s-2 t)\right)r_2\varpi_0}{\mt^2 \left(t-4
   \mt^2\right) \left(t (s+t)-4 s \mt^2\right)} \, , \\
    E_8^s =& \frac{4  \left(-12 \mt^2 s+8 \mt^2 t+3 s t+2 t^2\right)r_3 \varpi_0}{(4 \mt^2+s+t) (t (s+t)-4 \mt^2 s)}-\frac{32  \mt^2 r_3
   r_9G}{(s+t)^2 (4 \mt^2+s+t) (t (s+t)-4 \mt^2 s)} \, , \\
    E_8^t =& \frac{8   \left(4 \mt^2 s+(s+t)^2\right)r_3 r_9G}{t (s+t)^2 (4 \mt^2+s+t) (t (s+t)-4 \mt^2 s)}-\frac{4  s  (16 \mt^2+s)r_3\varpi_0}{(4\mt^2+s+t) (t (s+t)-4 \mt^2 s)} \, ,\\
    E_8^{\mt^2} =& \frac{8  r_3 r_9G}{\mt^2 \left(4 \mt^2+s+t\right) \left(4 s \mt^2-t (s+t)\right)}+\frac{8  s  \left(t (s+t)-2 \mt^2 (3 s+2 t)\right)r_3\varpi_0}{\mt^2 \left(4
   \mt^2+s+t\right) \left(4 s \mt^2-t (s+t)\right)} \, , \\
    E_9^s =& -\frac{4    (4 \mt^2 (t-2 s)+t (s+t))r_8r_9G}{s (s+t)^2 (s (s+t)-4 \mt^2 t) (t (s+t)-4 \mt^2 s)}-\\
    &-\frac{2   \left(-64 \mt^4 s t+4 \mt^2 \left(-4 s^3-2 s^2 t+5 s t^2+2 t^3\right)+s t (s+t) (4 s+3 t)\right)r_8\varpi_0}{s (s+t) (s (s+t)-4 \mt^2 t) (t (s+t)-4 \mt^2 s)} \, , \\
    E_9^t =& \frac{2  s  (16 \mt^2+s) (-4 \mt^2+s+t)r_8\varpi_0}{(s+t) (s (s+t)-4 \mt^2 t) (t (s+t)-4 \mt^2 s)}- \\
    &-\frac{4   (4 \mt^2 (s-2t)+s (s+t))r_8 r_9G}{t (s+t)^2 (s (s+t)-4 \mt^2 t) (t (s+t)-4 \mt^2 s)} \, ,\\
    E_9^{\mt^2} =& \frac{4   \left(-4 \mt^2+s+t\right)r_8 r_9G}{\mt^2 (s+t) \left(t (s+t)-4 s \mt^2\right) \left(s (s+t)-4 t \mt^2\right)}+ \\
    &+ \frac{2  \varpi_0 \left(4 \mt^2 \left(4
   s^2+s t-2 t^2\right)-3 s t (s+t)\right)r_8}{\mt^2 \left(4 s \mt^2-t (s+t)\right) \left(s (s+t)-4 t \mt^2\right)} \, , \\
    E_{10}^s =& -\frac{4   (4 \mt^2-t) (2 s+t)r_7r_9 G}{s t (s+t) (s t-4 \mt^2 (s+t)) (t (s+t)-4 \mt^2 s)}- \\
    &-\frac{2  \left(t^2 (8 \mt^2+3s)+2 s t (2 \mt^2+s)-8 \mt^2 s (8 \mt^2+s)\right)r_7 \varpi_0}{s (4 \mt^2 (s+t)-s t) (4 \mt^2 s-t (s+t))} \, , \\
    E_{10}^t =& \frac{4   (12 \mt^2 s+8 \mt^2 t-s t)r_7 r_9G}{t^2 (s+t) (s t-4 \mt^2 (s+t)) (t (s+t)-4 \mt^2 s)}\\
    &-\frac{2  s  (16\mt^2+s) (4 \mt^2+t)r_7\varpi_0}{t (s t-4 \mt^2 (s+t)) (t (s+t)-4 \mt^2 s)} \, ,\\
    E_{10}^{\mt^2} =& \frac{4   \left(4 \mt^2+t\right)r_7 r_9G}{t \mt^2 \left(t (s+t)-4 s \mt^2\right) \left(4 \mt^2 (s+t)-s t\right)} \\
    &+\frac{2  \left(3 s t (s+t)-4 \mt^2
   \left(s^2-5 s t-2 t^2\right)\right)r_7 \varpi_0}{\mt^2 \left(4 s \mt^2-t (s+t)\right) \left(4 \mt^2 (s+t)-s t\right)} \, , \\
    E_{11}^s =& \frac{32  r_9 r_{16}G}{s (s+t)^2 (t (s+t)-4 \mt^2 s)}+\frac{16   (t (3 s+2 t)-16 \mt^2 s)r_{16}\varpi_0}{s (s+t) (t (s+t)-4\mt^2 s)} \, , \\
    E_{11}^t =& \frac{16  s  (16 \mt^2+s)r_{16}\varpi_0}{t (s+t) (t (s+t)-4 \mt^2 s)}-\frac{32   (s+2 t)r_9 r_{16}G}{t^2 (s+t)^2 (t (s+t)-4 \mt^2 s)} \, ,\\
    E_{11}^{\mt^2} =& \frac{32  r_9 r_{16}G}{t \mt^2 (s+t) \left(t (s+t)-4 s \mt^2\right)}+\frac{16  (s+2 t)r_{16} \varpi_0}{4 s \mt^4-t \mt^2 (s+t)} \, , \\
    E_{12}^s =& -\frac{2 G^2}{s+t}+\frac{2  r_9 \varpi_0 G}{t (s+t)-4 \mt^2 s}+\frac{s^2  (8 \mt^2-t)\varpi_0^2}{8 \mt^2 s-2 t (s+t)} \, , \\
    E_{12}^t =& \frac{2 G^2 s}{s t+t^2}+\frac{s^2  (16 \mt^2+s)\varpi_0^2}{8 \mt^2 s-2 t (s+t)} \, ,\\
    E_{12}^{\mt^2} =& \frac{2   s r_9\varpi_0G}{4 s \mt^4-t \mt^2 (s+t)}+\frac{4 s^2  (s+2 t)\varpi_0^2}{t (s+t)-4 s \mt^2} \, , \\
    E_{13}^s =& -\frac{4 G^4}{s^2  (16 \mt^2+s)\varpi_0^2}+\frac{8  r_9G^3}{s  (s+t) (t (s+t)-4 \mt^2 s)\varpi_0}+ \\
    &+  \left(\frac{24 \mt^2}{4\mt^2 s-t (s+t)}+\frac{2}{16 \mt^2+s}-\frac{8}{s+t}+\frac{8}{s}\right)G^2-\\
    &-\frac{2   \left(-64 \mt^4 s-16 \mt^2 (s-2 t)(s+t)+3 s t (s+t)\right)r_9 \varpi_0G}{(s+t) (t (s+t)-4 \mt^2 s)^2}- \\
    &- \frac{s  \left(4096 \mt^8 s+2048 \mt^6 \left(s^2-4 t^2\right)+128\mt^4 s \left(s^2-3 s t-11 t^2\right)\right)\varpi_0^2}{4 (16 \mt^2+s) (t (s+t)-4 \mt^2 s)^2} + \\
    &+ \frac{s  \left(32 \mt^2 s^2 t (s+2 t)+s t^2 (s+t)^2\right)\varpi_0^2}{4 (16 \mt^2+s) (t (s+t)-4 \mt^2 s)^2} \, , \\
    E_{13}^t =& -\frac{8  r_9G^3}{t  (s+t) (t (s+t)-4 \mt^2 s)\varpi_0}+  \left(-\frac{8}{s+t}-\frac{8}{t}\right)G^2+\\
    &+\frac{2   s  \left(-64\mt^4 s+64 \mt^2 t (s+t)+3 s t (s+t)\right)r_9\varpi_0G}{t (s+t) (t (s+t)-4 \mt^2 s)^2}-\frac{8 \mt^2 s^2  (16 \mt^2+s) (s+2t)}{(t (s+t)-4 \mt^2 s)^2\varpi_0^2} \, , \\
    E_{13}^{\mt^2} =& \frac{4 G^4}{ \left(s^2 \mt^2+16 s \mt^4\right)\varpi_0^2}\\
    &+\frac{2  \left(64 s \mt^4-64 t \mt^2 (s+t)-3 s t (s+t)\right)G^2}{\mt^2 \left(16 \mt^2+s\right) \left(4s \mt^2-t (s+t)\right)} - \frac{32   s  (s+2 t)r_9\varpi_0G}{\left(t (s+t)-4 s \mt^2\right)^2}+ \\
    +&\frac{s^2  \left(128 s \mt^4 \left(s^2+5 s t+5t^2\right)+4096 s \mt^8+2048 \mt^6 (s+2 t)^2+s t^2 (s+t)^2\right)\varpi_0^2}{4 \left(16 \mt^2+s\right) \left(t \mt (s+t)-4 s \mt^3\right)2} \, .
\end{align*}

\bibliographystyle{JHEP} 
\bibliography{biblio} 

\end{document}